\documentclass[aps,prb,twocolumn]{revtex4}
\usepackage{bm}
\usepackage{tabularx}
\usepackage{ascmac}
\usepackage{amsmath}
\usepackage{wrapfig}
\usepackage{graphicx}
\usepackage{float}
\usepackage[FIGBOTCAP]{subfigure}
\usepackage{color}

\bmdefine{\bolds}{s}
\bmdefine{\boldS}{S}
\bmdefine{\boldso}{so}
\bmdefine{\boldSO}{SO}
\bmdefine{\boldi}{i}
\bmdefine{\boldj}{j}
\bmdefine{\boldtau}{\tau}
\bmdefine{\boldsigma}{\sigma}
\bmdefine{\boldl}{l}
\bmdefine{\boldnabla}{\nabla}
\bmdefine{\boldlambda}{\lambda}
\bmdefine{\boldx}{x}
\bmdefine{\boldX}{X}
\bmdefine{\boldk}{k}
\bmdefine{\boldK}{K}
\bmdefine{\boldp}{p}
\bmdefine{\boldq}{q}
\bmdefine{\boldQ}{Q}
\bmdefine{\boldr}{r}
\bmdefine{\boldj}{j}
\bmdefine{\boldA}{A}
\bmdefine{\boldzero}{0}
\bmdefine{\boldone}{1}
\bmdefine{\boldtwo}{2}
\bmdefine{\boldthree}{3}

\begin{document}


\title{
Microscopic theory of Dzyaloshinsky-Moriya interaction 
in pyrochlore oxides with spin-orbit coupling
}


\author{Naoya Arakawa}
\email{arakawa@hosi.phys.s.u-tokyo.ac.jp} 
\affiliation{
Center for Emergent Matter Science (CEMS), 
RIKEN, Wako, Saitama 351-0198, Japan}


\begin{abstract}
Pyrochlore oxides 
show several fascinating phenomena, 
such as the formation of heavy fermions and the thermal Hall effect. 
Although a key to understanding some phenomena 
may be the Dzyaloshinsky-Moriya (DM) interaction, 
its microscopic origin is unclear. 
To clarify the microscopic origin, 
we constructed a $t_{2g}$-orbital model 
with the kinetic energy, the trigonal-distortion potential, 
the multiorbital Hubbard interactions, 
and the $LS$ coupling, 
and derived the low-energy effective Hamiltonian for a $d^{1}$ Mott insulator 
with the weak $LS$ coupling. 
We first show that 
lack of the inversion center of each nearest-neighbor V-V bond 
causes the odd-mirror interorbital hopping integrals. 
Those are qualitatively different from the even-mirror hopping integrals, 
existing even with the inversion center. 
We next show that 
the second-order perturbation using the kinetic terms 
leads to the ferromagnetic and the antiferromagnetic superexchange interactions, 
whose competition is controllable by tuning the Hubbard interactions.   
Then, 
we show the most important result: 
the third-order perturbation terms using 
the combination of the even-mirror hopping integral, 
the odd-mirror hopping integral, 
and the $LS$ coupling 
causes the DM interaction 
due to the mirror-mixing effect, 
where those hopping integrals 
are necessary to obtain the antisymmetric kinetic exchange 
and the $LS$ coupling is necessary to excite 
the orbital angular momentum at one of two sites. 
We also show that 
the magnitude and sign of the DM interaction can be controlled 
by changing the positions of the O ions and 
the strength of the Hubbard interactions. 
We discuss the advantages 
in comparison with the phenomenological theory and Moriya's microscopic theory, 
applicability of our mechanism, 
and the similarities and differences between 
our case and the strong-$LS$-coupling case.

\end{abstract}


\date{\today}
\maketitle


\section{Introduction}
\begin{figure}[tb]
\includegraphics[width=84mm]{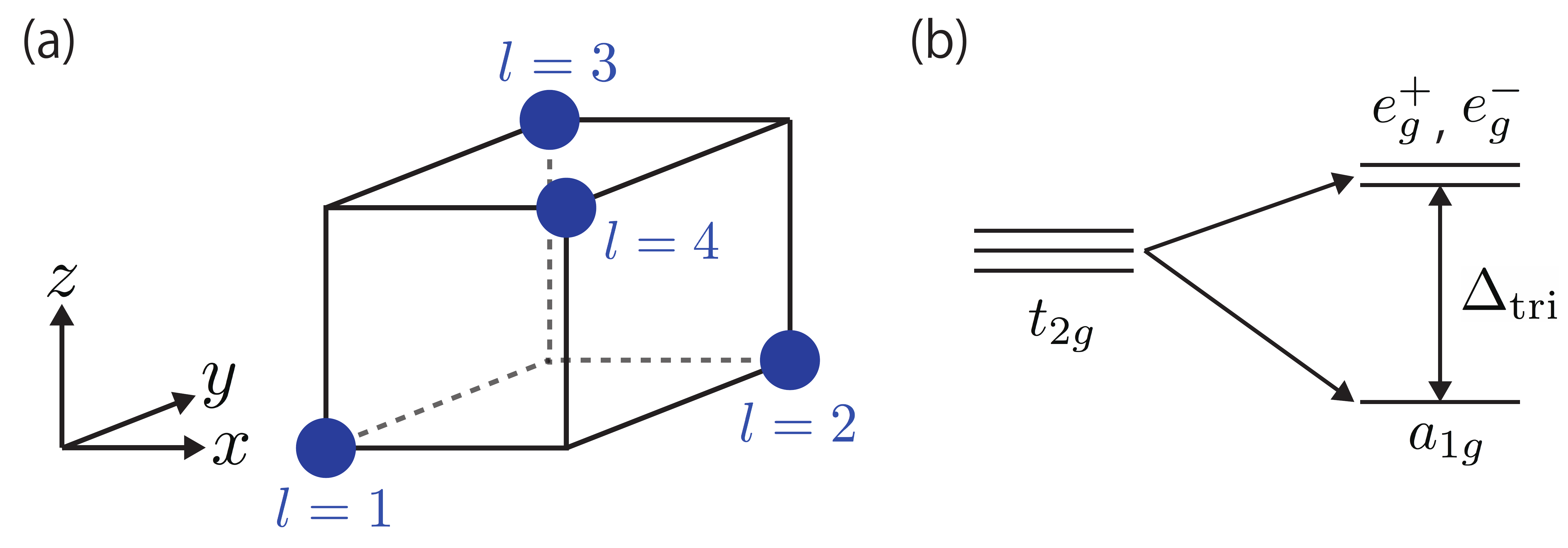}
\caption{Schematic pictures of (a) a tetrahedron of four V ions (blue circles), 
and (b) the splitting of the $t_{2g}$ orbitals 
under the trigonal-distortion potential, $\Delta_{\textrm{tri}}$.}
\label{fig1}
\end{figure}

The pyrochlore oxides~\cite{PyroReview} show several fascinating phenomena, 
and the electronic states of some pyrochlore oxides 
are categorized into 
a $t_{2g}$-orbital system with the tetrahedral sublattice structure 
under the trigonal-distortion potential. 
One of the fascinating phenomena is the formation of heavy fermions: 
in the paramagnetic metallic state of LiV$_{2}$O$_{4}$, 
the low-temperature coefficient of the electronic specific heat 
becomes about $0.42$J/mol-K$^{2}$, 
indicating the largest mass enhancement in transition-metal compounds~\cite{V-HF}. 
Another is the thermal-Hall effect: 
in the ferromagnetic (FM) Mott insulating state of Lu$_{2}$V$_{2}$O$_{7}$, 
the temperature gradient leads to the heat flow perpendicular to it~\cite{V-ThHall}. 
As an example of the pyrochlore oxides, 
let us consider the pyrochlore vanadates. 
The V ions form a network of corner-sharing tetrahedra, 
and each V ion and six O ions form a octahedron~\cite{PyroReview}; 
four V ions of a tetrahedron 
correspond to the sublattice degrees of freedom [see Fig. \ref{fig1}(a)], 
and the nearest-neighbor V ions are connected by an O ion. 
Then, 
the $t_{2g}$ orbitals of the V ions, 
i.e., the $d_{xz}$, $d_{yz}$, and $d_{xy}$ orbitals, 
give the main contributions to the bands near the Fermi level~\cite{
LDA-V124-Fujimori,LDA-V124-Anisimov,LDA-V227}. 
In addition, 
the trigonal distortion 
reduces the symmetry group around a V ion, 
and splits the $t_{2g}$ orbitals into the singlet $a_{1g}$ orbital 
and the doublet $e_{g}^{+}$ and $e_{g}^{-}$ orbitals [Fig. \ref{fig1}(b)]; 
the $a_{1g}$ orbital and the $e_{g}^{+}$ and $e_{g}^{-}$ orbitals 
correspond to the basis functions of 
the $A_{1g}$ and the $E_{g}$ irreducible representations, respectively. 

A key quantity to understand some properties of the pryrochlore oxides 
may be the Dzyaloshinsky-Moriya (DM) interaction. 
The DM interaction~\cite{DM1,DM2} has been believed to be realized 
in the pyrochlore oxides 
because of lack of the inversion center in 
each nearest-neighbor V-V bond of a tetrahedron~\cite{DMI-pyro}. 
This realization was pointed out by the phenomenological argument~\cite{DMI-pyro}: 
the authors of Ref. \onlinecite{DMI-pyro} showed 
the possible components of the DM interaction 
without the inversion center.  
This result may be correct because 
the fitting of the spin-wave dispersions obtained 
in the inelastic neutron scattering~\cite{InelaNeutron-V227} 
for Lu$_{2}$V$_{2}$O$_{7}$ suggests the finite DM interaction, 
although 
this fitting was carried out in a rough model and  
there is a controversy 
about the value of the DM interaction for Lu$_{2}$V$_{2}$O$_{7}$~\cite{LDA-LS-V227}. 
In addition, 
the results for CdCr$_{2}$O$_{4}$~\cite{InelaNeutron-Cr124,Theory-Cr124} 
and Sm$_{2}$Ir$_{2}$O$_{7}$~\cite{InelaNeutron-Ir227} 
suggest the finite DM interaction in pyrochlore oxides. 
Moreover, 
the emergence of the thermal-Hall effect may support 
the existence of the DM interaction 
because the necessity of the DM interaction 
was shown in several previous theoretical studies~\cite{V-ThHall,V-ThHall-Murakami,V-ThHallPRB}, 
although these theoretical studies neglected the sublattice degree of freedom 
in treating the DM interaction. 
 
It is important to clarify the microscopic origin of the DM interaction 
in pyrochlore oxides. 
This is because if we clarify the microscopic origin, 
we can understand the following four points. 
First, 
we can understand whether some components of the DM interaction 
are truly finite. 
That cannot be analyzed by the phenomenological argument~\cite{DM1} 
because the phenomenological theory 
just determines the permissible components 
from a symmetrical point of view. 
On the other hand, 
the microscopic theory, such as Moriya's microscopic theory~\cite{DM2}, 
determines the coefficients of the DM interaction, 
expressed in terms of the parameters of the noninteracting Hamiltonians 
and the interacting Hamiltonian. 
Second, 
the microscopic theory can clarify 
how the electronic structure of pyrochlore oxides 
leads to the DM interaction and 
how the magnitude and sign of the DM interaction are controlled by 
tuning the parameters of the system. 
Third, 
if we estimate the parameters of the noninteracting Hamiltonians and the interacting Hamiltonian 
by first-principles calculations such as the local-density approximation, 
we can determine the DM interaction appropriately including 
the material dependence of the electronic structure. 
Fourth, 
if we proceed to study the phenomena that may be related to the DM interaction, 
we can more deeply understand the physics of those phenomena. 

To clarify the microscopic origin of the DM interaction in pyrochlore oxides, 
we constructed a $t_{2g}$-orbital model with 
the appropriate treatments of the orbital degrees of freedom, 
lack of the inversion center of each V-V bond, 
the trigonal-distortion potential, 
the multiorbital electron correlation, 
and the spin-orbit coupling (SOC), 
and derived the low-energy effective Hamiltonian 
for a $d^{1}$ Mott insulator 
in the similar way for Moriya's microscopic theory~\cite{DM2}. 
We first show that 
due to lack of the inversion center, 
the indirect hopping integrals through the O $2p$ orbitals 
not only modify the values of the direct hopping integrals, 
which are even-mirror, 
but also induce the odd-mirror hopping integrals, 
which are missing only in the direct hopping integrals. 
The appearance of the odd-mirror hopping integrals is a microscopic effect 
of lack of the inversion center. 
We next show 
the FM and the antiferromagnetic (AF) superexchange interactions, 
derived by the second-order perturbation 
using two even-mirror hopping integrals or two odd-mirror hopping integrals. 
As the most important result, 
we show that 
the DM interaction arises from 
the mirror-mixing effect 
in the third-order perturbation 
using the combination of 
the even-mirror hopping integral, the odd-mirror hopping integral, 
and the $LS$ coupling. 
In this mirror-mixing effect, 
the role of those hopping integrals 
is to induce the antisymmetric kinetic exchange, 
and the role of the $LS$ coupling 
is to activate the orbital angular momentum at one of two sites. 
Those two roles are vital to get the DM-type antisymmetric exchange interactions 
in the weak SOC 
because in the nonperturbed states, 
the orbital angular momenta are quenched 
and because the combination of the antisymmetries of 
the kinetic exchange and the orbital angular momenta of two sites 
is necessary. 
In addition, from the equation of the coefficient of the DM interaction, 
we deduce how to control its magnitude and sign 
by tuning the parameters of the model. 
Then, we compare the present microscopic theory with 
the phenomenological theory~\cite{DMI-pyro} for the DM interaction in pyrochlore oxides, 
Moriya's microscopic theory~\cite{DM2}, 
and the previous microscopic theory~\cite{NA-Ir} in the strong SOC, 
and reveal the similarities and differences. 
We also argue the applicability of our mechanism 
to the DM interaction in solids with the weak SOC. 
   
In the remaining part of this paper, 
we explain how to construct the appropriate $t_{2g}$-orbital model 
for pyrochlore oxides, 
derive the low-energy effective Hamiltonian 
for a $d^{1}$ Mott insulator with the weak SOC, 
discuss the correspondences between 
our theory and several previous theories about the DM interaction 
and the applicability of our mechanism, 
and give the summary of our achievements. 
The construction of the $t_{2g}$-orbital model 
is explained in Sec. II. 
The model consists of four Hamiltonians, 
and the detail of each Hamiltonian is explained 
in each of Secs. II A, II B, II C, and II D, respectively. 
The low-energy effective Hamiltonian is derived in Sec. III. 
We derive the second-order perturbation terms in Sec. III A, 
and the third-order perturbation terms in Sec. III B. 
We also show the results of the rough estimations 
of the sign of the second-order terms, 
the ratio of the leading third-order term 
to the second-order terms, 
and the ratio of the secondary third-order term 
to the leading term 
in Sec. III A, III B 1, and III B 2, respectively. 
The correspondences with the previous theories and the applicability 
are discussed in Sec. IV. 
We summarize our results and their meanings in Sec. V.

\section{Model}
As an effective model of pyrochlore oxides, 
we introduce the following total Hamiltonian, $\hat{H}_{\textrm{tot}}$, 
with the chemical potential term, $\mu\hat{N}$: 
\begin{align}
\hat{H}_{\textrm{tot}}
-\mu\hat{N}
=\hat{H}_{\textrm{KE}}
+\hat{H}_{\textrm{tri}}
+\hat{H}_{\textrm{int}}
+\hat{H}_{LS}
-\mu\hat{N}.\label{eq:Htot}
\end{align}
Here $\hat{H}_{\textrm{KE}}$, $\hat{H}_{\textrm{tri}}$, $\hat{H}_{\textrm{int}}$, 
and $\hat{H}_{LS}$ 
represent the kinetic energy of the $t_{2g}$-orbital electrons, 
the trigonal-distortion potential, 
the $t_{2g}$-orbital Hubbard interactions,
the SOC for the $t_{2g}$-orbital electrons, respectively, 
and $\hat{N}$ is 
\begin{align}
\hat{N}=\sum\limits_{\boldi}
\sum\limits_{a=d_{xz},d_{yz},d_{xy}}
\sum\limits_{s=\uparrow,\downarrow}
\hat{c}^{\dagger}_{\boldi a s}
\hat{c}_{\boldi a s},
\end{align}
with the site index $\boldi$, 
the orbital index $a$, 
and the spin index $s$. 
$\hat{H}_{\textrm{KE}}$,  $\hat{H}_{\textrm{tri}}$, 
$\hat{H}_{\textrm{int}}$, and $\hat{H}_{LS}$ 
are explicitly shown 
in Secs. II A, II B, II C, and II D, respectively. 
$\mu$ is so determined that 
the electron number per site is $1$ for Lu$_{2}$V$_{2}$O$_{7}$ 
or $1.5$ for LiV$_{2}$O$_{4}$ for example. 

In comparison with several previous theoretical 
studies~\cite{V124-Tsune,V124-Ueda,V124-Hattori,IrPyro-Kim} 
of the pyrochlore oxides, 
a new point of this effective model 
is an appropriate treatment of lack of the inversion center of each nearest-neighbor V-V bond, 
resulting in an appropriate treatment of orbital and sublattice degrees of freedom 
(for the details see Sec. II A). 
It should be noted that 
although 
there is a previous study~\cite{V227-a1g} about 
the effects of lack of the inversion center on the hopping integrals for V ions, 
its authors did not consider 
the hopping integrals between the $t_{2g}$-orbital electrons 
arising from lack of the inversion center. 

In the following, 
we use the unit $\hbar=c=1$. 

\subsection{$\hat{H}_{\textrm{KE}}$}
To derive the kinetic energy of the $t_{2g}$-orbital electrons for pyrochlore oxides, 
we consider the nearest-neighbor hopping integrals for the V $t_{2g}$ orbitals 
and derive these by adopting the Slater-Koster method~\cite{Slater-Koster} 
to not only the direct hoppings 
but also the indirect hoppings through the O $2p$ orbitals. 
As we derive below, 
$\hat{H}_{\textrm{KE}}$ is given by
\begin{align}
\hat{H}_{\textrm{KE}}
=&\hat{H}_{0}+\hat{H}_{\textrm{odd}}\notag\\
=&\sum\limits_{<\boldi,\boldj>}
\sum\limits_{a,b=d_{xz},d_{yz},d_{xy}}
\sum\limits_{s=\uparrow, \downarrow}
t_{\boldi \boldj;ab}^{(\textrm{even})}
\hat{c}^{\dagger}_{\boldi a s}
\hat{c}_{\boldj b s}\notag\\
+
\sum\limits_{<\boldi,\boldj>}&
\sum\limits_{a,b=d_{xz},d_{yz},d_{xy}}
\sum\limits_{s=\uparrow, \downarrow}
t_{\boldi \boldj;ab}^{(\textrm{odd})}
\hat{c}^{\dagger}_{\boldi a s}
\hat{c}_{\boldj b s},\label{eq:HKE}
\end{align}
where 
$\textstyle\sum_{<\boldi,\boldj>}$ represents 
the summation between the nearest-neighbor sites, 
$\hat{H}_{0}$ represents 
the even-mirror hopping integrals, $t_{\boldi \boldj;ab}^{(\textrm{even})}$, 
existing even with the inversion center of each nearest-neighbor V-V bond, 
and 
$\hat{H}_{\textrm{odd}}$ represents 
the odd-mirror hopping integrals, $t_{\boldi \boldj;ab}^{(\textrm{odd})}$, 
appearing only 
with lack of the inversion center. 
The word ``even-mirror'' or ``odd-mirror'' mean that 
the mirror symmetry about the plane including the nearest-neighbor sites 
is even or odd, respectively; 
e.g., 
for a $xy$ plane including sublattices 1 and 2, 
the hopping integral between the $d_{xy}$ orbitals 
is even about the mirror symmetry of the $xy$ plane 
(i.e., $x \rightarrow x$, $y\rightarrow y$, and $z\rightarrow -z$) 
and the hopping integral between the $d_{xz}$ and $d_{xy}$ orbitals is odd 
because the former and latter behave like $xy\times xy\propto x^{2}y^{2}z^{0}$ 
and $xz\times xy\propto x^{2}y^{1}z^{1}$, respectively. 

We first derive the contributions from the direct hoppings 
between nearest-neighbor V ions. 
We derive only the hopping integrals between V ions 
at sublattices $1$ and $2$ [i.e. $l=1$ and $2$ in Fig. \ref{fig1}(a)]
because the others can be obtained from these 
by permuting the coordinates $x$, $y$, and $z$, defined in Fig. \ref{fig1}(a). 
For example, 
the direct hopping integrals between sublattices $1$ and $3$ 
are obtained by replacing $x$, $y$, and $z$ 
in the direct hopping integrals between sublattices $1$ and $2$ 
by $y$, $z$, and $x$, respectively, 
e.g., $t_{d_{xz}3d_{yz}1}^{(\textrm{direct})}
=t_{d_{yz}2 d_{xy}1}^{(\textrm{direct})}$. 
Adopting the Slater-Koster method~\cite{Slater-Koster} 
to the direct hopping integrals 
between V ions at sublattices $1$ and $2$, 
which are located at $(x,y,z)=(0,0,0)$ and $(1,1,0)$, respectively, 
we can express the hopping integrals for the $t_{2g}$-orbital electrons 
in terms of three Slater-Koster parameters, 
$V_{dd\sigma}$, $V_{dd\pi}$, and $V_{dd\delta}$: 
\begin{align}
&t_{d_{xy}2 d_{xy}1}^{(\textrm{direct})}=\frac{3}{4}V_{dd\sigma}+\frac{1}{4}V_{dd\delta},\label{eq:tdirect1}\\
&t_{d_{xz}2 d_{xz}1}^{(\textrm{direct})}=\frac{1}{2}V_{dd\pi}+\frac{1}{2}V_{dd\delta},\label{eq:tdirect2}\\
&t_{d_{yz}2 d_{yz}1}^{(\textrm{direct})}=t_{d_{xz}2 d_{xz}1}^{(\textrm{direct})},\label{eq:tdirect3}\\
&t_{d_{xy}2 d_{yz}1}^{(\textrm{direct})}=0,\label{eq:tdirect4}\\
&t_{d_{xy}2 d_{xz}1}^{(\textrm{direct})}=0,\label{eq:tdirect5}\\
&t_{d_{yz}2 d_{xz}1}^{(\textrm{direct})}=\frac{1}{2}V_{dd\pi}-\frac{1}{2}V_{dd\delta},\label{eq:tdirect6}\\
&t_{d_{xz}2 d_{yz}1}^{(\textrm{direct})}=t_{d_{yz}2 d_{xz}1}^{(\textrm{direct})},\label{eq:tdirect7}\\
&t_{d_{xz}2 d_{xy}1}^{(\textrm{direct})}=0,\label{eq:tdirect8}\\
&t_{d_{yz}2 d_{xy}1}^{(\textrm{direct})}=0.\label{eq:tdirect9}
\end{align}
Since all the finite direct hopping integrals are 
even about the mirror symmetry of the $xy$ plane, 
the direct hoppings lead only to the even-mirror hopping integrals. 

\begin{figure}[tb]
\includegraphics[width=80mm]{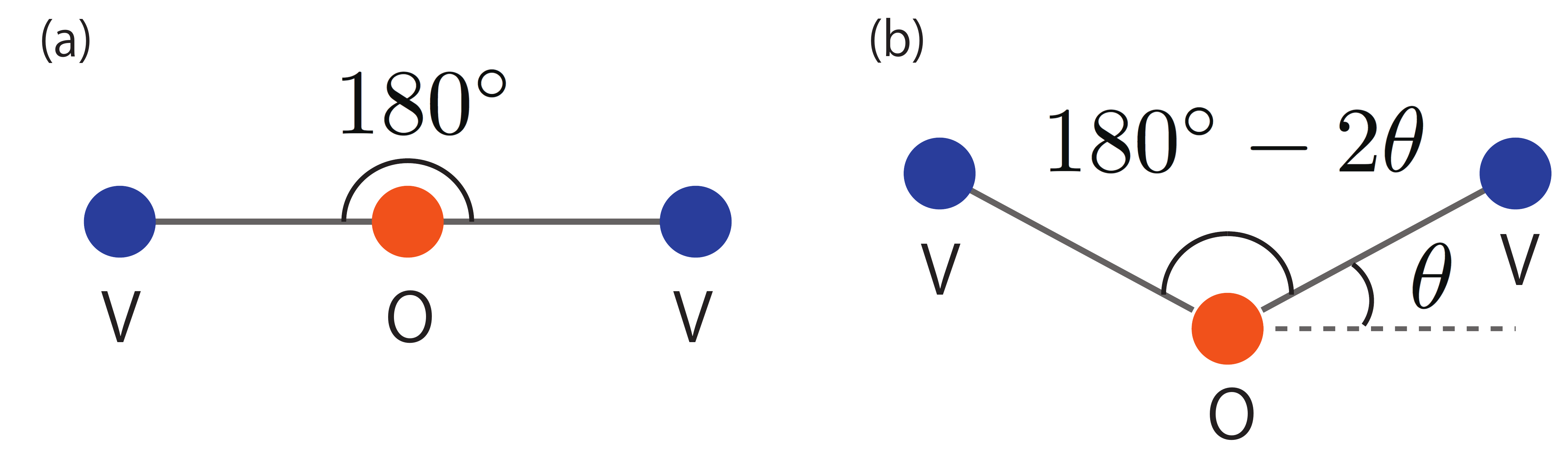}
\caption{Schematic pictures of a V-O-V bond 
(a) with and (b) without the inversion center. 
Blue and orange circles denote V and O ions, respectively. 
This angle deviation breaks 
the mirror symmetry about the plane including the V ions.}
\label{fig2}
\end{figure}
We next derive the contributions from the indirect hoppings through the O $2p$ orbitals. 
As we will show below, 
if we appropriately treat the effect of lack of the inversion center of 
each nearest-neighbor V-V bond, 
the indirect hoppings lead to the odd-mirror hopping integrals, 
which are missing in the direct hoppings. 
The position of the O ion of each V-O-V bond deviates from the center; 
i.e., each V-O-V angle deviates from $180^{\circ}$ 
[compare Figs. \ref{fig2}(a) and \ref{fig2}(b)]. 
In addition, 
this deviation results in the four-sublattice structure 
of a tetrahedron 
because V ions at sublattices $1$, $2$, $3$, and $4$ 
are different from each other; 
for example, 
if we compare the lower layer of a tetrahedron 
including sublattices $1$ and $2$ 
with the upper layer including sublattices $3$ and $4$, 
we see the difference between 
the V ions in the lower layer and the upper layer 
because the O ion for V ions of sublattices $1$ and $2$ 
(sublattices $3$ and $4$) is located below the lower layer (above the upper layer).   
To show that 
the indirect hoppings lead to the odd-mirror hopping integrals 
only with lack of the inversion center of each nearest-neighbor V-V bond, 
we consider the case in which each V-O-V angle slightly deviates from $180^{\circ}$, 
i.e., the angle is $180^{\circ}-2\theta$ [see Fig. \ref{fig2}(b)], 
and derive the indirect hopping integrals between V ions 
at sublattices $1$ and $2$ through the O ion. 
Because of the same reason for the derivation of the direct hopping integrals, 
we derive only the hopping integrals between sublattices $1$ and $2$. 
Since the indirect nearest-neighbor hopping integrals, $t_{a 2 b 1}^{(\textrm{indirect})}$, 
are given by 
\begin{align}
t_{a 2 b 1}^{(\textrm{indirect})}
=\sum\limits_{A=p_{x},p_{y},p_{z}}
\frac{V_{a2 A}V_{A b1}}{\Delta_{pd}},\label{eq:IndirectHop}
\end{align}
we need to derive the hybridizations between V $t_{2g}$ orbitals and O $2p$ orbitals, 
$V_{a2 A}$ and $V_{A b1}$, 
for a V-O-V bond including sublattices $1$ and $2$ 
by using the Slater-Koster method~\cite{Slater-Koster}. 
In Eq. (\ref{eq:IndirectHop}), 
we have neglected the orbital and $\theta$ dependence of $\Delta_{pd}$, 
the crystalline-electric-field energy difference 
between V $t_{2g}$ orbitals and O $2p$ orbitals, 
because this simplification is sufficient for our purpose, 
i.e., showing the appearance of the odd-mirror hopping integrals. 
$V_{A b1}$ are obtained by adopting the Slater-Koster method to 
the indirect hopping processes between the V ion at $(x,y,z)=(0,0,0)$ 
and the O ion at $(x,y,z)=(\frac{1}{2},\frac{1}{2},-\frac{1}{2}\tan \theta)$ 
with two Slater-Koster parameters, $V_{pd\sigma}$ and $V_{pd\pi}$: 
\begin{align}
&V_{p_{x} d_{xy}1}
=\frac{\sqrt{3}}{2\sqrt{2}}V_{pd\sigma},\label{eq:Vpd1}\\
&V_{p_{x} d_{yz}1}
=-\frac{\sqrt{3}}{2\sqrt{2}}\theta V_{pd\sigma}
+\frac{1}{\sqrt{2}}\theta V_{pd\pi},\label{eq:Vpd2}\\
&V_{p_{x} d_{xz}1}
=-\frac{\sqrt{3}}{2\sqrt{2}}\theta V_{pd\sigma},\label{eq:Vpd3}\\
&V_{p_{y} d_{xz}1}
=-\frac{\sqrt{3}}{2\sqrt{2}}\theta V_{pd\sigma}
+\frac{1}{\sqrt{2}}\theta V_{pd\pi},\label{eq:Vpd4}\\
&V_{p_{y} d_{yz}1}
=-\frac{\sqrt{3}}{2\sqrt{2}}\theta V_{pd\sigma},\label{eq:Vpd5}\\
&V_{p_{y} d_{xy}1}
=\frac{\sqrt{3}}{2\sqrt{2}}V_{pd\sigma},\label{eq:Vpd6}\\
&V_{p_{z} d_{xz}1}
=\frac{1}{\sqrt{2}}V_{pd\pi},\label{eq:Vpd7}\\
&V_{p_{z} d_{yz}1}
=\frac{1}{\sqrt{2}}V_{pd\pi},\label{eq:Vpd8}\\
&V_{p_{z} d_{xy}1}
=-\frac{\sqrt{3}}{2\sqrt{2}}\theta V_{pd\sigma}
+\frac{1}{\sqrt{2}}\theta V_{pd\pi},\label{eq:Vpd9}
\end{align}
where the $O(\theta^{2})$ terms are neglected. 
Similarly, we obtain $V_{a2 A}$:
\begin{align}
&V_{d_{xy}2 p_{x}}=-V_{p_{x} d_{xy}1},\label{eq:Vdp1}\\
&V_{d_{yz}2 p_{x}}=V_{p_{x} d_{yz}1},\label{eq:Vdp2}\\
&V_{d_{xz}2 p_{x}}=V_{p_{x} d_{xz}1},\label{eq:Vdp3}\\
&V_{d_{xz}2 p_{y}}=V_{p_{y} d_{xz}1},\label{eq:Vdp4}\\
&V_{d_{yz}2 p_{y}}=V_{p_{y} d_{yz}1},\label{eq:Vdp5}\\
&V_{d_{xy}2 p_{y}}=-V_{p_{y} d_{xy}1},\label{eq:Vdp6}\\
&V_{d_{xz}2 p_{z}}=-V_{p_{z} d_{xz}1},\label{eq:Vdp7}\\
&V_{d_{yz}2 p_{z}}=-V_{p_{z} d_{yz}1},\label{eq:Vdp8}\\
&V_{d_{xy}2 p_{z}}=V_{p_{z} d_{xy}1}.\label{eq:Vdp9}
\end{align}
Combining Eqs. (\ref{eq:Vpd1}){--}(\ref{eq:Vdp9}) with Eq. (\ref{eq:IndirectHop}), 
we obtain the indirect hopping integrals within the $O(\theta^{2})$ terms: 
\begin{align}
&t_{d_{xy}2 d_{xy}1}^{(\textrm{indirect})}
=-\frac{3V_{pd\sigma}^{2}}{4\Delta_{pd}},\label{eq:tind1}\\
&t_{d_{xz}2 d_{xz}1}^{(\textrm{indirect})}
=-\frac{V_{pd\pi}^{2}}{2\Delta_{pd}},\label{eq:tind2}\\
&t_{d_{yz}2 d_{yz}1}^{(\textrm{indirect})}
=t_{d_{xz}2 d_{xz}1}^{(\textrm{indirect})},\label{eq:tind3}\\
&t_{d_{xy}2 d_{yz}1}^{(\textrm{indirect})}
=
\theta\frac{3V_{pd\sigma}^{2}-2\sqrt{3}V_{pd\sigma}V_{pd\pi}+2V_{pd\pi}^{2}}{4\Delta_{pd}},\label{eq:tind4}\\
&t_{d_{xy}2 d_{xz}1}^{(\textrm{indirect})}
=t_{d_{xy}2 d_{yz}1}^{(\textrm{indirect})},\label{eq:tind5}\\
&t_{d_{yz}2 d_{xz}1}^{(\textrm{indirect})}
=-\frac{V_{pd\pi}^{2}}{2\Delta_{pd}},\label{eq:tind6}\\
&t_{d_{xz}2 d_{yz}1}^{(\textrm{indirect})}
=t_{d_{yz}2 d_{xz}1}^{(\textrm{indirect})},\label{eq:tind7}\\
&t_{d_{xz}2 d_{xy}1}^{(\textrm{indirect})}
=-t_{d_{xy}2 d_{xz}1}^{(\textrm{indirect})},\label{eq:tind8}\\
&t_{d_{yz}2 d_{xy}1}^{(\textrm{indirect})}
=-t_{d_{xy}2 d_{yz}1}^{(\textrm{indirect})}.\label{eq:tind9}
\end{align}
Those show that 
the deviation of the V-O-V angle from $180^{\circ}$ leads to 
$t_{d_{xy}2 d_{yz}1}^{(\textrm{indirect})}$, $t_{d_{xy}2 d_{xz}1}^{(\textrm{indirect})}$, 
$t_{d_{yz}2 d_{xy}1}^{(\textrm{indirect})}$, and $t_{d_{xz}2 d_{xy}1}^{(\textrm{indirect})}$ 
even if the deviation angle $\theta$ is very small. 
The most important difference between 
those hopping integrals and the others, 
existing even for $\theta=0^{\circ}$, 
is the mirror symmetry about the $xy$ plane including sublattices 1 and 2: 
those are odd-mirror, while the others are even-mirror.  
From a symmetrical point of view, 
the odd-mirror hopping integrals are permissible only 
without the inversion center 
because the mirror mixing occurs only without the inversion center. 
Although for not small $\theta$, the deviation of the V-O-V angle affects 
not only $t_{d_{xy}2 d_{yz}1}^{(\textrm{indirect})}$, $t_{d_{xy}2 d_{xz}1}^{(\textrm{indirect})}$, 
$t_{d_{yz}2 d_{xy}1}^{(\textrm{indirect})}$, and $t_{d_{xz}2 d_{xy}1}^{(\textrm{indirect})}$ 
but also the other indirect ones, 
for the higher-order effects of $\theta$ the symmetry of the hopping integrals remains the same. 
Thus, the above derivation is sufficient to show the appearance 
of the odd-mirror hopping integrals 
due to the indirect hoppings through the O $2p$ orbitals 
under the deviation of the V-O-V angle from $180^{\circ}$. 

\begin{figure}[tb]
\includegraphics[width=84mm]{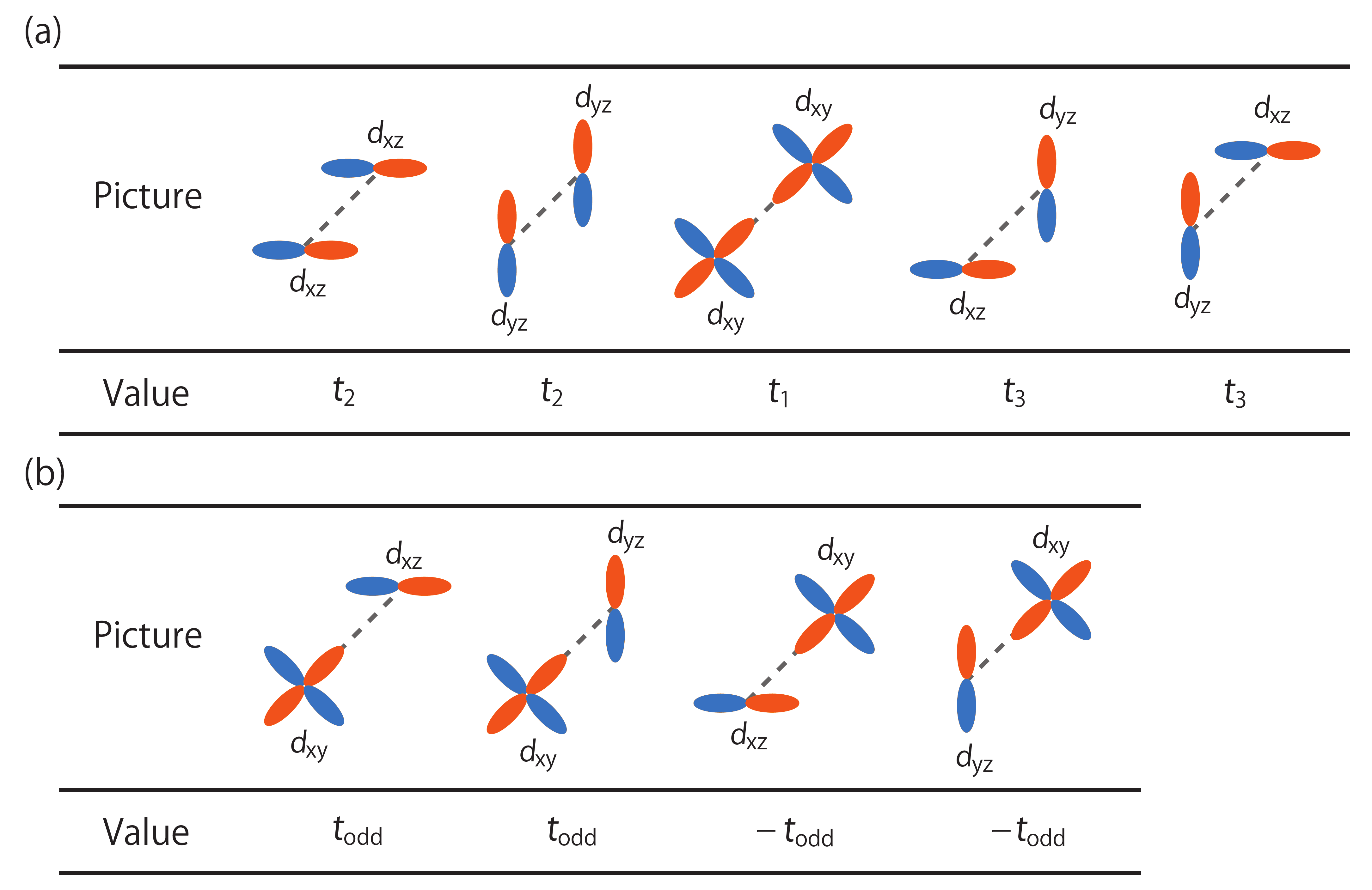}
\caption{
Schematic pictures of (a) the even-mirror hoppings of $\hat{H}_{0}$ 
and (b) the odd-mirror hoppings of $\hat{H}_{\textrm{odd}}$ 
for the nearest-neighbor V ions 
between sublatices $1$ and $2$ and the corresponding values. 
The color differences represent the sign differences of the wave function. 
}
\label{fig3}
\end{figure}

By combining the direct hopping integrals and the indirect ones, 
the kinetic energy of the $t_{2g}$-orbital electrons for pyrochlore oxides 
can be described by three parameters of $\hat{H}_{0}$, 
i.e., $t_{1}$, $t_{2}$, and $t_{3}$, 
and one parameter of $\hat{H}_{\textrm{odd}}$, i.e., $t_{\textrm{odd}}$. 
These parameters are related to the nearest-neighbor hopping integrals 
for the $t_{2g}$-orbital electrons. 
For example, 
those relations for the plane including sublattices $1$ and $2$ are given by 
\begin{align}
&t_{1}
=t_{d_{xy}2 d_{xy}1}^{(\textrm{direct})}+t_{d_{xy}2 d_{xy}1}^{(\textrm{indirect})},\label{eq:t1}\\
&t_{2}
=t_{d_{xz}2 d_{xz}1}^{(\textrm{direct})}+t_{d_{xz}2 d_{xz}1}^{(\textrm{indirect})}
=t_{d_{yz}2 d_{yz}1}^{(\textrm{direct})}+t_{d_{yz}2 d_{yz}1}^{(\textrm{indirect})},\label{eq:t2}\\
&t_{3}
=t_{d_{yz}2 d_{xz}1}^{(\textrm{direct})}+t_{d_{yz}2 d_{xz}1}^{(\textrm{indirect})}
=t_{d_{xz}2 d_{yz}1}^{(\textrm{direct})}+t_{d_{xz}2 d_{yz}1}^{(\textrm{indirect})},\label{eq:t3}\\
&t_{\textrm{odd}}
=t_{d_{xz}2 d_{xy}1}^{(\textrm{indirect})}
=t_{d_{yz}2 d_{xy}1}^{(\textrm{indirect})}\notag\\
&\ \ \ \ \
=-t_{d_{xy}2 d_{xz}1}^{(\textrm{indirect})}
=-t_{d_{xy}2 d_{yz}1}^{(\textrm{indirect})}.\label{eq:tISB}
\end{align}
Those relations are more easily seen from Fig. \ref{fig3}. 
By choosing $t_{1}$, $t_{2}$, $t_{3}$, and $t_{\textrm{odd}}$ appropriately, 
the Hamiltonian of Eq. (\ref{eq:HKE}) 
can describe the kinetic energy for any pyrochlore oxides 
as long as the main orbitals for low-energy excitations are the $t_{2g}$ orbitals. 

\subsection{$\hat{H}_{\textrm{tri}}$}
The effect of the trigonal distortion 
on the $t_{2g}$-orbital electrons 
can be described by $\hat{H}_{\textrm{tri}}$, 
\begin{align}
&\hat{H}_{\textrm{tri}}=
-\frac{\Delta_{\textrm{tri}}}{3}
\sum\limits_{\boldi}
\sum\limits_{a=d_{xz},d_{yz},d_{xy}}
\sum\limits_{b\neq a}
\sum\limits_{s=\uparrow,\downarrow}
\hat{c}^{\dagger}_{\boldi a s}\hat{c}_{\boldi b s}.\label{eq:Htri-t2g}
\end{align}
By diagonalizing $\hat{H}_{\textrm{tri}}$, 
we rewrite $\hat{H}_{\textrm{tri}}$ as 
\begin{align}
&\hat{H}_{\textrm{tri}}
=
-\frac{2\Delta_{\textrm{tri}}}{3}
\sum\limits_{\boldi}
\sum\limits_{s=\uparrow,\downarrow}
\hat{c}^{\dagger}_{\boldi a_{1g} s}\hat{c}_{\boldi a_{1g} s}\notag\\
&+\frac{\Delta_{\textrm{tri}}}{3}
\sum\limits_{\boldi}
\sum\limits_{s=\uparrow,\downarrow}
(\hat{c}^{\dagger}_{\boldi e_{g}^{+} s}\hat{c}_{\boldi e_{g}^{+} s}
+\hat{c}^{\dagger}_{\boldi e_{g}^{-} s}\hat{c}_{\boldi e_{g}^{-} s}),
\end{align}
where 
\begin{align}
&\hat{c}_{\boldi a_{1g} s}^{\dagger}
=\frac{1}{\sqrt{3}}
(\hat{c}_{\boldi d_{xz} s}^{\dagger}
+\hat{c}_{\boldi d_{yz} s}^{\dagger}
+\hat{c}_{\boldi d_{xy} s}^{\dagger}),\label{eq:c+a1g}\\
&\hat{c}_{\boldi e_{g}^{+} s}^{\dagger}
=\frac{1}{\sqrt{3}}
(\hat{c}_{\boldi d_{xz} s}^{\dagger}
+\omega\hat{c}_{\boldi d_{yz} s}^{\dagger}
+\omega^{2}\hat{c}_{\boldi d_{xy} s}^{\dagger}),\label{eq:c+eg+}\\
&\hat{c}_{\boldi e_{g}^{-} s}^{\dagger}
=\frac{1}{\sqrt{3}}
(\hat{c}_{\boldi d_{xz} s}^{\dagger}
+\omega^{2}\hat{c}_{\boldi d_{yz} s}^{\dagger}
+\omega\hat{c}_{\boldi d_{xy} s}^{\dagger}),\label{eq:c+eg-}
\end{align}
and 
\begin{align}
&\hat{c}_{\boldi a_{1g} s}
=\frac{1}{\sqrt{3}}
(\hat{c}_{\boldi d_{xz} s}
+\hat{c}_{\boldi d_{yz} s}
+\hat{c}_{\boldi d_{xy} s}),\label{eq:ca1g}\\
&\hat{c}_{\boldi e_{g}^{+} s}
=\frac{1}{\sqrt{3}}
(\hat{c}_{\boldi d_{xz} s}
+\omega^{2}\hat{c}_{\boldi d_{yz} s}
+\omega\hat{c}_{\boldi d_{xy} s}),\label{eq:ceg+}\\
&\hat{c}_{\boldi e_{g}^{-} s}
=\frac{1}{\sqrt{3}}
(\hat{c}_{\boldi d_{xz} s}
+\omega\hat{c}_{\boldi d_{yz} s}
+\omega^{2}\hat{c}_{\boldi d_{xy} s}),\label{eq:ceg-}
\end{align}
with $\omega=e^{-\frac{2\pi}{3}i}=-\frac{1}{2}-i\frac{\sqrt{3}}{2}$ 
and $\omega^{2}=e^{-\frac{4\pi}{3}i}=-\frac{1}{2}+i\frac{\sqrt{3}}{2}$. 
Thus, 
the effect of the trigonal distortion is splitting 
the energy level of the $t_{2g}$-orbital electron 
into the singlet $a_{1g}$ orbital 
and the doublet $e_{g}^{+}$ and $e_{g}^{-}$ orbitals, 
as shown in Fig. \ref{fig1}(b). 

Since the $a_{1g}$ orbital is the lower state 
for LiV$_{2}$O$_{4}$ and Lu$_{2}$V$_{2}$O$_{7}$ at low temperature, 
$\Delta_{\textrm{tri}} > 0$ is realized for these pyrochlore vanadates. 
In principle, 
we can switch the lower state 
from the $a_{1g}$ orbital to the $e_{g}^{+}$ and $e_{g}^{-}$ orbitals 
by controlling the trigonal distortion. 

If we express $\hat{H}_{0}$ and $\hat{H}_{\textrm{odd}}$ 
in terms of the creation and annihilation operators 
for the $a_{1g}$, $e_{g}^{+}$, and $e_{g}^{-}$ orbitals, 
we find the important difference between 
the even-mirror and the odd-mirror hopping integrals. 
Namely, 
the even-mirror hopping integrals for the $t_{2g}$ orbitals 
are expressed as 
the intraorbital and the interorbital hopping integrals 
for the $a_{1g}$, $e_{g}^{+}$, and $e_{g}^{-}$ orbitals, 
while 
the odd-mirror hopping integrals for the $t_{2g}$ orbitals 
are expressed as the interorbital hopping integrals 
for the $a_{1g}$, $e_{g}^{+}$, and $e_{g}^{-}$ orbitals. 
Those properties for the even-mirror and the odd-mirror hopping integrals 
are seen from, respectively, 
the intraorbital hopping integrals for the $d_{xz}$ orbital 
between sublattice $2$ at $\boldi=\boldtwo$ 
and subltattice $1$ at $\boldj=\boldone$,  
\begin{align}
&t_{2}\hat{c}_{\boldtwo d_{xz}s}^{\dagger}\hat{c}_{\boldone d_{xz}s}\notag\\
=&\frac{t_{2}}{3}
(\hat{c}^{\dagger}_{\boldtwo a_{1g}s}
+\hat{c}^{\dagger}_{\boldtwo e_{g}^{+}s}
+\hat{c}^{\dagger}_{\boldtwo e_{g}^{+}s})
(\hat{c}_{\boldone a_{1g}s}
+\hat{c}_{\boldone e_{g}^{+}s}
+\hat{c}_{\boldone e_{g}^{+}s}),
\end{align}
and 
the odd-mirror hopping integrals between sublattice $2$ at $\boldi=\boldtwo$ 
and subltattice $1$ at $\boldj=\boldone$,  
\begin{align}
&t_{\textrm{odd}}
(\hat{c}_{\boldtwo d_{yz}s}^{\dagger}\hat{c}_{\boldone d_{xy}s}
+\hat{c}_{\boldtwo d_{xz}s}^{\dagger}\hat{c}_{\boldone d_{xy}s}
-\hat{c}_{\boldtwo d_{xy}s}^{\dagger}\hat{c}_{\boldone d_{yz}s}
-\hat{c}_{\boldtwo d_{xy}s}^{\dagger}\hat{c}_{\boldone d_{xz}s})\notag\\
&=
t_{\textrm{odd}}
(
\omega^{2}\hat{c}^{\dagger}_{\boldtwo a_{1g}s}\hat{c}_{\boldone e_{g}^{+}s}
+\omega\hat{c}^{\dagger}_{\boldtwo a_{1g}s}\hat{c}_{\boldone e_{g}^{-}s}
-\omega\hat{c}^{\dagger}_{\boldtwo e_{g}^{+}s}\hat{c}_{\boldone a_{1g}s}\notag\\
&-\omega^{2}\hat{c}^{\dagger}_{\boldtwo e_{g}^{-}s}\hat{c}_{\boldone a_{1g}s}
).
\end{align}
Understanding this difference 
is useful to understand the physical meaning of the finite contributions 
in the low-energy effective Hamiltonian (e.g., see Figs. \ref{fig4} and \ref{fig5}). 

\subsection{$\hat{H}_{\textrm{int}}$}
$\hat{H}_{\textrm{int}}$ is given by four multiorbital Hubbard interactions, 
$U$, $U^{\prime}$, $J_{\textrm{H}}$, and $J^{\prime}$: 
\begin{align}
&\hat{H}_{\textrm{int}}
=
U\sum\limits_{\boldi}
\sum\limits_{a=d_{xz},d_{yz},d_{xy}}
\hat{c}_{\boldi a  \uparrow}^{\dagger}\hat{c}_{\boldi a  \downarrow}^{\dagger}
\hat{c}_{\boldi a  \downarrow}\hat{c}_{\boldi a  \uparrow}\notag\\
&+
U^{\prime}\sum\limits_{\boldi}
\sum\limits_{a=d_{xz},d_{yz},d_{xy}}
\sum\limits_{b >a}
\sum\limits_{s,s^{\prime}=\uparrow,\downarrow}
\hat{c}_{\boldi a  s}^{\dagger}\hat{c}_{\boldi b  s^{\prime}}^{\dagger}
\hat{c}_{\boldi b  s^{\prime}}\hat{c}_{\boldi a  s}\notag\\
&-
J_{\textrm{H}}
\sum\limits_{\boldi}
\sum\limits_{a=d_{xz},d_{yz},d_{xy}}
\sum\limits_{b >a}
\sum\limits_{s,s^{\prime}=\uparrow,\downarrow}
\hat{c}_{\boldi a  s}^{\dagger}\hat{c}_{\boldi b  s^{\prime}}^{\dagger}
\hat{c}_{\boldi b  s}\hat{c}_{\boldi a  s^{\prime}}\notag\\
&+
J^{\prime}\sum\limits_{\boldi}
\sum\limits_{a=d_{xz},d_{yz},d_{xy}}
\sum\limits_{b \neq a}
\hat{c}_{\boldi a  \uparrow}^{\dagger}\hat{c}_{\boldi a  \downarrow}^{\dagger}
\hat{c}_{\boldi b  \downarrow}\hat{c}_{\boldi b  \uparrow},\label{eq:Hint}
\end{align}
where $\textstyle\sum_{b > a}$ represents the restricted summation, 
$\textstyle\sum_{b > a}=\textstyle\sum_{b=d_{yz},d_{xy}}$ or $\textstyle\sum_{b=d_{xy}}$ or 
$0$ for $a=d_{xz}$ or $d_{yz}$ or $d_{xy}$, respectively. 
These four parameters reduce to two, $U$ and $J_{\textrm{H}}$, 
if we use $J^{\prime}=J_{\textrm{H}}$ and $U^{\prime}=U-2J_{\textrm{H}}$. 

For the derivations of the low-energy effective Hamiltonian of 
the $d^{1}$ Mott insulator in Sec. III, 
we rewrite $\hat{H}_{\textrm{int}}$ in terms of the irreducible representations 
of the $d^{2}$ states, where two electrons exist per site: 
\begin{align}
\hat{H}_{\textrm{int}}
=\sum\limits_{\boldi}
\sum\limits_{\Gamma}\sum\limits_{g_{\Gamma}}
U_{\Gamma}|\boldi;\Gamma,g_{\Gamma}\rangle 
\langle \boldi;\Gamma,g_{\Gamma}|.\label{eq:Hint-irrep}
\end{align}
Here $\Gamma$ represents the irreducible representations, 
$g_{\Gamma}$ represents the degeneracy, 
$U_{\Gamma}$ are given by 
\begin{align}
&U_{A_{1}}=U+2J^{\prime},\label{eq:U-Irrep1}\\
&U_{E}=U-J^{\prime},\label{eq:U-Irrep2}\\
&U_{T_{1}}=U^{\prime}-J_{\textrm{H}},\label{eq:U-Irrep3}\\
&U_{T_{2}}=U^{\prime}+J_{\textrm{H}},\label{eq:U-Irrep4}
\end{align}
and 
$|\boldi;\Gamma,g_{\Gamma}\rangle$ are 
$|\boldi;\Gamma,g_{\Gamma}\rangle=\hat{X}_{\boldi \Gamma g_{\Gamma}}^{\dagger}|0,0\rangle$ 
with 
\begin{align}
&
\hat{X}_{\boldi A_{1}}^{\dagger}
=
\frac{1}{\sqrt{3}}
(\hat{c}^{\dagger}_{\boldi d_{xz}\uparrow}\hat{c}^{\dagger}_{\boldi d_{xz}\downarrow}
+\hat{c}^{\dagger}_{\boldi d_{yz}\uparrow}\hat{c}^{\dagger}_{\boldi d_{yz}\downarrow}
+\hat{c}^{\dagger}_{\boldi d_{xy}\uparrow}\hat{c}^{\dagger}_{\boldi d_{xy}\downarrow})
,\label{eq:basis-Irrep1}\\
&
\hat{X}_{\boldi E u}^{\dagger}
=
\sqrt{\frac{2}{3}}
(-\hat{c}^{\dagger}_{\boldi d_{xz}\uparrow}\hat{c}^{\dagger}_{\boldi d_{xz}\downarrow}
+\frac{1}{2}\hat{c}^{\dagger}_{\boldi d_{yz}\uparrow}\hat{c}^{\dagger}_{\boldi d_{yz}\downarrow}
+\frac{1}{2}\hat{c}^{\dagger}_{\boldi d_{xy}\uparrow}\hat{c}^{\dagger}_{\boldi d_{xy}\downarrow})
,\label{eq:basis-Irrep2}\\
&
\hat{X}_{\boldi  E v}^{\dagger}
=
\frac{1}{\sqrt{2}}
(\hat{c}^{\dagger}_{\boldi d_{yz}\uparrow}\hat{c}^{\dagger}_{\boldi d_{yz}\downarrow}
-\hat{c}^{\dagger}_{\boldi d_{xy}\uparrow}\hat{c}^{\dagger}_{\boldi d_{xy}\downarrow}),\label{eq:basis-Irrep3}\\
&
\hat{X}_{\boldi  T_{1} \zeta_{+}}^{\dagger}
=
\hat{c}^{\dagger}_{\boldi d_{xz}\uparrow}\hat{c}^{\dagger}_{\boldi d_{yz}\uparrow},\label{eq:basis-Irrep4}\\
&
\hat{X}_{\boldi T_{1} \zeta_{-}}^{\dagger}
=
\hat{c}^{\dagger}_{\boldi d_{xz}\downarrow}\hat{c}^{\dagger}_{\boldi d_{yz}\downarrow},\label{eq:basis-Irrep5}\\
&
\hat{X}_{\boldi T_{1} \zeta_{0}}^{\dagger}
=
\frac{1}{\sqrt{2}}
(\hat{c}^{\dagger}_{\boldi d_{xz} \uparrow}\hat{c}^{\dagger}_{\boldi d_{yz}\downarrow}
+\hat{c}^{\dagger}_{\boldi d_{xz}\downarrow}\hat{c}^{\dagger}_{\boldi d_{yz}\uparrow}),\label{eq:basis-Irrep6}\\
&
\hat{X}_{\boldi T_{2} \zeta_{0}}^{\dagger}
=
\frac{1}{\sqrt{2}}
(\hat{c}^{\dagger}_{\boldi d_{xz}\uparrow}\hat{c}^{\dagger}_{\boldi d_{yz}\downarrow}
-\hat{c}^{\dagger}_{\boldi d_{xz}\downarrow}\hat{c}^{\dagger}_{\boldi d_{yz}\uparrow}),\label{eq:basis-Irrep7}\\
&
\hat{X}_{\boldi T_{1} \xi_{+}}^{\dagger}
=
\hat{c}^{\dagger}_{\boldi d_{xz}\uparrow}\hat{c}^{\dagger}_{\boldi d_{xy}\uparrow},\label{eq:basis-Irrep8}\\
&
\hat{X}_{\boldi T_{1} \xi_{-}}^{\dagger}
=
\hat{c}^{\dagger}_{\boldi d_{xz}\downarrow}\hat{c}^{\dagger}_{\boldi d_{xy}\downarrow},\label{eq:basis-Irrep9}\\
&
\hat{X}_{\boldi T_{1} \xi_{0}}^{\dagger}
=
\frac{1}{\sqrt{2}}
(\hat{c}^{\dagger}_{\boldi d_{xz}\uparrow}\hat{c}^{\dagger}_{\boldi d_{xy}\downarrow}
+\hat{c}^{\dagger}_{\boldi d_{xz}\downarrow}\hat{c}^{\dagger}_{\boldi d_{xy}\uparrow}),\label{eq:basis-Irrep10}\\
&
\hat{X}_{\boldi T_{2} \xi_{0}}^{\dagger}
=
\frac{1}{\sqrt{2}}
(\hat{c}^{\dagger}_{\boldi d_{xz}\uparrow}\hat{c}^{\dagger}_{\boldi d_{xy}\downarrow}
-\hat{c}^{\dagger}_{\boldi d_{xz}\downarrow}\hat{c}^{\dagger}_{\boldi d_{xy}\uparrow}),\label{eq:basis-Irrep11}\\
&
\hat{X}_{\boldi T_{1} \eta_{+}}^{\dagger}
=
\hat{c}^{\dagger}_{\boldi d_{yz}\uparrow}\hat{c}^{\dagger}_{\boldi d_{xy}\uparrow},\label{eq:basis-Irrep12}\\
&
\hat{X}_{\boldi T_{1} \eta_{-}}^{\dagger}
=
\hat{c}^{\dagger}_{\boldi d_{yz}\downarrow}\hat{c}^{\dagger}_{\boldi d_{xy}\downarrow},\label{eq:basis-Irrep13}\\
&
\hat{X}_{\boldi T_{1} \eta_{0}}^{\dagger}
=
\frac{1}{\sqrt{2}}
(\hat{c}^{\dagger}_{\boldi d_{yz}\uparrow}\hat{c}^{\dagger}_{\boldi d_{xy}\downarrow}
+\hat{c}^{\dagger}_{\boldi d_{yz}\downarrow}\hat{c}^{\dagger}_{\boldi d_{xy}\uparrow}),\label{eq:basis-Irrep14}\\
&
\hat{X}_{\boldi T_{2}\eta_{0}}^{\dagger}
=
\frac{1}{\sqrt{2}}
(\hat{c}^{\dagger}_{\boldi d_{yz}\uparrow}\hat{c}^{\dagger}_{\boldi d_{xy}\downarrow}
-\hat{c}^{\dagger}_{\boldi d_{yz}\downarrow}\hat{c}^{\dagger}_{\boldi d_{yx}\uparrow}).\label{eq:basis-Irrep15}
\end{align}

\subsection{$\hat{H}_{LS}$}
$\hat{H}_{LS}$ is given by the atomic SOC, 
the so-called $LS$ coupling, 
of the $t_{2g}$-orbital electrons: 
\begin{align}
\hat{H}_{\textrm{SOC}}
=
&\frac{i\lambda_{LS}}{2}
\sum\limits_{\boldi}
\sum\limits_{s}
\textrm{sgn}(s)
(\hat{c}_{\boldi d_{yz}  s}^{\dagger}\hat{c}_{\boldi d_{xz}  s}
-\hat{c}_{\boldi d_{xz} s}^{\dagger}\hat{c}_{\boldi d_{yz}  s})\notag\\
+
&\frac{i\lambda_{LS}}{2}
\sum\limits_{\boldi}
\sum\limits_{s}
(\hat{c}_{\boldi d_{xz} s}^{\dagger}\hat{c}_{\boldi d_{xy} -s}
-\hat{c}_{\boldi d_{xy} -s}^{\dagger}\hat{c}_{\boldi d_{xz} s})\notag\\
-&\frac{\lambda_{LS}}{2}
\sum\limits_{\boldi}
\sum\limits_{s}
\textrm{sgn}(s)
(\hat{c}_{\boldi d_{yz} s}^{\dagger}\hat{c}_{\boldi d_{xy}  -s}
-\hat{c}_{\boldi d_{xy}  s}^{\dagger}\hat{c}_{\boldi d_{yz}  -s}),\label{eq:HLS-t2g}
\end{align}
with $\lambda_{LS}$, the coupling constant, 
$\textrm{sgn}(\uparrow)=1$, $\textrm{sgn}(\downarrow)=-1$, 
$-s=\downarrow$ for $s=\uparrow$, and $-s=\uparrow$ for $s=\downarrow$.  
The $LS$ coupling is appropriate to 
take account of the SOC of electrons in solid. 
This is because the SOC of electrons 
arises from the relativistic effect near the nucleus, 
i.e. the correction to the nonrelativistic treatment 
is necessary only when an electron approaches the nucleus. 
Actually, 
we can describe even the antisymmetric SOC 
for an inversion-symmetry-broken quasi-two-dimensional system 
near a surface or an interface 
by using $\hat{H}_{LS}$ and 
the appropriate hopping-integral terms 
such as $\hat{H}_{\textrm{odd}}$~\cite{Yanase-ISB,Mizoguchi-SHE}; 
this successful description holds not only for the effective single-orbital system, 
which is sufficiently described by a Rashba-type SOC~\cite{Rashba}, 
but also the $t_{2g}$-orbital system, 
which is not described by the Rashba-type SOC. 
The key to the successful description is the appropriate treatment 
of orbital degrees of freedom of the SOC and the kinetic energy. 

$\hat{H}_{LS}$ causes not only the interorbital excitations for the $t_{2g}$ orbitals 
but also the excitations between the $a_{1g}$ orbital and the $e_{g}^{+}$ or $e_{g}^{-}$ orbitals. 
The former is directly seen from Eq. (\ref{eq:HLS-t2g}), 
and the latter can be seen by rewriting Eq. (\ref{eq:HLS-t2g}) 
in terms of the creation and the annihilation operators of 
the $a_{1g}$, $e_{g}^{+}$, and $e_{g}^{-}$ orbitals, 
Eqs. (\ref{eq:c+a1g}){--}(\ref{eq:ceg-}). 
The rewritten expression of Eq. (\ref{eq:HLS-t2g}) becomes
\begin{widetext}
\begin{align}
\hat{H}_{LS}
=
&\frac{i\lambda_{LS}}{6}
\sum\limits_{\boldi}
\sum\limits_{s}
\textrm{sgn}(s)
\Bigl[
(\omega^{2}-1)
(\hat{c}_{\boldi e_{g}^{+} s}^{\dagger}\hat{c}_{\boldi a_{1g}  s}
-\hat{c}_{\boldi a_{1g}  s}^{\dagger}\hat{c}_{\boldi e_{g}^{-}  s})
+(\omega-1)(\hat{c}_{\boldi e_{g}^{-}  s}^{\dagger}\hat{c}_{\boldi a_{1g}  s}
-\hat{c}_{\boldi a_{1g} s}^{\dagger}\hat{c}_{\boldi e_{g}^{+} s})
\Bigr]\notag\\
&+
\frac{i\lambda_{LS}}{6}
\sum\limits_{\boldi}
\sum\limits_{s}
\Bigl[
(1-\omega)
(\hat{c}_{\boldi e_{g}^{+} s}^{\dagger}\hat{c}_{\boldi a_{1g}  -s}
-\hat{c}_{\boldi a_{1g}  s}^{\dagger}\hat{c}_{\boldi e_{g}^{-}  -s})
+(1-\omega^{2})
(\hat{c}_{\boldi e_{g}^{-}  s}^{\dagger}\hat{c}_{\boldi a_{1g}  -s}
-\hat{c}_{\boldi a_{1g}  s}^{\dagger}\hat{c}_{\boldi e_{g}^{+}  -s})
\Bigr]\notag\\
&
-\frac{\lambda_{LS}}{6}
\sum\limits_{\boldi}
\sum\limits_{s}
\textrm{sgn}(s)
(\omega^{2}-\omega)
(\hat{c}_{\boldi e_{g}^{+}  s}^{\dagger}\hat{c}_{\boldi a_{1g}  -s}
+\hat{c}_{\boldi a_{1g}  s}^{\dagger}\hat{c}_{\boldi e_{g}^{+}  -s}
-\hat{c}_{\boldi e_{g}^{-}  s}^{\dagger}\hat{c}_{\boldi a_{1g}  -s}
-\hat{c}_{\boldi a_{1g}  s}^{\dagger}\hat{c}_{\boldi e_{g}^{-}  -s})
\notag\\
&+(\textrm{others}).\label{eq:HLS-a1g,eg}
\end{align}
\end{widetext}
Here we have shown only the interorbital excitations between the $a_{1g}$ orbital 
and the $e_{g}^{+}$ or $e_{g}^{-}$ orbital explicitly, 
and the other terms have been written as $(\textrm{others})$. 
This is because those explicitly shown terms are 
sufficient for the derivation of the low-energy effective Hamiltonian, 
explained in Sec. III.  

In addition, 
$\hat{H}_{LS}$ connects the different irreducible representations 
of the $d^{2}$ states for $\hat{H}_{\textrm{int}}$. 
This can be shown by multiplying $\hat{H}_{LS}$ and $|\boldi;\Gamma,g_{\Gamma}\rangle$; 
for example,  
\begin{align}
\hat{H}_{LS}|\boldi;A_{1} \rangle
=&
-\frac{i\lambda_{LS}}{\sqrt{3}}
(|\boldi;T_{1},\xi_{+} \rangle
-|\boldi;T_{1},\xi_{-} \rangle)\notag\\
&+
\frac{\lambda_{LS}}{\sqrt{3}}
(|\boldi;T_{1},\eta_{+} \rangle
+|\boldi;T_{1},\eta_{-} \rangle)\notag\\
&-\frac{i\lambda_{LS}\sqrt{2}}{\sqrt{3}}
(|\boldi;T_{1},\zeta_{0} \rangle,\label{eq:HLS-Hint-1}\\
\hat{H}_{LS}|\boldi;E, u \rangle
=&
\frac{i\lambda_{LS}}{2\sqrt{6}}
(|\boldi;T_{1},\xi_{+} \rangle
-|\boldi;T_{1},\xi_{-} \rangle)\notag\\
&+
\frac{\lambda_{LS}}{\sqrt{6}}
(|\boldi;T_{1},\eta_{+} \rangle
+|\boldi;T_{1},\eta_{-} \rangle)\notag\\
&+\frac{i\lambda_{LS}}{2\sqrt{3}}
|\boldi;T_{1},\zeta_{0} \rangle,\label{eq:HLS-Hint-2}\\
\hat{H}_{LS}|\boldi;T_{1}, \eta_{\pm} \rangle
=&
\mp\frac{i\lambda_{LS}}{2}
|\boldi;T_{1}, \xi_{\pm} \rangle
+
\frac{\lambda_{LS}}{\sqrt{3}}
|\boldi;A_{1} \rangle\notag\\
&+
\frac{\lambda_{LS}}{\sqrt{6}}
|\boldi;E,u \rangle\notag\\
&-
\frac{i\lambda_{LS}}{2\sqrt{2}}
(|\boldi;T_{1},\zeta_{0} \rangle
\mp 
|\boldi;T_{2},\zeta_{0} \rangle),\label{eq:HLS-Hint-3}
\end{align}
\begin{align}
\hat{H}_{LS}|\boldi;T_{2}, \eta_{0} \rangle
=&
-\frac{i\lambda_{LS}}{2}|\boldi;T_{1}, \xi_{0} \rangle\notag\\
&-\frac{i\lambda_{LS}}{2\sqrt{2}} 
(|\boldi;T_{1},\zeta_{+}\rangle - 
|\boldi;T_{1},\zeta_{-}\rangle ).\label{eq:HLS-Hint-4}
\end{align}
Thus, 
the $LS$ coupling causes the excitations between 
the different $d^{2}$ multiplets; 
here we have assumed that 
$J_{\textrm{H}}$ and $J^{\prime}$ are finite. 
As we will see in Sec. III B 2, 
this effect of the $LS$ coupling leads to a new contribution to 
the low-energy effective Hamiltonian, 
which is missing in Moriya's microscopic theory~\cite{DM2}. 

Before going into the derivation of the low-energy effective Hamiltonian, 
we will see the important differences between 
the SOC expressed by the $LS$ coupling 
and the SOC expressed by the spin-gauge potential. 
If we consider the SOC in a continuum, i.e., a system without lattice, 
the SOC can be expressed in terms of the $2\times 2$ spin-gauge potential~\cite{review-Tokura}, 
$(A_{\textrm{S}})_{\alpha}(r)=\textstyle\sum_{\beta}A_{\alpha}^{\beta}(r)\sigma^{\beta}$ 
($\alpha,\beta=x,y,z$) with the Pauli matrices $\sigma^{\beta}$: 
\begin{align}
\hat{H}_{\textrm{SOC}}
=&\frac{e}{2m^{2}}(\boldnabla V(r)\times \bolds)\cdot \boldp\notag\\
=&\frac{e}{2m^{2}}\boldA_{\textrm{S}}(r)\cdot \boldp.
\end{align}
This expression is also rewritten in terms of 
$A_{\alpha}^{\beta}(r)$ and the spin current~\cite{review-Tokura}. 
In the spin-gauge-potential expression, 
the spin and the site dependence of the SOC is included, 
and the orbital dependence is neglected. 
Thus, the differences between this expression and 
the $LS$-coupling expression 
are about the site and the orbital dependence of the SOC. 
About the site dependence, 
only the onsite component of the SOC is sufficient 
to analyze the effects of the SOC in solids, i.e., the systems with lattice. 
This is because the relativistic effect on the electrons in solids 
should be considered only near the nucleus, 
and because even an effectively off-site SOC, 
such as a single-orbital Rashba SOC~\cite{Rashba}, 
can be described by the SOC expressed by the $LS$ coupling 
and the appropriate kinetic terms~\cite{Yanase-ISB,Mizoguchi-SHE}. 
About the orbital dependence, 
its treatment in the spin-gauge-potential expression 
is insufficient for solids. 
This is because 
the orbital dependence arises from the orbital angular momentum, 
and because the orbital angular momentum plays significant roles in solids 
even with quenching of the orbital angular momentum; 
the examples of the significant roles are 
the anisotropies of the exchange interactions~\cite{DM2}. 
Actually, in Sec. III B 1, 
we show 
the important role of the orbital angular momentum 
in the intermediate states 
in the third-order perturbation terms.
Since we focus on not continua but solids 
and a continuum cannot be connected to a solid due to a crucial difference 
in translational symmetry, 
the SOC can be more appropriately described by the $LS$ coupling 
than by the spin-gauge potential. 
The reason why the expression of the spin-gauge potential becomes insufficient 
can be understood that 
we cannot choose any specific gauge about spins 
in the similar way to the magnetic field expressed by the vector potential and charge current, 
because the spins are nonconserved quantities due to 
the coupling of spin and orbital angular momenta; 
i.e., gauge invariance about spins does not exist.

\section{Low-energy effective Hamiltonian}
In this section, 
we derive the low-energy effective Hamiltonian for a $d^{1}$ Mott insulator 
in pyrochlore oxides 
by using the similar perturbation theory to Moriya's microscopic theory~\cite{DM2}, 
and show the results of the rough estimations about the derived coefficients.  
In this derivation, 
we consider the $d^{1}$ case for our model, 
corresponding to case of Lu$_{2}$V$_{2}$O$_{7}$ for example, 
and use three assumptions. 
One is that 
the Hubbard interactions are so large that 
the system becomes the $d^{1}$ Mott insulator. 
Another is that 
the trigonal-distortion potential 
is larger than 
the terms of $\hat{H}_{\textrm{KE}}$ and $\hat{H}_{LS}$. 
Due to those assumptions, 
the ground state of our model is the Mott insulator, 
in which one electron occupies the $a_{1g}$ orbital at each site. 
Furthermore, 
we can use $\hat{H}_{\textrm{int}}$ and $\hat{H}_{\textrm{tri}}$ as the non-perturbative terms, 
and treat $\hat{H}_{\textrm{KE}}$ and $\hat{H}_{LS}$ perturbatively. 
The other assumption is that
$U_{\Gamma}$ for all $\Gamma$ are larger than $\Delta_{\textrm{tri}}$. 
Due to this, 
we can more easily treat the effects of the non-perturbative terms 
because we can approximate 
$\frac{1}{-\frac{4\Delta_{\textrm{tri}}}{3}-(\hat{H}_{\textrm{int}}+\hat{H}_{\textrm{tri}})}$ 
as $\frac{1}{-\hat{H}_{\textrm{int}}}$. 
If such approximation is not used, 
the perturbation calculations become very difficult 
because the $d^{2}$ states diagonalizing $\hat{H}_{\textrm{int}}$ are not equal 
to the $d^{2}$ states diagonalizing $\hat{H}_{\textrm{tri}}$. 
If we represents those three assumptions as equations, 
the first one is 
$U,U^{\prime}\gg |t_{1}|, |t_{2}|, |t_{3}|, |t_{\textrm{odd}}|, \lambda_{LS}$, 
the second one is 
$\Delta_{\textrm{tri}}\gg |t_{1}|, |t_{2}|, |t_{3}|, |t_{\textrm{odd}}|, \lambda_{LS}$, 
and the third one is $U_{\Gamma}\gg \Delta_{\textrm{tri}}$. 
If the inequality signs of these conditions hold, 
the low-energy effective Hamiltonian derived under the three assumptions 
remains the leading term. 
Thus, the derived low-energy effective Hamiltonian is a ubiquitous model 
for $d^{1}$ pyrochlore oxides 
with the strong electron-electron interaction, 
the possitive trigonal-distortion potential, 
and the weak SOC. 

The derivations of the low-energy effective Hamiltonian 
and rough estimations are explained as follows. 
In Sec. III A, 
we calculate the second-order perturbation terms for our model 
by using $\hat{H}_{\textrm{KE}}$ twice; 
the term by using $\hat{H}_{LS}$ twice is not shown 
because that does not give the exchange interactions 
but gives the interaction between the charge-density operators. 
This calculation is application of Anderson's theory~\cite{Anderson} to our model. 
We also estimate the sign of the coefficient of the second-order terms 
as a function of $\frac{J_{\textrm{H}}}{U}$ within the $O(\frac{J_{\textrm{H}}}{U})$ terms 
in four limiting cases.  
In Sec. III B, 
we calculate the third-order perturbation terms for our model 
by using $\hat{H}_{\textrm{KE}}$ twice and $\hat{H}_{LS}$ once; 
the terms by using $\hat{H}_{\textrm{KE}}$ once and $\hat{H}_{LS}$ twice are zero. 
The finite terms come from three terms, 
$\hat{H}_{\textrm{3rd}}^{\textrm{KE}-\textrm{KE}-LS}$,  
$\hat{H}_{\textrm{3rd}}^{LS-\textrm{KE}-\textrm{KE}}$, 
and $\hat{H}_{\textrm{3rd}}^{\textrm{KE}-LS-\textrm{KE}}$
(for each definition, see Sec. III B). 
As we will discuss in detail in Sec. IV, 
only $\hat{H}_{\textrm{3rd}}^{\textrm{KE}-\textrm{KE}-LS}$ 
and $\hat{H}_{\textrm{3rd}}^{LS-\textrm{KE}-\textrm{KE}}$ 
are taken into account in Moriya's theory~\cite{DM2}, and  
the present theory can take account of 
not only the contribution included in Moriya's theory~\cite{DM2} 
but also the contribution neglected, $\hat{H}_{\textrm{3rd}}^{\textrm{KE}-LS-\textrm{KE}}$.  
In addition to the derivations, 
within the $O(\frac{J_{\textrm{H}}}{U})$ terms, 
we estimate
the ratio of the coefficient of 
$\hat{H}_{\textrm{3rd}}^{\textrm{KE}-\textrm{KE}-LS}+\hat{H}_{\textrm{3rd}}^{LS-\textrm{KE}-\textrm{KE}}$ 
to that of the second-order terms 
as a function of the ratio of $t_{\textrm{odd}}$ to 
$A=t_{1}+t_{2}+t_{3}$ or $B=t_{2}+t_{3}$
 the sum of the even-mirror hopping integrals 
in four cases, differing from the four cases for the second-order terms, 
and 
the ratio of the coefficient of $\hat{H}_{\textrm{3rd}}^{\textrm{KE}-LS-\textrm{KE}}$ 
to that of $\hat{H}_{\textrm{3rd}}^{\textrm{KE}-\textrm{KE}-LS}+\hat{H}_{\textrm{3rd}}^{LS-\textrm{KE}-\textrm{KE}}$
as a function of $\frac{\Delta_{\textrm{tri}}}{U}$ in the same four cases 
as those for the second-order terms. 
The former and latter results are shown in Secs. III B 1 and III B 2, 
respectively. 

\subsection{Second-order perturbation terms}
Before the actual derivation of the contributions 
from the second-order perturbation terms, 
we explain what we should do. 
The second-order perturbation term, $\hat{H}_{\textrm{2nd}}$, is generally given by~\cite{JJSakurai} 
\begin{align}
\hat{H}_{\textrm{2nd}}
=\langle \textrm{f}|\hat{H}_{\textrm{p}}
\frac{\hat{\phi}}{E_{0}-\hat{H}_{\textrm{np}}}
\hat{H}_{\textrm{p}}|\textrm{i}\rangle 
\times 
|\textrm{f}\rangle \langle \textrm{i}|.\label{eq:2nd-general}
\end{align}
Here $\hat{H}_{\textrm{p}}$ represents the perturbation terms, 
$\hat{H}_{\textrm{np}}$ represents the non-perturbation terms, 
$|\textrm{i}\rangle$ and $|\textrm{f}\rangle$ represent 
the ground states only with $\hat{H}_{\textrm{np}}$, 
$E_{0}$ is the ground-state energy, 
and $\hat{\phi}$ is the projection operator excluding the same-energy state~\cite{JJSakurai}. 
In our case, 
Eq. (\ref{eq:2nd-general}) between sublattices $1$ and $2$ 
at $\boldi=\boldone$ and $\boldtwo$, respectively, becomes
\begin{widetext}
\begin{align}
(\hat{H}_{\textrm{2nd}})_{\boldone \boldtwo}
=&
\sum\limits_{s_{1},s_{2},s_{3},s_{4}}
\langle a_{1g}^{s_{3}}, a_{1g}^{s_{4}}|
\hat{H}_{\textrm{KE}}
\frac{\hat{\phi}}{-\frac{4\Delta_{\textrm{tri}}}{3}-(\hat{H}_{\textrm{int}}+\hat{H}_{\textrm{tri}})}
\hat{H}_{\textrm{KE}}
|a_{1g}^{s_{1}},a_{1g}^{s_{2}}\rangle 
\times
| a_{1g}^{s_{3}}, a_{1g}^{s_{4}}\rangle \langle a_{1g}^{s_{1}},a_{1g}^{s_{2}}|\notag\\
\approx 
&
\sum\limits_{\boldi=\boldone,\boldtwo}
\sum\limits_{s_{1},s_{2},s_{3},s_{4}}
\sum\limits_{\Gamma}
\sum\limits_{g_{\Gamma}}
\frac{
\langle a_{1g}^{s_{3}}, a_{1g}^{s_{4}}|
\hat{H}_{\textrm{KE}}
|\boldi;\Gamma,g_{\Gamma}\rangle
\langle \boldi;\Gamma,g_{\Gamma}|
\hat{H}_{\textrm{KE}}
|a_{1g}^{s_{1}},a_{1g}^{s_{2}}\rangle} 
{-E_{\Gamma}}
| a_{1g}^{s_{3}}, a_{1g}^{s_{4}}\rangle \langle a_{1g}^{s_{1}},a_{1g}^{s_{2}}|,\label{eq:H2nd}
\end{align} 
\end{widetext}
with $|a_{1g}^{s_{1}},a_{1g}^{s_{2}}\rangle = 
\hat{c}_{\boldone a_{1g}  s_{1}}^{\dagger}\hat{c}_{\boldtwo a_{1g} s_{2}}^{\dagger}|0,0\rangle$. 
Here we have neglected the second-order term for $\hat{H}_{LS}$ 
because that does not give the exchange interactions, 
as described.  
The operator part $| a_{1g}^{s_{3}}, a_{1g}^{s_{4}}\rangle \langle a_{1g}^{s_{1}},a_{1g}^{s_{2}}|$ 
can be expressed in terms of the charge-density and the spin operators for the $a_{1g}$ orbital, 
e.g. $| a_{1g}^{\uparrow}, a_{1g}^{\downarrow}\rangle \langle a_{1g}^{\uparrow},a_{1g}^{\downarrow}|
=(\frac{1}{2}\hat{n}_{\boldone}+\hat{s}_{\boldone}^{z})
(\frac{1}{2}\hat{n}_{\boldtwo}-\hat{s}_{\boldtwo}^{z})$ 
or 
$| a_{1g}^{\uparrow}, a_{1g}^{\downarrow}\rangle \langle a_{1g}^{\downarrow},a_{1g}^{\uparrow}|
=\hat{s}_{\boldone}^{+}\hat{s}_{\boldtwo}^{-}$. 
Thus, 
what we should do is to express the right-hand side of Eq. (\ref{eq:H2nd}) 
in terms of those operators, 
$t_{1}$, $t_{2}$, $t_{3}$, $t_{\textrm{odd}}$, 
$U$, $U^{\prime}$, $J_{\textrm{H}}$, and $J^{\prime}$. 

\begin{figure*}[tb]
\includegraphics[width=160mm]{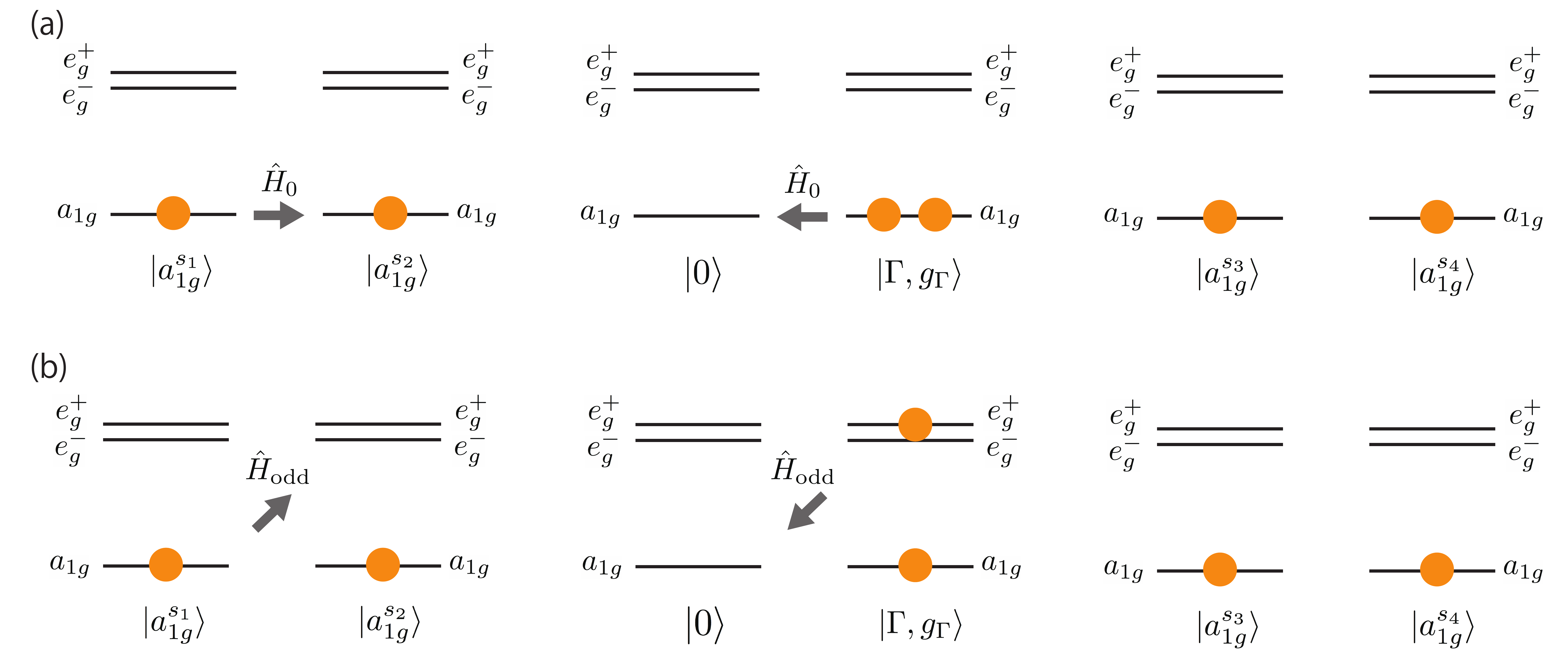}
\caption{
Schematic pictures of examples of the processes of the second-order perturbation 
by (a) using $\hat{H}_{0}$ and (b) using $\hat{H}_{\textrm{odd}}$. 
Yellow circles denote the electrons, and 
gray arrows denote the perturbations. 
}
\label{fig4}
\end{figure*}

We can calculate the right-hand side of Eq. (\ref{eq:H2nd}) for our model 
in the similar way for Ref. \onlinecite{NA-Ir}. 
As we will describe the detail of the calculation in Appendix A, 
we can express Eq. (\ref{eq:H2nd}) as follows:
\begin{align}
(\hat{H}_{\textrm{2nd}})_{\boldone \boldtwo}
=&J^{\textrm{AF}}
(\frac{1}{4}\hat{n}_{\boldone}\hat{n}_{\boldtwo}
-\hat{\bolds}_{\boldone}\cdot \hat{\bolds}_{\boldtwo})\notag\\
&+J^{\textrm{FM}}
(\frac{3}{4}\hat{n}_{\boldone}\hat{n}_{\boldtwo}
+\hat{\bolds}_{\boldone}\cdot \hat{\bolds}_{\boldtwo}),\label{eq:superexchange}
\end{align} 
with 
\begin{align}
&J^{\textrm{AF}}
=-\frac{4}{27}\frac{(t_{1}+2t_{2}+2t_{3})^{2}}{U+2J^{\prime}}
-\frac{8}{27}\frac{(t_{1}-t_{2}-t_{3})^{2}+9t_{\textrm{odd}}^{2}}{U-J^{\prime}}\notag\\
&\ \ \ \
-\frac{4}{9}\frac{(t_{1}+t_{2}+t_{3})^{2}+2(t_{2}+t_{3})^{2}
+3t_{\textrm{odd}}^{2}}{U^{\prime}+J_{\textrm{H}}}
,\label{eq:JAF}\\
&J^{\textrm{FM}}
=-\frac{4}{9}\frac{(t_{1}-t_{2}-t_{3})^{2}+9t_{\textrm{odd}}^{2}}{U^{\prime}-J_{\textrm{H}}}.
\label{eq:JFM}
\end{align}
Some of the processes of the second-order perturbation are shown in 
Figs. \ref{fig4}(a) and \ref{fig4}(b). 
Thus, the second-order perturbation terms give 
the AF and the FM superexchange interactions. 
Since the superexchange interactions between sublattices $1$ and $2$ are 
isotropic about the spin operators, 
the superexchange interactions between the other two sublattices are the same 
except the sign changes of $t_{3}$ and $t_{\textrm{odd}}$.  

\begin{figure}[tb]
\includegraphics[width=64mm]{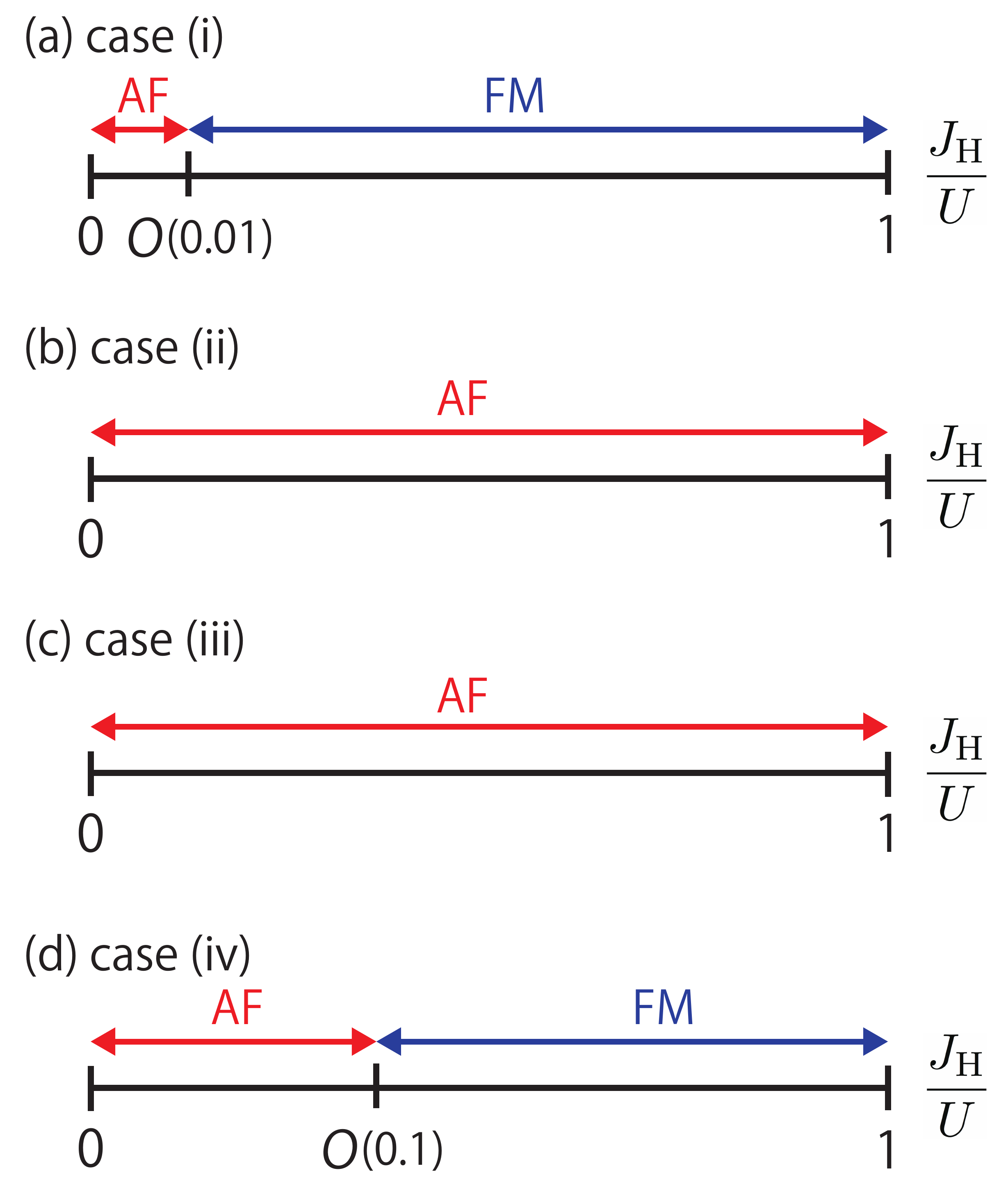}
\caption{Relation between $\frac{J_{\textrm{H}}}{U}$ and the sign of $J^{\textrm{FM}}-J^{\textrm{AF}}$ 
in the four limiting cases without the $t_{\textrm{odd}}^{2}$ terms 
within the $O(\frac{J_{\textrm{H}}}{U})$ terms 
for $J^{\prime}=J_{\textrm{H}}$ and $U^{\prime}=U-2J_{\textrm{H}}$. 
$J^{\textrm{FM}}-J^{\textrm{AF}}$ becomes AF or FM in the red-line region or the blue-line region, 
respectively. 
Since $J^{\textrm{FM}}-J^{\textrm{AF}}$ is expressed by the power series of $\frac{J_{\textrm{H}}}{U}$, 
the results are meaningful only for $\frac{J_{\textrm{H}}}{U}<1$.}
\label{figAd1}
\end{figure}

Before the derivations of the DM interaction, 
we see the relation between $J_{\textrm{H}}$ and the sign of $J^{\textrm{FM}}-J^{\textrm{AF}}$ 
in a rough calculation in order to understand 
when the dominant superexchange interactions become FM or AF. 
In this rough calculation, 
we use simple relations of the Slater-Koster parameters~\cite{Harrison} 
($V_{dd\sigma}=-2V_{dd\pi}$, $V_{dd\delta}=0$, and $V_{pd\sigma}=-2V_{pd\pi}$), 
and set $J^{\prime}=J_{\textrm{H}}$ and $U^{\prime}=U-2J_{\textrm{H}}$; 
then, 
we analyze the sign of $J^{\textrm{FM}}-J^{\textrm{AF}}$ 
within the $O(\frac{J_{\textrm{H}}}{U})$ terms in four limiting cases, 
(i) $|V_{dd\pi}|\gg |\frac{V_{pd\pi}^{2}}{\Delta_{pd}}|$, 
(ii) $|V_{dd\pi}|\ll |\frac{V_{pd\pi}^{2}}{\Delta_{pd}}|$, 
(iii) $V_{dd\pi}\sim -\frac{V_{pd\pi}^{2}}{\Delta_{pd}}$, 
and (iv) $V_{dd\pi}\sim \frac{V_{pd\pi}^{2}}{\Delta_{pd}}$. 
Those limiting cases are sufficient for qualitative analyses 
of the $J_{\textrm{H}}$ dependence of $J^{\textrm{FM}}-J^{\textrm{AF}}$ 
because as results of the simple relations, 
the direct and the indirect hopping integrals 
can be expressed by $V_{dd\pi}$ and $\frac{V_{pd\pi}^{2}}{\Delta_{pd}}$, 
respectively, 
and because in combination with $J^{\prime}=J_{\textrm{H}}$ and $U^{\prime}=U-2J_{\textrm{H}}$, 
$J^{\textrm{FM}}-J^{\textrm{AF}}$ can be expressed in terms of 
$V_{dd\pi}$, $\frac{V_{pd\pi}^{2}}{\Delta_{pd}}$, $U$, and $J_{\textrm{H}}$. 
To estimate the order of critical $\frac{J_{\textrm{H}}}{U}$ 
for the boundary between the AF and FM interactions in each limiting case, 
we assume that 
$\theta$ is so small that the terms of the odd-mirror hopping integrals are negligible; 
if we include the terms, 
the critical $J_{\textrm{H}}/U$ becomes smaller 
because the terms assist 
the FM interactions [see Eq. (\ref{eq:superexchange-simpler})]. 
Setting $J^{\prime}=J_{\textrm{H}}$ and $U^{\prime}=U-2J_{\textrm{H}}$ 
in Eqs. (\ref{eq:JAF}) and (\ref{eq:JFM}) 
and expanding $J^{\textrm{FM}}-J^{\textrm{AF}}$ in a power series about $\frac{J_{\textrm{H}}}{U}$ 
within the $O(\frac{J_{\textrm{H}}}{U})$ terms, 
we rewrite $J^{\textrm{FM}}-J^{\textrm{AF}}$ as 
\begin{align}
J^{\textrm{FM}}-J^{\textrm{AF}}
=\frac{4}{9U}\{(A+B)^{2}-2(\frac{J_{\textrm{H}}}{U})
[(A-2B)^{2}+9t_{\textrm{odd}}^{2}]\},\label{eq:superexchange-simpler} 
\end{align}
with 
$A=t_{1}+t_{2}+t_{3}$ and $B=t_{2}+t_{3}$.  
Then, 
by 
neglecting the $t_{\textrm{odd}}^{2}$ terms, 
using the simple relations of the Slater-Koster parameters 
and considering the leading terms of $J^{\textrm{FM}}-J^{\textrm{AF}}$ 
in each limiting case, 
the leading terms in cases (i), (ii), (iii), and (iv) are given by
\begin{align}
&J^{\textrm{FM}}-J^{\textrm{AF}}
\sim
\frac{1}{9U}V_{dd\pi}^{2}
(1-50\frac{J_{\textrm{H}}}{U}),\\
&J^{\textrm{FM}}-J^{\textrm{AF}}
\sim
\frac{100}{9U}(\frac{V_{pd\pi}^{2}}{\Delta_{pd}})^{2}
(1-\frac{8}{25}\frac{J_{\textrm{H}}}{U}),\\
&J^{\textrm{FM}}-J^{\textrm{AF}}
\sim
\frac{121}{9U}V_{dd\pi}^{2}
(1-\frac{2}{121}\frac{J_{\textrm{H}}}{U}),
\end{align}
and 
\begin{align}
&J^{\textrm{FM}}-J^{\textrm{AF}}
\sim
\frac{9}{U}V_{dd\pi}^{2}
(1-2\frac{J_{\textrm{H}}}{U}),
\end{align}
respectively. 
Thus, 
the order of the critical $\frac{J_{\textrm{H}}}{U}$ 
becomes $O(0.01)$ for case (i), $O(1)$ for case (ii), 
$O(10)$ for case (iii), 
and $O(0.1)$ for case (iv); 
those results are summarized in Fig. \ref{figAd1}.    
Since case (iv) may be the most realistic situation 
as pyrochlore vanadates~\cite{LDA-V124-Fujimori}, 
the superexchange interactions for the $d^{1}$ Mott insulator in the pyrochlore vanadates 
are FM for a realistic value of $\frac{J_{\textrm{H}}}{U}$. 

\subsection{Third-order perturbation terms}
As for the second-order perturbation terms, 
we start to prescribe what we should calculate 
for the third-order perturbation terms. 
The third-order perturbation terms for our model are given by
\begin{align}
\hat{H}_{\textrm{3rd}}
=&\hat{H}_{\textrm{3rd}}^{\textrm{KE}-\textrm{KE}-LS}
+\hat{H}_{\textrm{3rd}}^{LS-\textrm{KE}-\textrm{KE}}
+\hat{H}_{\textrm{3rd}}^{\textrm{KE}-LS-\textrm{KE}}\notag\\
=& 
\sum\limits_{n_{1},n_{2}}
\frac{\langle \textrm{f}|\hat{H}_{\textrm{KE}}|n_{1}\rangle 
\langle n_{1}|\hat{\phi}\hat{H}_{\textrm{KE}}|n_{2}\rangle 
\langle n_{2}|\hat{\phi}\hat{H}_{LS}|\textrm{i}\rangle}
{(E_{0}-E_{n_{1}})(E_{0}-E_{n_{2}})}|\textrm{f}\rangle \langle \textrm{i}|\notag\\
+&
\sum\limits_{n_{1},n_{2}}
\frac{\langle \textrm{f}|\hat{H}_{LS}|n_{2}\rangle 
\langle n_{2}|\hat{\phi}\hat{H}_{\textrm{KE}}|n_{1}\rangle 
\langle n_{1}|\hat{\phi}\hat{H}_{\textrm{KE}}|\textrm{i}\rangle}
{(E_{0}-E_{n_{2}})(E_{0}-E_{n_{1}})}|\textrm{f}\rangle \langle \textrm{i}|\notag\\
+&
\sum\limits_{n_{3},n_{4}}\frac{\langle \textrm{f}|\hat{H}_{\textrm{KE}}|n_{3}\rangle 
\langle n_{3}|\hat{\phi}\hat{H}_{LS}|n_{4}\rangle 
\langle n_{4}|\hat{\phi}\hat{H}_{\textrm{KE}}|\textrm{i}\rangle}
{(E_{0}-E_{n_{3}})(E_{0}-E_{n_{4}})}|\textrm{f}\rangle \langle \textrm{i}|,\label{eq:3rd-total}
\end{align}
where $E_{n_{i}}$  represent the energies of $|n_{i}\rangle$; 
as described, 
the terms using $\hat{H}_{\textrm{KE}}$ once and $\hat{H}_{LS}$ twice become zero.  
We will carry out the detailed calculation only for 
the third-order perturbation terms between sublattices $1$ and $2$, 
$(\hat{H}_{\textrm{3rd}})_{\boldone \boldtwo}$, 
because the other third-order perturbation terms 
can be derived from the result of $(\hat{H}_{\textrm{3rd}})_{\boldone \boldtwo}$ 
by permuting $x$, $y$, and $z$; 
for example, 
the terms between sublattices $1$ and $3$, 
$(\hat{H}_{\textrm{3rd}})_{\boldone \boldthree}$, is obtained 
by replacing 
$(\hat{s}_{\boldone}^{x},\hat{s}_{\boldone}^{y},\hat{s}_{\boldone}^{z})$ and 
$(\hat{s}_{\boldtwo}^{x}, \hat{s}_{\boldtwo}^{y}, \hat{s}_{\boldtwo}^{z})$ 
in $(\hat{H}_{\textrm{3rd}})_{\boldone \boldtwo}$ 
by 
$(\hat{s}_{\boldone}^{y}, \hat{s}_{\boldone}^{z}, \hat{s}_{\boldone}^{x})$ 
and $(\hat{s}_{\boldthree}^{y}, \hat{s}_{\boldthree}^{z}, \hat{s}_{\boldthree}^{x})$, respectively. 
The important differences among the three terms of $(\hat{H}_{\textrm{3rd}})_{\boldone \boldtwo}$ 
in Eq. (\ref{eq:3rd-total}) are about the intermediate states: 
in the first and the second term, 
$|n_{1}\rangle$ belongs to 
one of the $d^{2}$-$d^{0}$ or the $d^{0}$-$d^{2}$ states, 
such as $\hat{X}_{\boldone  A_{1}}^{\dagger}|0,0\rangle$, 
and $|n_{2}\rangle$ belongs to one of the $d^{1}$-$d^{1}$ states 
for one $a_{1g}$-orbital state and one $e_{g}^{+}$- or $e_{g}^{-}$-orbital state, 
such as $|a_{1g}^{s_{1}},e_{g}^{+;s_{2}}\rangle 
=\hat{c}^{\dagger}_{\boldone a_{1g} s_{1}}\hat{c}^{\dagger}_{\boldtwo e_{g}^{+} s_{2}}|0,0\rangle$; 
in the third term, 
$|n_{3}\rangle$ and $|n_{4} \rangle$ belong to the $d^{2}$-$d^{0}$ or the $d^{0}$-$d^{2}$ states. 
Thus, 
in the first and the second term, 
$\hat{H}_{LS}$ causes the excitations between the $a_{1g}$ orbital 
and the $e_{g}^{+}$ or $e_{g}^{-}$ orbital; 
in the third term, 
$\hat{H}_{LS}$ causes the excitations between the different-energy 
irreducible representations of the $d^{2}$ states for $\hat{H}_{\textrm{int}}$. 

In the following, 
we will calculate each term of 
$(\hat{H}_{\textrm{3rd}})_{\boldone \boldtwo}$ in Eq. (\ref{eq:3rd-total}). 

\subsubsection{$\hat{H}_{\textrm{3rd}}^{\textrm{KE}-\textrm{KE}-LS}$ and 
$\hat{H}_{\textrm{3rd}}^{LS-\textrm{KE}-\textrm{KE}}$}

We first calculate $\hat{H}_{\textrm{3rd}}^{\textrm{KE}-\textrm{KE}-LS}$ and 
$\hat{H}_{\textrm{3rd}}^{LS-\textrm{KE}-\textrm{KE}}$ 
for two sites of sublattices $1$ and $2$. 
Since $\hat{H}_{\textrm{3rd}}^{LS-\textrm{KE}-\textrm{KE}}$ is calculated 
from $\hat{H}_{\textrm{3rd}}^{LS-\textrm{KE}-\textrm{KE}}
=(\hat{H}_{\textrm{3rd}}^{\textrm{KE}-\textrm{KE}-LS})^{\dagger}$, 
we explain the derivation only for $\hat{H}_{\textrm{3rd}}^{\textrm{KE}-\textrm{KE}-LS}$. 
The derivation is divided into four steps 
because $\hat{H}_{\textrm{3rd}}^{\textrm{KE}-\textrm{KE}-LS}$ 
can be decomposed into the product of two operators: 
\begin{align}
\hat{H}_{\textrm{3rd}}^{\textrm{KE}-\textrm{KE}-LS}
=&
\sum\limits_{g_{e}=+,-}
\hat{H}_{\textrm{exch}}^{a_{1g}\textrm{-}e_{g}^{g_{e}}}
\hat{H}_{\textrm{excit}}^{e_{g}^{g_{e}}\textrm{-}a_{1g}},\label{eq:3rd-K-K-L}
\end{align}
where
\begin{align}
&\hat{H}_{\textrm{exch}}^{a_{1g}\textrm{-}e_{g}^{g_{e}}}
=\sum\limits_{n_{1}}
\frac{\langle \textrm{f}|\hat{H}_{\textrm{KE}}|n_{1}\rangle 
\langle n_{1}|\hat{\phi}\hat{H}_{\textrm{KE}}|n_{2}\rangle}
{(E_{0}-E_{n_{1}})}|\textrm{f}\rangle \langle n_{2}|,\label{eq:Hexch}
\end{align}
and 
\begin{align}
&\hat{H}_{\textrm{excit}}^{e_{g}^{g_{e}}\textrm{-}a_{1g}}
=
\frac{
\langle n_{2}|\hat{\phi}\hat{H}_{\textrm{LS}}|\textrm{i}\rangle}
{(E_{0}-E_{n_{2}})}|n_{2}\rangle \langle \textrm{i}|,\label{eq:Hexcit}
\end{align} 
with 
$|\textrm{i}\rangle=|a_{1g}^{s_{1}},a_{1g}^{s_{2}}\rangle$, 
$|\textrm{f}\rangle=|a_{1g}^{s_{3}},a_{1g}^{s_{4}}\rangle$,  
$|n_{2}\rangle=|a_{1g}^{s_{1}},e_{g}^{g_{e};s_{2}^{\prime}}\rangle$ 
or $|e_{g}^{g_{e};s_{1}^{\prime}},a_{1g}^{s_{2}}\rangle$, 
and $|n_{1}\rangle =|\boldi;\Gamma,g_{\Gamma}\rangle$. 
The first step is to derive $\hat{H}_{\textrm{exch}}^{a_{1g}\textrm{-}e_{g}^{g_{e}}}$. 
This is similar for the derivation of the second-order perturbation terms. 
The second step is to derive $\hat{H}_{\textrm{excit}}^{e_{g}^{g_{e}}\textrm{-}a_{1g}}$. 
This is the calculation of the finite matrix elements of $\hat{H}_{LS}$ 
for the excitations between the $a_{1g}$ orbital and the $e_{g}^{+}$ or $e_{g}^{-}$ orbital. 
The third step is to combine those two results by using Eq. (\ref{eq:3rd-K-K-L}). 
The fourth step is to combine the results for 
$\hat{H}_{\textrm{3rd}}^{\textrm{KE}-\textrm{KE}-LS}$ and 
$\hat{H}_{\textrm{3rd}}^{LS-\textrm{KE}-\textrm{KE}}$. 

We begin with the first step. 
In the similar way for the second-order perturbation terms, 
we can calculate $\hat{H}_{\textrm{exch}}^{a_{1g}\textrm{-}e_{g}^{g_{e}}}$ for two sites 
of sublattices $1$ and $2$ for our model 
(for the details, see Appendix B): 
\begin{widetext}
\begin{align}
&(\hat{H}_{\textrm{exch}}^{a_{1g}\textrm{-}e_{g}^{g_{e}}})_{\boldone\boldtwo}\notag\\
=&
\tilde{J}_{\textrm{S}}^{\textrm{AF}}\omega^{n(g_{e})}
[(\frac{1}{4}\hat{n}_{\boldone}\hat{o}_{\boldtwo}^{g_{e}}
-\hat{\bolds}_{\boldone}\cdot \hat{\boldso}_{\boldtwo}^{g_{e}})
+(\frac{1}{4}\hat{o}_{\boldone}^{g_{e}}\hat{n}_{\boldtwo}
-\hat{\boldso}_{\boldone}^{g_{e}}\cdot \hat{\bolds}_{\boldtwo})]
+
\tilde{J}^{\textrm{FM}}_{\textrm{S}}\omega^{n(g_{e})}
[(\frac{3}{4}\hat{n}_{\boldone}\hat{o}_{\boldtwo}^{g_{e}}
+\hat{\bolds}_{\boldone}\cdot \hat{\boldso}_{\boldtwo}^{g_{e}})
+(\frac{3}{4}\hat{o}_{\boldone}^{g_{e}}\hat{n}_{\boldtwo}
+\hat{\boldso}_{\boldone}^{g_{e}}\cdot \hat{\bolds}_{\boldtwo})]\notag\\
&+
\tilde{J}^{\textrm{AF}}_{\textrm{A}}\omega^{n(g_{e})}
[(\frac{1}{4}\hat{n}_{\boldone}\hat{o}_{\boldtwo}^{g_{e}}
-\hat{\bolds}_{\boldone}\cdot \hat{\boldso}_{\boldtwo}^{g_{e}})
-(\frac{1}{4}\hat{o}_{\boldone}^{g_{e}}\hat{n}_{\boldtwo}
-\hat{\boldso}_{\boldone}^{g_{e}}\cdot \hat{\bolds}_{\boldtwo})]
+
\tilde{J}^{\textrm{FM}}_{\textrm{A}}\omega^{n(g_{e})}
[(\frac{3}{4}\hat{n}_{\boldone}\hat{o}_{\boldtwo}^{g_{e}}
+\hat{\bolds}_{\boldone}\cdot \hat{\boldso}_{\boldtwo}^{g_{e}})
-(\frac{3}{4}\hat{o}_{\boldone}^{g_{e}}\hat{n}_{\boldtwo}
+\hat{\boldso}_{\boldone}^{g_{e}}\cdot \hat{\bolds}_{\boldtwo})],\label{eq:Hexch-explicit}
\end{align}
\end{widetext}
with 
\begin{align}
&\hat{o}_{\boldi}^{g_{e}}
=\sum\limits_{s=\uparrow,\downarrow}
\hat{c}_{\boldi a_{1g} s}^{\dagger}\hat{c}_{\boldi e_{g}^{g_{e}} s},\\
&\hat{\boldso}_{\boldi}^{g_{e}}
=\frac{1}{2}\sum\limits_{s,s^{\prime}=\uparrow,\downarrow}
\hat{c}_{\boldi a_{1g} s}^{\dagger}(\boldsigma)_{ss^{\prime}}\hat{c}_{\boldi e_{g}^{g_{e}} s^{\prime}},\\
&\tilde{J}^{\textrm{AF}}_{\textrm{S}}
=
-\frac{4}{27}
\frac{(t_{1}+2t_{2}+2t_{3})(t_{1}-t_{2}-t_{3})}{U+2J^{\prime}}\notag\\
&-\frac{2}{27}
\frac{(t_{1}-t_{2}-t_{3})(4t_{1}+2t_{2}+2t_{3})+9t_{\textrm{odd}}^{2}}{U-J^{\prime}}\notag\\
&-\frac{1}{9}
\frac{-4(t_{2}+t_{3})^{2}+(t_{1}+t_{2}+t_{3})^{2}-6t_{\textrm{odd}}^{2}}{U^{\prime}+J_{\textrm{H}}},
\label{eq:J-AFS}\\
&\tilde{J}^{\textrm{FM}}_{\textrm{S}}
=
-\frac{1}{9}
\frac{(t_{1}-t_{2}-t_{3})^{2}}{U^{\prime}-J_{\textrm{H}}}\label{eq:J-FMS},\\
&\tilde{J}^{\textrm{AF}}_{\textrm{A}}
=
\frac{4}{9}
\frac{(t_{1}+2t_{2}+2t_{3})t_{\textrm{odd}}}{U+2J^{\prime}}
+\frac{2}{9}
\frac{(t_{1}-t_{2}-t_{3})t_{\textrm{odd}}}{U-J^{\prime}}\notag\\
&-\frac{1}{3}
\frac{(t_{1}-3t_{2}-3t_{3})t_{\textrm{odd}}}{U^{\prime}+J_{\textrm{H}}}\label{eq:J-AFA},\\
&\tilde{J}^{\textrm{FM}}_{\textrm{A}}
=
-\frac{(t_{1}+t_{2}+t_{3})t_{\textrm{odd}}}{U^{\prime}-J_{\textrm{H}}}\label{eq:J-FMA},\\
&\omega^{n(g_{e})}
=
\begin{cases}
\omega^{2} & \textrm{for}\ g_{e}=+ \\
\omega & \textrm{for}\ g_{e}=-
\end{cases}.\label{eq:n-def}
\end{align}

Equation (\ref{eq:Hexch-explicit}) shows three important properties of 
$\hat{H}_{\textrm{exch}}^{a_{1g}\textrm{-}e_{g}^{g_{e}}}$. 
First, 
as in the case with the second-order perturbation terms, 
$\hat{H}_{\textrm{exch}}^{a_{1g}\textrm{-}e_{g}^{g_{e}}}$ 
has the FM-type and the AF-type interactions. 
Second, 
in contrast to the second-order perturbation terms, 
$\hat{H}_{\textrm{exch}}^{a_{1g}\textrm{-}e_{g}^{g_{e}}}$ 
includes the orbital-density operator, $\hat{o}_{\boldi}^{g_{e}}$, 
and the spin-orbital-combined operator, $\hat{\boldso}_{\boldi}^{g_{e}}$. 
The orbital-density operator or the spin-orbital-combined operator is 
essentially different from, respectively, 
the charge-density operator, $\hat{n}_{\boldi}$, 
or the spin operator, $\hat{\bolds}_{\boldi}$, 
because the charge-density and the spin operators should be defined 
for the product of the creation and annihilation operators 
of an electron for the same orbital; 
$\hat{n}_{\boldi}$ and $\hat{\bolds}_{\boldi}$ for the $a_{1g}$-orbital electrons 
are defined as 
\begin{align}
\hat{n}_{\boldi}=\sum\limits_{s=\uparrow,\downarrow}
\hat{c}^{\dagger}_{\boldi a_{1g} s}\hat{c}_{\boldi a_{1g}s},
\end{align} 
and 
\begin{align}
\hat{\bolds}_{\boldi}
=\frac{1}{2}\sum\limits_{s,s^{\prime}=\uparrow,\downarrow}
\hat{c}^{\dagger}_{\boldi a_{1g} s}
(\boldsigma)_{ss^{\prime}}\hat{c}_{\boldi a_{1g}s^{\prime}},
\end{align}
respectively. 
Due to this property, 
$\hat{H}_{\textrm{exch}}^{a_{1g}\textrm{-}e_{g}^{g_{e}}}$ can not be regarded 
as the simple superexchange interactions 
such as the second-order perturbation terms. 
Third, 
$\hat{H}_{\textrm{exch}}^{a_{1g}\textrm{-}e_{g}^{g_{e}}}$ possesses 
not only the symmetric terms, 
proportional to $\tilde{J}^{\textrm{AF}}_{\textrm{S}}$ or $\tilde{J}^{\textrm{FM}}_{\textrm{S}}$, 
but also the antisymmetric terms, 
proportional to $\tilde{J}^{\textrm{AF}}_{\textrm{A}}$ or $\tilde{J}^{\textrm{FM}}_{\textrm{A}}$. 
Here the terms for $\tilde{J}^{\textrm{AF}}_{\textrm{S}}$ and $\tilde{J}^{\textrm{FM}}_{\textrm{S}}$ 
have been referred to as the symmetric terms 
because those are symmetric about site indices, $\boldone$ and $\boldtwo$; 
the terms for $\tilde{J}^{\textrm{AF}}_{\textrm{A}}$ and $\tilde{J}^{\textrm{FM}}_{\textrm{A}}$, 
which are antisymmetric about the site indices, 
have been referred to as the antisymmetric terms. 

In the second step, 
we calculate $\hat{H}_{\textrm{excit}}^{e_{g}^{g_{e}}\textrm{-}a_{1g}}$ 
from Eq. (\ref{eq:Hexcit}) using Eq. (\ref{eq:HLS-a1g,eg}). 
The calculated result is 
\begin{widetext}
\begin{align}
\hat{H}_{\textrm{excit}}^{e_{g}^{g_{e}}\textrm{-}a_{1g}}
=&
-\frac{\lambda_{LS}}{6\Delta_{\textrm{tri}}}
|a_{1g}^{\uparrow},e_{g}^{g_{e};\uparrow}\rangle 
\{
i(\omega^{n(g_{e})}-1)
\langle a_{1g}^{\uparrow},a_{1g}^{\uparrow}|
+[i(1-\omega^{m(g_{e})})-(\omega^{n(g_{e})}-\omega^{m(g_{e})})]
\langle a_{1g}^{\uparrow},a_{1g}^{\downarrow}|
\}\notag\\
-&
\frac{\lambda_{LS}}{6\Delta_{\textrm{tri}}}
|e_{g}^{g_{e};\uparrow},a_{1g}^{\uparrow}\rangle 
\{
i(\omega^{n(g_{e})}-1)
\langle a_{1g}^{\uparrow},a_{1g}^{\uparrow}|
+[i(1-\omega^{m(g_{e})})-(\omega^{n(g_{e})}-\omega^{m(g_{e})})]
\langle a_{1g}^{\downarrow},a_{1g}^{\uparrow}|
\}\notag\\
-&
\frac{\lambda_{LS}}{6\Delta_{\textrm{tri}}}
|a_{1g}^{\uparrow},e_{g}^{g_{e};\downarrow}\rangle 
\{
[i(1-\omega^{m(g_{e})})+(\omega^{n(g_{e})}-\omega^{m(g_{e})})]
\langle a_{1g}^{\uparrow},a_{1g}^{\uparrow}|
-i(\omega^{n(g_{e})}-1)
\langle a_{1g}^{\uparrow},a_{1g}^{\downarrow}|
\}\notag\\
-&\frac{\lambda_{LS}}{6\Delta_{\textrm{tri}}}
|e_{g}^{g_{e};\uparrow},a_{1g}^{\downarrow}\rangle 
\{
i(\omega^{n(g_{e})}-1)
\langle a_{1g}^{\uparrow},a_{1g}^{\downarrow}|
+[i(1-\omega^{m(g_{e})})-(\omega^{n(g_{e})}-\omega^{m(g_{e})})]
\langle a_{1g}^{\downarrow},a_{1g}^{\downarrow}|
\}\notag\\
-&
\frac{\lambda_{LS}}{6\Delta_{\textrm{tri}}}
|a_{1g}^{\downarrow},e_{g}^{g_{e};\uparrow}\rangle 
\{
i(\omega^{n(g_{e})}-1)
\langle a_{1g}^{\downarrow},a_{1g}^{\uparrow}|
+[i(1-\omega^{m(g_{e})})-(\omega^{n(g_{e})}-\omega^{m(g_{e})})]
\langle a_{1g}^{\downarrow},a_{1g}^{\downarrow}|
\}\notag\\
-&\frac{\lambda_{\textrm{LS}}}{6\Delta_{\textrm{tri}}}
|e_{g}^{g_{e};\downarrow},a_{1g}^{\uparrow}\rangle 
\{
[i(1-\omega^{m(g_{e})})+(\omega^{n(g_{e})}-\omega^{m(g_{e})})]
\langle a_{1g}^{\uparrow},a_{1g}^{\uparrow}|
-i(\omega^{n(g_{e})}-1)
\langle a_{1g}^{\downarrow},a_{1g}^{\uparrow}|
\}\notag\\
-&
\frac{\lambda_{LS}}{6\Delta_{\textrm{tri}}}
|a_{1g}^{\downarrow},e_{g}^{g_{e};\downarrow}\rangle 
\{
[i(1-\omega^{m(g_{e})})+(\omega^{n(g_{e})}-\omega^{m(g_{e})})]
\langle a_{1g}^{\downarrow},a_{1g}^{\uparrow}|
-i(\omega^{n(g_{e})}-1)
\langle a_{1g}^{\downarrow},a_{1g}^{\downarrow}|
\}\notag\\
-&\frac{\lambda_{LS}}{6\Delta_{\textrm{tri}}}
|e_{g}^{g_{e};\downarrow},a_{1g}^{\downarrow}\rangle 
\{
[i(1-\omega^{m(g_{e})})+(\omega^{n(g_{e})}-\omega^{m(g_{e})})]
\langle a_{1g}^{\uparrow},a_{1g}^{\downarrow}|
-i(\omega^{n(g_{e})}-1)
\langle a_{1g}^{\downarrow},a_{1g}^{\downarrow}|
\},\label{eq:Hexcit-explicit}
\end{align}
\end{widetext}
with
\begin{align}
&\omega^{m(g_{e})}
=
\begin{cases}
\omega & \textrm{for}\ g_{e}=+ \\
\omega^{2} & \textrm{for}\ g_{e}=-
\end{cases}.\label{eq:m-def}
\end{align}

Equation (\ref{eq:Hexcit-explicit}) clearly shows that 
the $LS$ coupling induces the excitations between the $a_{1g}$ orbital 
and the $e_{g}^{+}$ or $e_{g}^{-}$ orbital 
at one of the two sites 
in the terms of $\hat{H}_{\textrm{3rd}}^{\textrm{KE}-\textrm{KE}-LS}$. 
More precisely, 
the terms proportional to $i(\omega^{n(g_{e})}-1)$ in Eq. (\ref{eq:Hexcit-explicit}) 
are induced 
by the $z$ components of the $LS$ coupling, 
i.e., the $LS$ couplings between the $d_{xz}$ and $d_{yz}$ orbitals; 
the terms proportional to $i(1-\omega^{m(g_{e})})$ 
are induced by the $x$ components, 
i.e., the $LS$ couplings between the $d_{xz}$ and $d_{xy}$ orbitals; 
the terms proportional to $(\omega^{n(g_{e})}-\omega^{m(g_{e})})$ 
are induced by the $y$ components, 
i.e., the $LS$ couplings between the $d_{yz}$ and $d_{xy}$ orbitals. 

In the third step, 
by combining Eqs. (\ref{eq:Hexch-explicit}) and (\ref{eq:Hexcit-explicit}) 
with Eq. (\ref{eq:3rd-K-K-L}), 
we obtain $(\hat{H}_{\textrm{3rd}}^{\textrm{KE}-\textrm{KE}-LS})_{\boldone\boldtwo}$: 
\begin{widetext}
\begin{align}
(\hat{H}_{\textrm{3rd}}^{\textrm{KE}-\textrm{KE}-LS})_{\boldone\boldtwo}
=
&
-\sum\limits_{g_{e}=+,-}
\frac{\lambda_{LS}}{6\Delta_{\textrm{tri}}}
i(\omega^{n(g_{e})}-1)
[2\tilde{J}^{\textrm{FM}}_{\textrm{S}}
(\hat{n}_{\boldone}\hat{S}_{\boldtwo}^{z}+\hat{S}_{\boldone}^{z}\hat{n}_{\boldtwo})
-(\tilde{J}^{\textrm{FM}}_{\textrm{A}}+\tilde{J}^{\textrm{AF}}_{\textrm{A}})
(\hat{S}_{\boldone}^{z}\hat{n}_{\boldtwo}-\hat{n}_{\boldone}\hat{S}_{\boldtwo}^{z})]\notag\\
&-
\sum\limits_{g_{e}=+,-}
\frac{\lambda_{LS}}{3\Delta_{\textrm{tri}}}
(\omega^{n(g_{e})}-1)
(\tilde{J}^{\textrm{FM}}_{\textrm{A}}-\tilde{J}^{\textrm{AF}}_{\textrm{A}})
(\hat{S}_{\boldone}^{x}\hat{S}_{\boldtwo}^{y}-\hat{S}_{\boldone}^{y}\hat{S}_{\boldtwo}^{x})\notag\\
&-
\sum\limits_{g_{e}=+,-}
\frac{\lambda_{LS}}{6\Delta_{\textrm{tri}}}
i(1-\omega^{m(g_{e})})
[2\tilde{J}^{\textrm{FM}}_{\textrm{S}}
(\hat{n}_{\boldone}\hat{S}_{\boldtwo}^{x}+\hat{S}_{\boldone}^{x}\hat{n}_{\boldtwo})
+(\tilde{J}^{\textrm{FM}}_{\textrm{A}}+\tilde{J}^{\textrm{AF}}_{\textrm{A}})
(\hat{n}_{\boldone}\hat{S}_{\boldtwo}^{x}-\hat{S}_{\boldone}^{x}\hat{n}_{\boldtwo})]\notag\\
&-
\sum\limits_{g_{e}=+,-}
\frac{\lambda_{LS}}{3\Delta_{\textrm{tri}}}
(1-\omega^{m(g_{e})})
(\tilde{J}^{\textrm{FM}}_{\textrm{A}}-\tilde{J}^{\textrm{AF}}_{\textrm{A}})
(\hat{S}_{\boldone}^{y}\hat{S}_{\boldtwo}^{z}
-\hat{S}_{\boldone}^{z}\hat{S}_{\boldtwo}^{y})\notag\\
&-
\sum\limits_{g_{e}=+,-}
\frac{\lambda_{LS}}{6\Delta_{\textrm{tri}}}
i(\omega^{m(g_{e})}-\omega^{n(g_{e})})
[2\tilde{J}^{\textrm{FM}}_{\textrm{S}}
(\hat{n}_{\boldone}\hat{S}_{\boldtwo}^{y}+\hat{S}_{\boldone}^{y}\hat{n}_{\boldtwo})
+(\tilde{J}^{\textrm{FM}}_{\textrm{A}}+\tilde{J}^{\textrm{AF}}_{\textrm{A}})
(\hat{n}_{\boldone}\hat{S}_{\boldtwo}^{y}-\hat{S}_{\boldone}^{y}\hat{n}_{\boldtwo})]\notag\\
&-
\sum\limits_{g_{e}=+,-}
\frac{\lambda_{LS}}{3\Delta_{\textrm{tri}}}
(\omega^{m(g_{e})}-\omega^{n(g_{e})})
(\tilde{J}^{\textrm{FM}}_{\textrm{A}}-\tilde{J}^{\textrm{AF}}_{\textrm{A}})
(\hat{S}_{\boldone}^{z}\hat{S}_{\boldtwo}^{x}-\hat{S}_{\boldone}^{x}\hat{S}_{\boldtwo}^{z}),
\label{eq:H-K-K-L-explicit}
\end{align}
\end{widetext}
where 
we have used the relations, 
\begin{widetext}
\begin{align}
&2(\frac{1}{4}\hat{n}_{\boldone}\hat{o}_{\boldtwo}^{g_{e}}
-\hat{\bolds}_{\boldone}\cdot \hat{\boldso}_{\boldtwo}^{g_{e}})
=
|a_{1g}^{\uparrow},a_{1g}^{\downarrow}\rangle 
\langle a_{1g}^{\uparrow},e_{g}^{g_{e};\downarrow}|
-|a_{1g}^{\uparrow},a_{1g}^{\downarrow}\rangle 
\langle a_{1g}^{\downarrow},e_{g}^{g_{e};\uparrow}|
-|a_{1g}^{\downarrow},a_{1g}^{\uparrow}\rangle 
\langle a_{1g}^{\uparrow},e_{g}^{g_{e};\downarrow}|
+|a_{1g}^{\downarrow},a_{1g}^{\uparrow}\rangle 
\langle a_{1g}^{\downarrow},e_{g}^{g_{e};\uparrow}|,\\
&2(\frac{1}{4}\hat{o}_{\boldone}^{g_{e}}\hat{n}_{\boldtwo}
-\hat{\boldso}_{\boldone}^{g_{e}}\cdot \hat{\bolds}_{\boldtwo})
=
|a_{1g}^{\uparrow},a_{1g}^{\downarrow}\rangle 
\langle e_{g}^{g_{e};\uparrow},a_{1g}^{\downarrow}|
-|a_{1g}^{\uparrow},a_{1g}^{\downarrow}\rangle 
\langle e_{g}^{g_{e};\downarrow},a_{1g}^{\uparrow}|
-|a_{1g}^{\downarrow},a_{1g}^{\uparrow}\rangle 
\langle e_{g}^{g_{e};\uparrow},a_{1g}^{\downarrow}|
+|a_{1g}^{\downarrow},a_{1g}^{\uparrow}\rangle 
\langle e_{g}^{g_{e};\downarrow},a_{1g}^{\uparrow}|,\\
&2(\frac{3}{4}\hat{n}_{\boldone}\hat{o}_{\boldtwo}^{g_{e}}
+\hat{\bolds}_{\boldone}\cdot \hat{\boldso}_{\boldtwo}^{g_{e}})
=
2
|a_{1g}^{\uparrow},a_{1g}^{\uparrow}\rangle 
\langle a_{1g}^{\uparrow},e_{g}^{g_{e};\uparrow}|
+2
|a_{1g}^{\downarrow},a_{1g}^{\downarrow}\rangle 
\langle a_{1g}^{\downarrow},e_{g}^{g_{e};\downarrow}|
+
|a_{1g}^{\uparrow},a_{1g}^{\downarrow}\rangle 
\langle a_{1g}^{\uparrow},e_{g}^{g_{e};\downarrow}|
+|a_{1g}^{\uparrow},a_{1g}^{\downarrow}\rangle 
\langle a_{1g}^{\downarrow},e_{g}^{g_{e};\uparrow}|\notag\\
&+|a_{1g}^{\downarrow},a_{1g}^{\uparrow}\rangle 
\langle a_{1g}^{\uparrow},e_{g}^{g_{e};\downarrow}|
+|a_{1g}^{\downarrow},a_{1g}^{\uparrow}\rangle 
\langle a_{1g}^{\downarrow},e_{g}^{g_{e};\uparrow}|,\\
&2(\frac{3}{4}\hat{o}_{\boldone}^{g_{e}}\hat{n}_{\boldtwo}
+\hat{\boldso}_{\boldone}^{g_{e}}\cdot \hat{\bolds}_{\boldtwo})
=
2|a_{1g}^{\uparrow},a_{1g}^{\uparrow}\rangle 
\langle e_{g}^{g_{e};\uparrow},a_{1g}^{\uparrow}|
+2|a_{1g}^{\downarrow},a_{1g}^{\downarrow}\rangle 
\langle e_{g}^{g_{e};\downarrow},a_{1g}^{\downarrow}|
+|a_{1g}^{\uparrow},a_{1g}^{\downarrow}\rangle 
\langle e_{g}^{g_{e};\uparrow},a_{1g}^{\downarrow}|
+|a_{1g}^{\uparrow},a_{1g}^{\downarrow}\rangle 
\langle e_{g}^{g_{e};\downarrow},a_{1g}^{\uparrow}|\notag\\
&+|a_{1g}^{\downarrow},a_{1g}^{\uparrow}\rangle 
\langle e_{g}^{g_{e};\uparrow},a_{1g}^{\downarrow}|
+|a_{1g}^{\downarrow},a_{1g}^{\uparrow}\rangle 
\langle e_{g}^{g_{e};\downarrow},a_{1g}^{\uparrow}|.
\end{align}
\end{widetext}

Using the relations between each term of Eq. (\ref{eq:Hexcit-explicit}) 
and each component of the $LS$ coupling, 
described below Eq. (\ref{eq:m-def}), 
we see from Eq. (\ref{eq:H-K-K-L-explicit}) 
that 
the $z$, $x$, and $y$ components of the $LS$ coupling 
lead to, respectively, 
the $z$, $x$, and $y$ components 
of the DM interaction in Eq. (\ref{eq:H-K-K-L-explicit}). 

\begin{figure*}[tb]
\includegraphics[width=180mm]{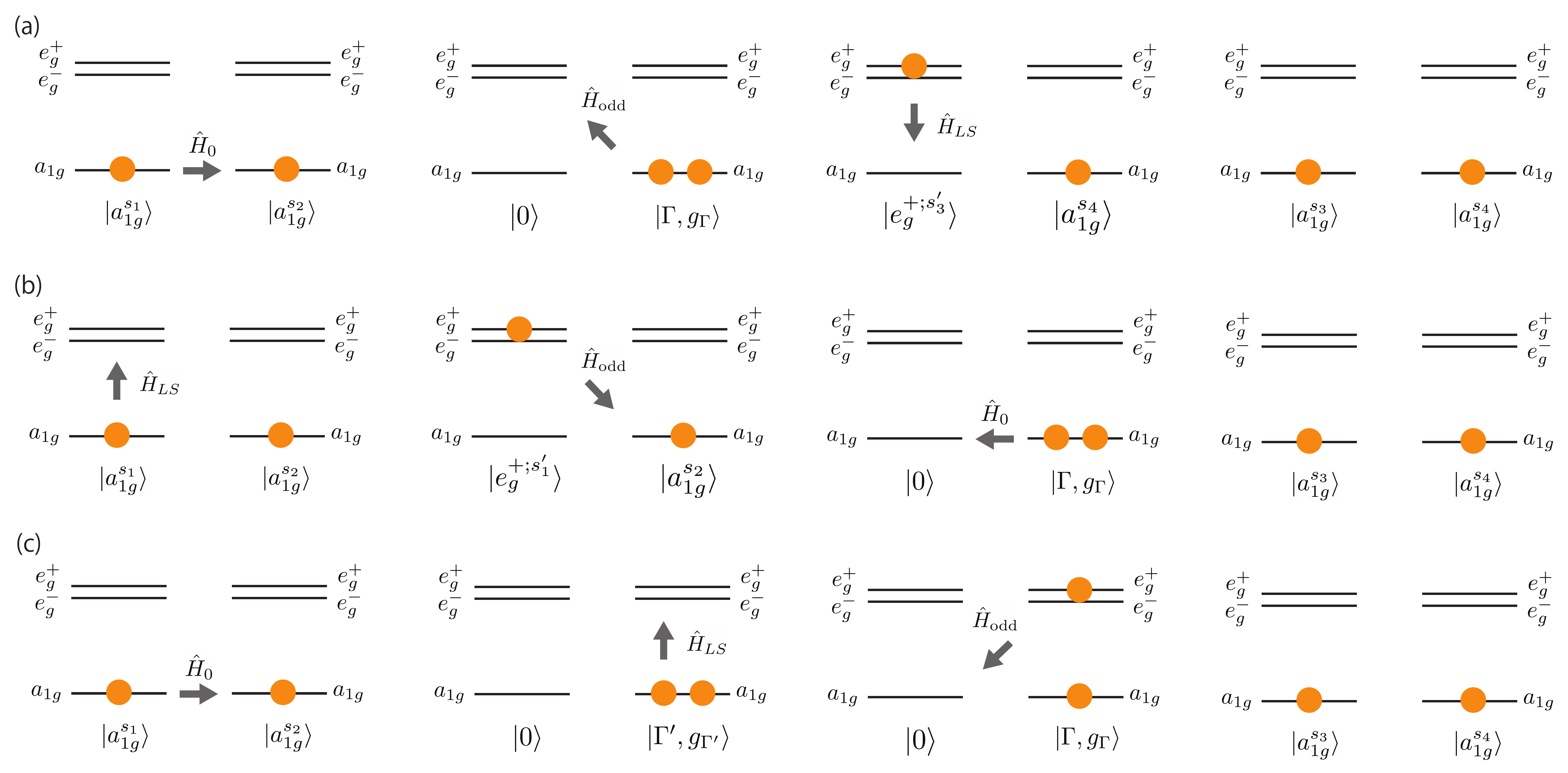}
\caption{
Schematic pictures of examples of the processes of 
(a) $\hat{H}_{\textrm{3rd}}^{\textrm{KE}-\textrm{KE}-LS}$, 
(b) $\hat{H}_{\textrm{3rd}}^{LS-\textrm{KE}-\textrm{KE}}$, and 
(c) $\hat{H}_{\textrm{3rd}}^{\textrm{KE}-LS-\textrm{KE}}$. 
Yellow circles denote the electrons, 
and gray arrows denote the perturbations. 
}
\label{fig5}
\end{figure*}

Since 
$(\hat{H}_{\textrm{3rd}}^{LS-\textrm{KE}-\textrm{KE}})_{\boldone\boldtwo}$ 
is calculated from Eq. (\ref{eq:H-K-K-L-explicit}) 
by using the equality, 
$\hat{H}_{\textrm{3rd}}^{LS-\textrm{KE}-\textrm{KE}}
=(\hat{H}_{\textrm{3rd}}^{\textrm{KE}-\textrm{KE}-LS})^{\dagger}$, 
we obtain 
$\hat{H}_{\textrm{3rd}}^{\textrm{KE}-\textrm{KE}-LS}
+\hat{H}_{\textrm{3rd}}^{LS-\textrm{KE}-\textrm{KE}}$ 
after taking the summation about $g_{e}=+,-$ 
using Eqs. (\ref{eq:n-def}) and (\ref{eq:m-def}): 
\begin{align}
&(\hat{H}_{\textrm{3rd}}^{\textrm{KE}-\textrm{KE}-LS})_{\boldone\boldtwo}
+(\hat{H}_{\textrm{3rd}}^{LS-\textrm{KE}-\textrm{KE}})_{\boldone\boldtwo}\notag\\
=&-D(\hat{S}_{\boldone}^{y}\hat{S}_{\boldtwo}^{z}-\hat{S}_{\boldone}^{z}\hat{S}_{\boldtwo}^{y})
+D(\hat{S}_{\boldone}^{z}\hat{S}_{\boldtwo}^{x}-\hat{S}_{\boldone}^{x}\hat{S}_{\boldtwo}^{z}),
\label{eq:HDM-a1g-eg}
\end{align}
with
\begin{align}
D
=&-\frac{2\lambda_{LS}}{\Delta_{\textrm{tri}}}
(\tilde{J}^{\textrm{FM}}_{\textrm{A}}-\tilde{J}^{\textrm{AF}}_{\textrm{A}})\notag\\
=&\frac{2\lambda_{LS}t_{\textrm{odd}}}{\Delta_{\textrm{tri}}}
[\frac{t_{1}+t_{2}+t_{3}}{U^{\prime}-J_{\textrm{H}}}
-\frac{4}{9}\frac{t_{1}+2t_{2}+2t_{3}}{U+2J^{\prime}}\notag\\
&-\frac{2}{9}\frac{t_{1}-t_{2}-t_{3}}{U-J^{\prime}}
+\frac{1}{3}\frac{t_{1}-3t_{2}-3t_{3}}{U^{\prime}+J_{\textrm{H}}}].\label{eq:D-a1g-eg}
\end{align}
This antisymmetric exchange interaction is the DM interaction; 
as we will discuss in Sec. IV, 
the symmetry of the finite components is consistent 
with the phenomenological argument~\cite{DMI-pyro} based on Moriya's rule~\cite{DM2}. 
Examples of the finite contributions to $\hat{H}_{\textrm{3rd}}^{\textrm{KE}-\textrm{KE}-LS}$ and 
$\hat{H}_{\textrm{3rd}}^{LS-\textrm{KE}-\textrm{KE}}$ 
are schematically shown in Figs. \ref{fig5}(a) and \ref{fig5}(b), respectively. 

\begin{figure}[tb]
\includegraphics[width=86mm]{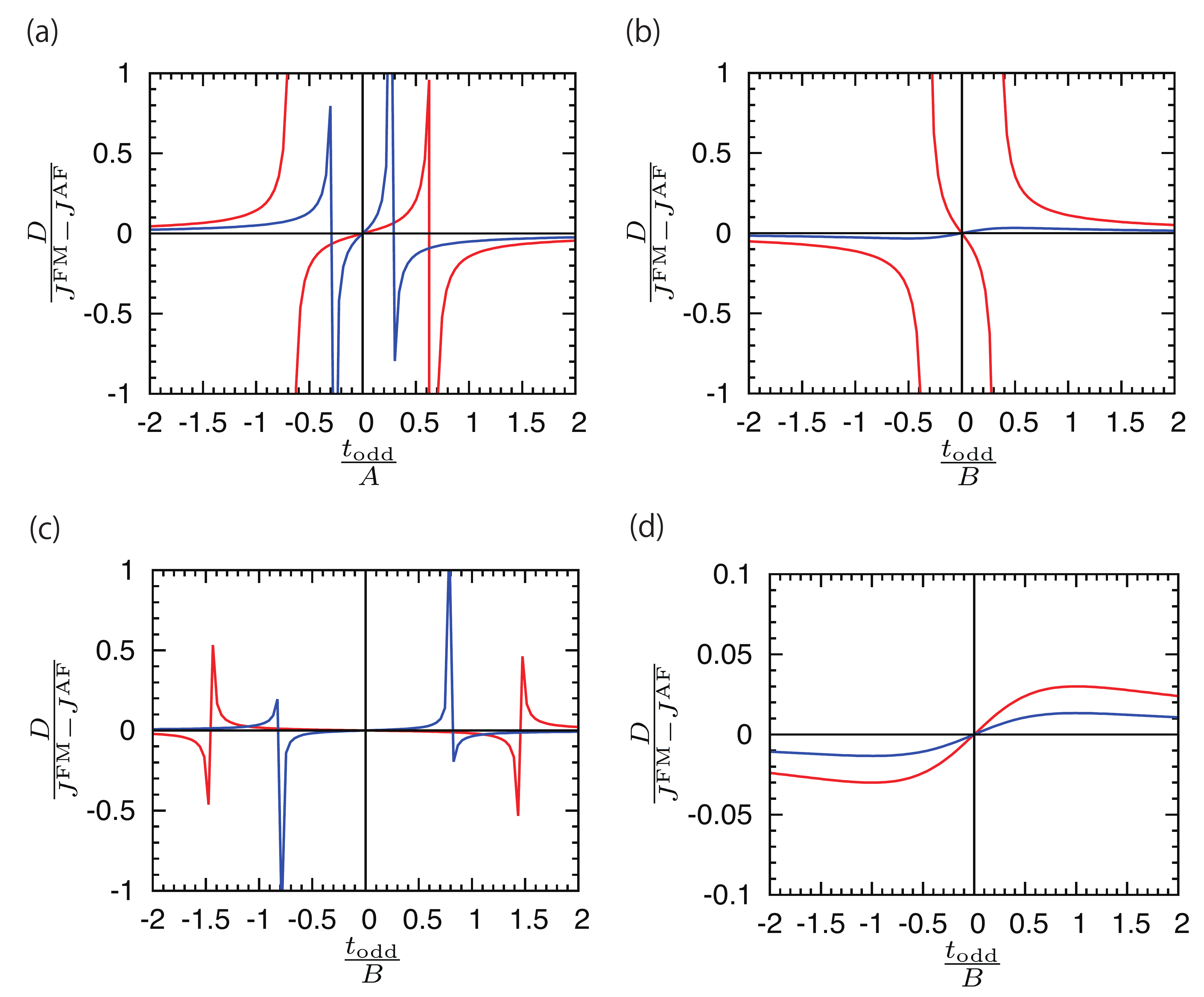}
\caption{
$\frac{D}{J^{\textrm{FM}}-J^{\textrm{AF}}}$ 
as a function of $\frac{t_{\textrm{odd}}}{A}$ in (a) $A\gg B$ 
or $\frac{t_{\textrm{odd}}}{B}$ in (b) $A\ll B$ or (c) $A\sim B$ or (d) $A\sim -B$ 
for $\frac{J_{\textrm{H}}}{U}=0.1$ (red lines) and $0.3$ (blue lines). 
$\frac{D}{J^{\textrm{FM}}-J^{\textrm{AF}}}$ were estimated by 
considering the $O(\frac{J_{\textrm{H}}}{U})$ terms 
at $J^{\prime}=J_{\textrm{H}}$ and $U^{\prime}=U-2J_{\textrm{H}}$ 
and setting $\lambda_{LS}=0.03$ eV and $\Delta_{\textrm{tri}}=1$ eV 
in Eqs. (\ref{eq:D-a1g-eg_rough-case1}){--}(\ref{eq:D-a1g-eg_rough-case4}). 
Those four cases are different from the four limiting cases 
considered in Figs. \ref{figAd1} and \ref{figAd3}.}
\label{figAd2}
\end{figure}

Before turning to the derivation of $\hat{H}_{\textrm{3rd}}^{\textrm{KE}-LS-\textrm{KE}}$, 
we remark on several properties seen from the derived DM interaction 
because those are useful to understand 
the microscopic origin and ways to control it. 
First, from Eq. (\ref{eq:H-K-K-L-explicit}) and its dagger terms 
with Eqs. (\ref{eq:J-AFS}){--}(\ref{eq:J-FMA}), 
we deduce that 
the DM interaction in 
$\hat{H}_{\textrm{3rd}}^{\textrm{KE}-\textrm{KE}-LS}$ and 
$\hat{H}_{\textrm{3rd}}^{LS-\textrm{KE}-\textrm{KE}}$ 
arises from 
the combination of the antisymmetric terms of $\hat{H}_{\textrm{exch}}^{a_{1g}\textrm{-}e_{g}^{g_{e}}}$ 
and the terms of $\hat{H}_{\textrm{exch}}^{e_{g}^{g_{e}}\textrm{-}a_{1g}}$.  
This highlights the importance of the mirror-mixing effect 
due to the combination of the antisymmetric kinetic exchange 
using the even-mirror hopping integrals and the odd-mirror hopping integral 
and the excitation of the $LS$ coupling between the $a_{1g}$ orbital 
and the $e_{g}^{+}$ or $e_{g}^{-}$ orbital 
at one of the two sites. 
By a simpler argument, 
we can deduce from Eq. (\ref{eq:HDM-a1g-eg}) 
the importance of the multiorbital properties and the mirror-mixing effect 
in the DM interaction, 
as we see in Appendix C. 
Then, the combination of the antisymmetric kinetic exchange 
and the excitation of the $LS$ coupling 
is vital to obtain the DM interaction 
in the weak SOC 
because both give the antisymmetry between the two sites 
and because the combination of those antisymmetries 
is necessary to obtain the finite matrix elements 
of the perturbation terms 
between $|\textrm{i}\rangle$ and $|\textrm{f}\rangle$ 
in $\hat{H}_{\textrm{3rd}}^{\textrm{KE}-\textrm{KE}-LS}$ and 
$\hat{H}_{\textrm{3rd}}^{LS-\textrm{KE}-\textrm{KE}}$. 
In addition, 
the importance of the mirror-mixing effect 
shows the importance of the activation of 
the orbital angular momentum at one of the two sites 
in the intermediate states in the perturbation terms. 
This is because the orbital angular momentum is quenched 
in the non-perturbed states, i.e., $|\textrm{i}\rangle$ and $|\textrm{f}\rangle$, 
and because $\hat{H}_{\textrm{exch}}^{a_{1g}\textrm{-}e_{g}^{g_{e}}}$ 
and $\hat{H}_{\textrm{exch}}^{e_{g}^{g_{e}}\textrm{-}a_{1g}}$ 
cause the excitations between the $a_{1g}$ orbital and the $e_{g}^{+}$ or $e_{g}^{-}$ orbital 
at one of the two sites (see Fig. \ref{fig5}); 
in the degenerate $e_{g}^{+}$ and $e_{g}^{-}$ orbitals, 
the orbital angular momentum is active, i.e., nonquenched. 
This important effect about the orbital angular momentum 
cannot be taken into account if the SOC is expressed in terms of 
the spin-gauge potential (see Sec. II D).
As we will discuss in more detail in Sec. IV, 
the importance of the combination of the even-mirror and the odd-mirror hopping integral 
was not revealed in Moriya's microscopic theory~\cite{DM2} 
due to choosing a single parameter of the hopping integrals, 
and this microscopic origin of the DM interaction 
highlights 
the microscopic role of lack of the inversion center 
and 
the similarity to the microscopic origin  
in the strong SOC~\cite{NA-Ir}.  
Then, 
Eq. (\ref{eq:D-a1g-eg}) shows that 
the coefficient of the DM interaction in the weak SOC  
is given by the product of a ratio of $\lambda_{LS}$ to $\Delta_{\textrm{tri}}$ 
and the difference between the FM and the AF antisymmetric exchange interaction of 
$\hat{H}_{\textrm{exch}}^{a_{1g}\textrm{-}e_{g}^{g_{e}}}$. 
Thus, we can control the magnitude and sign of the DM interaction 
by tuning the relative strength of the FM and the AF exchange interactions 
(e.g., as a result of changing $J_{\textrm{H}}$) 
or by changing the magnitude and sign of $t_{\textrm{odd}}$ 
(as a result of changing the position of O ions). 
Those properties are clearly seen from Figs. \ref{figAd2}(a){--}(d); 
those figures are about 
the results of the rough estimation of $\frac{D}{J^{\textrm{FM}}-J^{\textrm{AF}}}$ 
within the $O(\frac{J_{\textrm{H}}}{U})$ terms at $J^{\prime}=J_{\textrm{H}}$ 
and $U^{\prime}=U-2J_{\textrm{H}}$ 
in four cases, 
(a) $A\gg B$, (b) $A\ll B$, (c) $A\sim B$, and (d) $A\sim -B$, 
which are different from the four limiting cases considered in Sec. III A and III B 2. 
The detail of this estimation is described in Appendix D. 

\subsubsection{$\hat{H}_{\textrm{3rd}}^{\textrm{KE}-LS-\textrm{KE}}$}

We turn to the calculation of $\hat{H}_{\textrm{3rd}}^{\textrm{KE}-LS-\textrm{KE}}$ 
for two sites of sublattices $1$ and $2$. 
Since $\langle n_{3}|\hat{H}_{LS}|n_{4}\rangle$ 
in the third term of Eq. (\ref{eq:3rd-total}) 
represents the matrix elements 
between the different irreducible representations for the $d^{2}$ states 
for $\hat{H}_{\textrm{int}}$, 
$\hat{H}_{\textrm{3rd}}^{\textrm{KE}-LS-\textrm{KE}}$ 
can be obtained by using the results 
of $\langle n_{4}|\hat{H}_{\textrm{KE}}|\textrm{i}\rangle$ 
and $\langle \textrm{f}|\hat{H}_{\textrm{KE}}|n_{3}\rangle$ 
in the second-order perturbation terms 
with $\langle n_{3}|\hat{H}_{LS}|n_{4}\rangle$, 
such as Eqs. (\ref{eq:HLS-Hint-1}) and (\ref{eq:HLS-Hint-2}). 
By combining the contributions from every irreducible representation, 
$(\hat{H}_{\textrm{3rd}}^{\textrm{KE}-LS-\textrm{KE}})_{\boldone\boldtwo}$ becomes
\begin{align}
&(\hat{H}_{\textrm{3rd}}^{\textrm{KE}-LS-\textrm{KE}})_{\boldone\boldtwo}\notag\\
=&
-D^{\prime}(\hat{S}_{\boldone}^{y}\hat{S}_{\boldtwo}^{z}
-\hat{S}_{\boldone}^{z}\hat{S}_{\boldtwo}^{y})
+D^{\prime}(\hat{S}_{\boldone}^{z}\hat{S}_{\boldtwo}^{x}-\hat{S}_{\boldone}^{x}\hat{S}_{\boldtwo}^{z}),
\label{eq:HDM-CoulombMultiplet}
\end{align}
with 
\begin{align}
D^{\prime}
=
&\frac{8\lambda_{LS}}{9}
\frac{t_{\textrm{odd}}(t_{1}+2t_{2}+2t_{3})}{(U^{\prime}-J_{\textrm{H}})(U+2J^{\prime})}\notag\\
&+\frac{4\lambda_{LS}}{9}
\frac{t_{\textrm{odd}}(-t_{1}+4t_{2}+4t_{3})}{(U^{\prime}+J_{\textrm{H}})(U^{\prime}-J_{\textrm{H}})}\notag\\
&+\frac{8\lambda_{LS}}{9}
\frac{t_{\textrm{odd}}(t_{1}-t_{2}-t_{3})}{9(U^{\prime}-J_{\textrm{H}})(U-J^{\prime})},
\label{eq:D-CoulombMultiplet} 
\end{align}
as derived in Appendix E. 
An example of the finite contributions to 
$(\hat{H}_{\textrm{3rd}}^{\textrm{KE}-LS-\textrm{KE}})_{\boldone\boldtwo}$ 
is schematically shown in Fig. \ref{fig5}(c). 

\begin{figure}[tb]
\includegraphics[width=84mm]{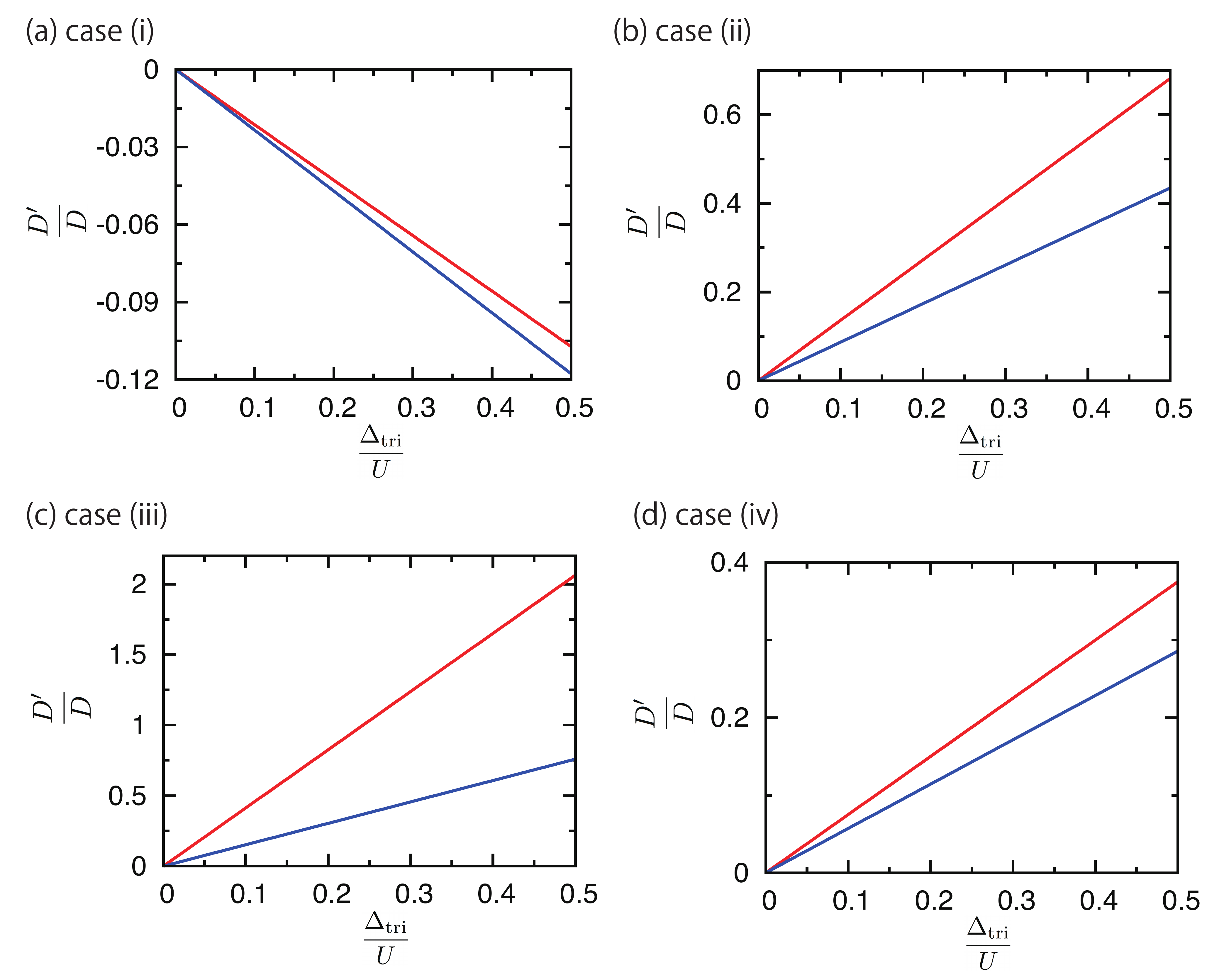}
\caption{$\frac{\Delta_{\textrm{tri}}}{U}$ dependence of $\frac{D^{\prime}}{D}$ 
in the four limiting cases considered in Fig. \ref{figAd1} 
for $\frac{J_{\textrm{H}}}{U}=0.1$ (red lines) and $0.3$ (blue lines).}
\label{figAd3}
\end{figure}

The DM interaction of Eq. (\ref{eq:HDM-CoulombMultiplet}) 
has the same symmetry as Eq. (\ref{eq:HDM-a1g-eg}), 
and gives another contribution to the DM interaction between sublattices $1$ and $2$. 
As discussed in Sec. IV, 
this contribution was missing in Moriya's theory~\cite{DM2} 
because that considered 
a special case of the Hubbard interactions, 
i.e. $U^{\prime}=U$ and $J_{\textrm{H}}=J^{\prime}=0$, 
in which 
the irreducible representations merged into one state. 
However, 
the contribution using the inter-$d^{2}$-multiplet excitations 
is much smaller than the contribution 
using the interorbital excitations 
between the $a_{1g}$ orbital and the $e_{g}^{+}$ or $e_{g}^{-}$ orbital 
in our considered case 
because $\frac{D^{\prime}}{D}\sim O(\frac{\Delta_{\textrm{tri}}}{U_{\Gamma}})$ is negligible 
in $U_{\Gamma}\gg \Delta_{\textrm{tri}}$. 
If $\Delta_{\textrm{tri}}$ becomes not much smaller than $U_{\Gamma}$, 
we should consider not only the contribution from 
$\hat{H}_{\textrm{3rd}}^{\textrm{KE}-\textrm{KE}-LS}+\hat{H}_{\textrm{3rd}}^{LS-\textrm{KE}-\textrm{KE}}$ 
but also the contribution from $\hat{H}_{\textrm{3rd}}^{\textrm{KE}-LS-\textrm{KE}}$. 
More precise estimation of $\frac{D^{\prime}}{D}$ can be carried out 
by the rough calculation which is similar to that 
for $J^{\textrm{FM}}-J^{\textrm{AF}}$ in Sec. III A. 
Namely, 
$\frac{D^{\prime}}{D}$ within the $O(\frac{J_{\textrm{H}}}{U})$ terms 
for $J^{\prime}=J_{\textrm{H}}$ and $U^{\prime}=U-2J_{\textrm{H}}$ is given by
\begin{align}
\frac{D^{\prime}}{D}
=\frac{\Delta_{\textrm{tri}}}{U}
\frac{(A+B)(1+2\frac{J_{\textrm{H}}}{U})}{(A-2B)+6A\frac{J_{\textrm{H}}}{U}},
\end{align}
and 
the leading terms in the cases (a), (b), (c), and (d) 
considered in Sec. III A are, respectively, 
\begin{align}
&\frac{D^{\prime}}{D}
=-\frac{\Delta_{\textrm{tri}}}{U}
\frac{1+2\frac{J_{\textrm{H}}}{U}}{5+6\frac{J_{\textrm{H}}}{U}},\\
&\frac{D^{\prime}}{D}
=\frac{5\Delta_{\textrm{tri}}}{2U}
\frac{1+2\frac{J_{\textrm{H}}}{U}}{1+12\frac{J_{\textrm{H}}}{U}},\\
&\frac{D^{\prime}}{D}
=-\frac{11\Delta_{\textrm{tri}}}{U}
\frac{1+2\frac{J_{\textrm{H}}}{U}}{1-42\frac{J_{\textrm{H}}}{U}},
\end{align}
and
\begin{align}
\frac{D^{\prime}}{D}
=\frac{\Delta_{\textrm{tri}}}{U}
\frac{1+2\frac{J_{\textrm{H}}}{U}}{1+6\frac{J_{\textrm{H}}}{U}}.
\end{align}
The leading terms depend on $\frac{\Delta_{\textrm{tri}}}{U}$ 
in the way shown in Fig. \ref{figAd3}. 
If we set $\Delta_{\textrm{tri}}=1$ eV and $U=3$ eV (i.e., 
$\frac{\Delta_{\textrm{tri}}}{U}\sim 0.3$) in Fig. \ref{figAd3}(d), 
we find that the effect of the $D^{\prime}$ term 
is a small magnitude increase of the coefficient. 

As the case with the remarks in Sec. III B 1, 
we deduce several properties of the derived DM interaction 
from Eqs. (\ref{eq:HDM-CoulombMultiplet}) and (\ref{eq:D-CoulombMultiplet}). 
First, 
the mirror-mixing effect is important even for 
this derived DM interaction 
because the numerators of $D^{\prime}$ 
consist of the products of the even-mirror and odd-mirror hopping integrals: 
the DM interaction using the inter-$d^{2}$-multiplet excitations 
can be also understood as the mirror-mixing effect. 
The similarity between $\hat{H}_{\textrm{3rd}}^{\textrm{KE}-\textrm{KE}-LS}+
\hat{H}_{\textrm{3rd}}^{LS-\textrm{KE}-\textrm{KE}}$ and 
$\hat{H}_{\textrm{3rd}}^{\textrm{KE}-LS-\textrm{KE}}$ 
can be more easily seen by rewriting $\hat{H}_{\textrm{3rd}}^{\textrm{KE}-LS-\textrm{KE}}$ as
\begin{align}
&\hat{H}_{\textrm{3rd}}^{\textrm{KE}-LS-\textrm{KE}}\notag\\
=&
\frac{1}{2}\sum\limits_{n_{3},n_{4}}
\frac{\langle \textrm{f}|\hat{H}_{\textrm{KE}}|n_{3}\rangle 
\langle n_{3}|\hat{\phi}\hat{H}_{LS}|n_{4}\rangle 
\langle n_{4}|\hat{\phi}\hat{H}_{\textrm{KE}}|\textrm{i}\rangle}
{(E_{0}-E_{n_{3}})(E_{0}-E_{n_{4}})}|\textrm{f}\rangle \langle \textrm{i}|\notag\\
&+
\frac{1}{2}\sum\limits_{n_{3},n_{4}}
\frac{\langle \textrm{f}|\hat{H}_{\textrm{KE}}|n_{4}\rangle 
\langle n_{4}|\hat{\phi}\hat{H}_{LS}|n_{3}\rangle 
\langle n_{3}|\hat{\phi}\hat{H}_{\textrm{KE}}|\textrm{i}\rangle}
{(E_{0}-E_{n_{4}})(E_{0}-E_{n_{3}})}|\textrm{f}\rangle \langle \textrm{i}|. 
\label{eq:H3rd-KLK-decomp}
\end{align}
Namely, 
the relation between the first and the second term of Eq. (\ref{eq:H3rd-KLK-decomp}) 
is similar to the relation between $\hat{H}_{\textrm{3rd}}^{\textrm{KE}-\textrm{KE}-LS}$ 
and $\hat{H}_{\textrm{3rd}}^{LS-\textrm{KE}-\textrm{KE}}$. 
Then, 
the magnitude and sign of $D^{\prime}$ can be controlled 
by varying $J_{\textrm{H}}$ or $t_{\textrm{odd}}$.

\section{Discussion}

In this section, 
we discuss four points 
in order to clarify the meanings of our achievements. 
We first compare our results of the DM interaction derived microscopically 
with the result derived phenomenologically~\cite{DMI-pyro}, 
and show what are new findings of our study. 
We next compare our microscopic theory formulated in this paper 
and Moriya's microscopic theory~\cite{DM2}, 
and deduce the similarities and differences between them. 
Then, 
we argue the general applicability of our mechanism 
to the DM interaction in solids with the weak SOC. 
Finally, 
by comparison with the previous result 
in the strong SOC~\cite{NA-Ir}, 
we address the similarities and differences 
of the effects of the $LS$ coupling 
on the low-energy effective Hamiltonian. 

We begin with the comparison with the phenomenological theory~\cite{DMI-pyro} 
based on Moriya's rule~\cite{DM2}. 
By adopting Moriya's rule to the symmetry of the pyrochlore crystal, 
we can determine which terms of the DM interaction are permissible 
under the symmetry. 
Since the result of this phenomenological theory results from the symmetry, 
the finite components of the DM interaction obtained in the microscopic theory 
should be the same as that obtained in the phenomenological theory. 
Our results of the finite components are consistent with 
the results of the phenomenological theory~\cite{DMI-pyro}. 
Then, 
the phenomenological theory cannot reveal the microscopic origin of the DM interaction. 
On the other hand, 
our theory reveals the microscopic origin: 
the microscopic origin is 
the mirror-mixing effect 
by using the antisymmetric kinetic exchange of 
the even-mirror hopping and the odd-mirror hopping 
and the different-energy excitations of the $LS$ coupling. 
In that effect, 
the role of the inversion-center lacking for each V-V bond 
is to induce the odd-mirror hopping integrals; 
since the even-mirror hopping integrals exist even with the inversion center, 
the inversion-center lacking is essential to obtain 
the antisymmetric kinetic exchange. 
Moreover, 
our theory can explain the microscopic reason why 
the mirror symmetry plays important roles 
in the phenomenological theory: 
the reason is that 
which plane's mirror symmetry is broken 
is linked with the kind of the permissible hopping integrals, 
which is important to discuss the mirror-mixing effect. 
Then, we highlight the close relation between the DM interaction 
and the sublattice structure for pyrochlore oxides: 
if we consider the DM interaction in pyrochlore oxides, 
we should simultaneously take account of the four-sublattice structure, 
shown in Fig. \ref{fig1}(a), 
because the inversion-center lacking causes 
not only the odd-mirror hopping 
but also the difference among four V ions in a tetrahedron 
as a result of the differences in the displacement of the O ion 
of each V-O-V bond. 
This relation provides 
an important restriction on the theory 
analyzing the DM interaction in pyrochlore oxides. 

We next clarify the similarities and differences between 
the present theory and Moriya's microscopic theory~\cite{DM2}. 
In Moriya's microscopic theory, 
the SOC, one parameter of the kinetic term, 
and one parameter of the interaction term are considered, 
the third-order perturbation terms 
corresponding to 
$\hat{H}_{\textrm{3rd}}^{\textrm{KE}-\textrm{KE}-LS}
+\hat{H}_{\textrm{3rd}}^{LS-\textrm{KE}-\textrm{KE}}$ 
in the present theory are calculated, 
and the DM interaction is obtained by using the excitations between 
the non-degenerate orbitals due to the SOC; 
the one parameter of the kinetic term 
corresponds to parameterizing all the hopping integrals by a single parameter; 
the one parameter of the interaction term is $U$, 
corresponding to a special case with $U^{\prime}=U$ and $J_{\textrm{H}}=J^{\prime}=0$. 
On the other hand, 
the present theory includes the much more details 
of the kinetic term and interaction term, 
calculates not only $\hat{H}_{\textrm{3rd}}^{\textrm{KE}-\textrm{KE}-LS}
+\hat{H}_{\textrm{3rd}}^{LS-\textrm{KE}-\textrm{KE}}$ 
but also another third-order perturbation term, 
$\hat{H}_{\textrm{3rd}}^{\textrm{KE}-LS-\textrm{KE}}$, 
and obtains two different contributions to the DM interaction;  
the contribution from another is finite 
except the special case with $U^{\prime}=U$ and $J_{\textrm{H}}=J^{\prime}=0$. 
Roughly speaking, 
by setting $t_{1}=t_{2}=t_{3}=t_{\textrm{odd}}$, $U^{\prime}=U$, and $J_{\textrm{H}}=J^{\prime}=0$, 
our theory reduces to Moriya's microscopic theory. 
Thus, the present theory reveals another contribution to the DM interaction 
within the third-order perturbation  
using the kinetic term twice and the SOC once 
as a result of the difference in the four Hubbard interactions. 
In addition, 
the present theory clarifies 
the microscopic origin why the inversion-center lacking is necessary 
to obtain the DM interaction in the weak SOC 
as a result of the difference in the hopping integrals: 
the inversion-center lacking is necessary to 
get the antisymmetric kinetic exchange by using 
the even-mirror hopping once and 
the odd-mirror hopping once. 
Moreover, 
the present theory finds the dependence of the coefficients of the DM interaction 
on the typical four multiorbital Hubbard interactions. 
In particular, 
the simple relation between the difference between 
the FM and AF antisymmetric exchange interactions 
and the coefficient obtained for $\hat{H}_{\textrm{3rd}}^{\textrm{KE}-\textrm{KE}-LS}
+\hat{H}_{\textrm{3rd}}^{LS-\textrm{KE}-\textrm{KE}}$ 
is revealed. 
Those achievements develop 
our understanding of the physics of the DM interaction 
and suggest how to control it. 

Then, 
we argue the general applicability of our mechanism. 
As we demonstrate below, 
our mechanism can explain the emergence of the DM interaction in any solids 
for the weak SOC 
if both the even-mirror hopping integral and the odd-mirror hopping integral 
are permissible 
and if there are at least two nondegenerate orbitals 
which are connected by the $LS$ coupling. 
To demonstrate this generality, 
we consider two cases, 
a $d$-electron system in a cubic symmetry as a high-symmetry case 
and 
a $t_{2g}$-electron system in a tetragonal symmetry as a low-symmetry case. 
In the cubic symmetry, 
the five $d$ orbitals split into the low-energy $t_{2g}$ orbitals 
and the high-energy $e_{g}$ orbitals. 
Whichever partially occupied cases [e.g., $(t_{2g})^{1}$ or $(t_{2g})^{3}$ configuration] 
we consider, 
the emergence of the DM interaction can be explained 
as the mirror-mixing effect 
if the odd-mirror hopping integral between the $t_{2g}$ and $e_{g}$ orbitals is permissible. 
This is because 
the $LS$ coupling connects the $t_{2g}$ and $e_{g}$ orbitals 
and because the mirror-mixing effect using the even-mirror hopping integral, 
the odd-mirror hopping integral between the $t_{2g}$ and $e_{g}$ orbitals, 
and the $LS$ coupling between them leads to the DM interaction, 
whose magnitude is either 
$O(\frac{t^{(\textrm{even})}t^{(\textrm{odd})}\lambda_{LS}}{U\Delta_{\textrm{cub}}})$ 
for $U>\Delta_{\textrm{cub}}$ 
or 
$O(\frac{t^{(\textrm{even})}t^{(\textrm{odd})}\lambda_{LS}}{\Delta_{\textrm{cub}}^{2}})$ 
for $U<\Delta_{\textrm{cub}}$, 
with $\Delta_{\textrm{cub}}$, the energy difference between the $t_{2g}$ and $e_{g}$ orbitals. 
This cubic case for the $(t_{2g})^{3}$ configuration 
corresponds to, for example, 
CdCr$_{2}$O$_{4}$~\cite{InelaNeutron-Cr124} at high temperature; 
here, 
we have neglected the effects of the trigonal-distortion potential 
because that may be small in AB$_{2}$O$_{4}$-type pyrochlore oxides 
(e.g., $\sim 0.1$ eV for LiV$_{2}$O$_{4}$~\cite{LDA-V124-Anisimov}). 
Then, 
in the tetragonal symmetry, 
the $t_{2g}$ orbitals, which are degenerate in the cubic symmetry, 
split into the $d_{xy}$ orbital and the degenerate $d_{xz}$ and $d_{yz}$ orbitals. 
The degenerate $d_{xz}$ and $d_{yz}$ orbitals are low-energy 
for the $c$ axis longer than the $a$ and $b$ axes (i.e., $c> a=b$), 
while the $d_{xy}$ orbital is low-energy for the shorter $c$ axis (i.e., $c < a=b$). 
The DM interaction 
in this tetragonal case 
can be also understood as the mirror-mixing effect 
using the even-mirror hopping integral, 
the odd-mirror hopping integral between the $d_{xy}$ orbital 
and the $d_{xz}$ or $d_{yz}$ orbital, 
and the $LS$ coupling between them; 
the magnitude is either 
$O(\frac{t^{(\textrm{even})}t^{(\textrm{odd})}\lambda_{LS}}{U\Delta_{\textrm{tetra}}})$ 
for $U>\Delta_{\textrm{tetra}}$ 
or 
$O(\frac{t^{(\textrm{even})}t^{(\textrm{odd})}\lambda_{LS}}{\Delta_{\textrm{tetra}}^{2}})$ 
for $U<\Delta_{\textrm{tetra}}$, 
with $\Delta_{\textrm{tetra}}$, the energy difference between the $d_{xy}$ orbital 
and the $d_{xz}$ or $d_{yz}$ orbital. 
For example, 
this tetragonal case with $c>a=b$ or $c<a=b$ for the $(t_{2g})^{3}$ configuration 
corresponds to CdCr$_{2}$O$_{4}$~\cite{InelaNeutron-Cr124} at low temperature 
or ZnCr$_{2}$O$_{4}$~\cite{Zn-Cr124} at low temperature, 
respectively 
(the effect of the trigonal distortion is neglected). 
Moreover, 
in the similar way, 
we can understand the DM interaction 
in not only other $d$-electron systems 
but also $p$-electron or $f$-electron systems 
with the weak SOC. 
This is because even for the DM interaction in those systems, 
the multiorbital properties and the mirror-mixing effect 
remain important to get the two characteristic properties of the DM interaction, 
as in the case of Appendix C. 
Thus, 
our mechanism provides the general mechanism for the DM interaction 
in solids with the weak SOC. 

Finally, we compare 
the present theory in the weak-SOC system 
with the previous microscopic theory~\cite{NA-Ir} 
in the strong-SOC system. 
In the latter theory, 
the low-energy effective model is derived for a $d^{5}$ Mott insulator 
in a quasi-two-dimensional $t_{2g}$-orbital Hubbard model 
on a square lattice with the inversion-symmetry breaking 
of an $ab$ plane, 
and the DM interaction is obtained in the second-order perturbation, 
i.e., the same order of magnitude as the superexchange interaction. 
The difference between the two theories 
is the order of the perturbation to obtain the DM interaction. 
This can be understood as the difference in the effect of the $LS$ coupling: 
in a weak-SOC system, 
the effect of the $LS$ coupling can be treated perturbatively, 
and the orbital angular momentum is quenched; 
in a strong-SOC system, 
the nonperturbative treatment of the $LS$ coupling becomes necessary, and 
the effects of the $LS$ coupling causes the formation of the pseudospin  
as a result of the addition of the spin and orbital angular momenta. 
Namely, 
in the weak-SOC system, 
the combination of the second-order perturbation 
using the kinetic terms and the one-shot perturbation of the $LS$ coupling 
is necessary to obtain the DM interaction 
because 
the $LS$ coupling activates the orbital angular momentum, 
which is quenched in the nonperturbed states, 
at one of the two sites; 
in the strong-SOC system, 
the second-order perturbation 
using the kinetic terms is sufficient 
because the orbital angular momenta are not quenched 
and are not conserved quantities; 
i.e., the antisymmetry between 
the orbital angular momenta at two sites 
can be realized even in the nonperturbed states for the pseudospin. 
Then, 
the similarity between the two theories is the origin of the DM interaction, 
which is the mirror-mixing effect. 
Actually, 
the coefficient of the DM interaction 
in the strong-SOC system 
is given by 
the antisymmetric kinetic exchange, using the even-mirror hopping once 
and the odd-mirror hopping once; 
in the quasi-two-dimensional $t_{2g}$-orbital Hubbard model 
on a square lattice, 
the inversion-symmetry breaking of an $ab$ plane 
induces the odd-mirror nearest-neighbor hoppings 
between the $d_{yz}$ and $d_{xy}$ orbitals along the $x$ direction 
and between the $d_{xz}$ and $d_{xy}$ orbitals along the $y$ direction 
due to the similar mechanism for the present theory 
(to see the similarity, compare the derivations in Sec. II A and 
Sec. I of the Supplemental Material of Ref. \onlinecite{Mizoguchi-SHE}). 
The above comparisons show that 
the DM interaction in solids can be understood in a unified way 
in the microscopic theories for the multiorbital models with the $LS$ coupling, 
in which the effects of the inversion-center lacking are appropriately treated 
in the kinetic energy. 

\section{Summary}
In summary, 
we constructed the $t_{2g}$-orbital model for pyrochlore oxides, 
derived the low-energy effective Hamiltonian 
for the $d^{1}$ Mott insulator with the weak SOC, 
and clarified 
the microscopic origin of the DM interaction. 
First, 
the $t_{2g}$-orbital model 
was constructed by 
considering four terms: 
(1) the kinetic energy for not only the even-mirror hopping integrals, 
which exist even with the inversion center of each nearest-neighbor V-V bond, 
but also the odd-mirror hopping integrals, 
which appear only without the inversion center 
due to the indirect hoppings through the O-$2p$ orbitals, 
(2) the trigonal-distortion potential, 
(3) the $t_{2g}$-orbital Hubbard interactions, 
and (4) the $LS$ coupling. 
The main difference between this and 
the previous $t_{2g}$-orbital models~\cite{V124-Tsune,V124-Ueda,V124-Hattori,IrPyro-Kim} 
is the existence of the odd-mirror hopping integrals. 
Then, 
the low-energy effective Hamiltonian 
for the $d^{1}$ Mott insulator, where 
one electron occupied the $a_{1g}$ orbital per site, 
was derived from the second-order perturbation terms 
using the nearest-neighbor hopping integrals twice 
and from the third-order perturbation terms 
using the hopping integrals twice and the $LS$ coupling once; 
the second-order perturbation terms give 
the FM and the AF superexchange interactions, 
and the third-order perturbation terms give 
the DM interactions, 
whose symmetry is the same for the phenomenological theory~\cite{DMI-pyro}. 
The derived DM interactions consist of two contributions: 
one, corresponding to the DM interaction considered 
in Moriya's microscopic theory~\cite{DM2}, 
arises from the combination 
of the interorbital excitation of the $LS$ coupling 
between the $a_{1g}$ and the $e_{g}^{+}$ or $e_{g}^{-}$ orbital 
and the antisymmetric kinetic exchange, using the even-mirror hopping once 
and the odd-mirror hopping once; 
the other, 
which was missing in Moriya's microscopic theory~\cite{DM2}, 
arises from the combination of the different-energy $d^{2}$-multiplet excitation 
of the $LS$ coupling, 
and the even-mirror hopping integral and the odd-mirror hopping integral. 
The latter contribution appears except 
the special case with $U^{\prime}=U$ and $J_{\textrm{H}}=J^{\prime}=0$; 
the former contribution 
is dominant as long as the interactions are larger 
than the trigonal-distortion potential. 
The coefficients of those contributions revealed 
not only the importance of the mirror-mixing effect, 
but also the methods to control the DM interaction. 
One method is to tune the magnitude and sign of the odd-mirror hopping integral 
by changing the positions of the O ions; 
another is to tune the AF and the FM interactions 
by changing the $t_{2g}$-orbital Hubbard interactions. 

Those achievements develop our understanding of the DM interaction 
in the weak SOC. 
In addition, 
our results showed  
the restriction on the theories studying the DM interaction 
in pyrochlore oxides: 
for such study, 
we should appropriately treat the four-sublattice structure of the V ions. 
Moreover, 
since we can apply the present theory to other multiorbital systems 
with the weak SOC, 
the present theory provides the general formalism 
to study the DM interaction in solids with the weak SOC.


\begin{acknowledgments}
I thank H. Tsunetsugu 
for pointing out that 
the essential symmetry in discussing the hopping integral 
under the inversion-center lacking is 
the mirror symmetry, and indicating the additional analyses 
about the sign of $J^{\textrm{FM}}-J^{\textrm{AF}}$, $\frac{D}{J^{\textrm{FM}}-J^{\textrm{AF}}}$, 
and $\frac{D^{\prime}}{D}$. 
I also thank K. Riedl 
for useful discussions about Ref. \onlinecite{LDA-LS-V227}.  
\end{acknowledgments}

\appendix

\section{Derivation of Eq. (\ref{eq:superexchange})}
In this appendix, 
we derive Eq. (\ref{eq:superexchange}) from Eq. (\ref{eq:H2nd}) for our model. 
Equation (\ref{eq:superexchange}) is derived by calculating the contribution 
from each $|\boldi;\Gamma,g_{\Gamma}\rangle$ in Eq. (\ref{eq:H2nd}), 
and the calculation from each contribution 
is similar to the calculation for a single-orbital system~\cite{Anderson}. 
First, 
the contribution from $|\boldi;\Gamma,g_{\Gamma}\rangle
=|\boldi;A_{1}\rangle$ becomes
\begin{align}
-\frac{4}{27}
\frac{(t_{1}+2t_{2}+2t_{3})^{2}}{U+2J^{{\prime}}}
(\frac{1}{4}\hat{n}_{\boldone}\hat{n}_{\boldtwo}
-\hat{\bolds}_{\boldone}\cdot\hat{\bolds}_{\boldtwo}).\label{eq:H2nd12-1}
\end{align} 
In this calculation, 
we first calculated the finite terms of 
$\langle \boldi;A_{1}|\hat{H}_{0}+\hat{H}_{\textrm{odd}}|a_{1g}^{s_{1}},a_{1g}^{s_{2}}\rangle$ 
for $\boldi=\boldone, \boldtwo$ and $s_{1},s_{2}=\uparrow,\downarrow$, 
and then expressed the contribution 
by using the finite matrix elements, 
$E_{A_{1}}=U+2J^{\prime}$, and 
the corresponding operator part 
$| a_{1g}^{s_{3}}, a_{1g}^{s_{4}}\rangle \langle a_{1g}^{s_{1}},a_{1g}^{s_{2}}|$. 
By the similar calculations, 
we obtain the contributions from 
the other $|\boldi;\Gamma,g_{\Gamma}\rangle$: 
the contributions from $|\boldi;\Gamma,g_{\Gamma}\rangle
=|\boldi;E,u\rangle$, $|\boldi;E,v\rangle$, 
$|\boldi;T_{1},\xi_{\pm}\rangle$, 
$|\boldi;T_{1},\eta_{\pm}\rangle$,  
$|\boldi;T_{2},\zeta_{0}\rangle$, 
$|\boldi;T_{1},\xi_{0}\rangle$,
$|\boldi;T_{2},\xi_{0}\rangle$, 
$|\boldi;T_{1},\eta_{0}\rangle$,
and $|\boldi;T_{2},\eta_{0}\rangle$ are, respectively, 
\begin{align}
&-\frac{2}{27}
\frac{(t_{1}-t_{2}-t_{3})^{2}+9t_{\textrm{odd}}^{2}}{U-J^{\prime}}
(\frac{1}{4}\hat{n}_{\boldone}\hat{n}_{\boldtwo}
-\hat{\bolds}_{\boldone}\cdot\hat{\bolds}_{\boldtwo}),\label{eq:H2nd12-2}\\
&-\frac{2}{9}
\frac{(t_{1}-t_{2}-t_{3})^{2}+9t_{\textrm{odd}}^{2}}{U-J^{\prime}}
(\frac{1}{4}\hat{n}_{\boldone}\hat{n}_{\boldtwo}
-\hat{\bolds}_{\boldone}\cdot\hat{\bolds}_{\boldtwo}),\label{eq:H2nd12-3}\\
&-\frac{2}{9}
\frac{(t_{1}-t_{2}-t_{3})^{2}+9t_{\textrm{odd}}^{2}}{U^{\prime}-J_{\textrm{H}}}
(\frac{1}{2}\hat{n}_{\boldone}\pm \hat{s}_{\boldone}^{z})
(\frac{1}{2}\hat{n}_{\boldtwo}\pm \hat{s}_{\boldtwo}^{z}),\label{eq:H2nd12-4}\\
&-\frac{2}{9}
\frac{(t_{1}-t_{2}-t_{3})^{2}+9t_{\textrm{odd}}^{2}}{U^{\prime}-J_{\textrm{H}}}
(\frac{1}{2}\hat{n}_{\boldone}\pm \hat{s}_{\boldone}^{z})
(\frac{1}{2}\hat{n}_{\boldtwo}\pm \hat{s}_{\boldtwo}^{z}),\label{eq:H2nd12-5}\\
&-\frac{8}{9}
\frac{(t_{2}+t_{3})^{2}+t_{\textrm{odd}}^{2}}{U^{\prime}+J_{\textrm{H}}}
(\frac{1}{4}\hat{n}_{\boldone}\hat{n}_{\boldtwo}
-\hat{\bolds}_{\boldone}\cdot\hat{\bolds}_{\boldtwo}),\label{eq:H2nd12-6}
\end{align}
\begin{align}
&-\frac{2}{9}
\frac{(t_{1}-t_{2}-t_{3})^{2}+9t_{\textrm{odd}}^{2}}{U^{\prime}-J_{\textrm{H}}}
(\frac{1}{4}\hat{n}_{\boldone}\hat{n}_{\boldtwo}
-2\hat{s}_{\boldone}^{z}\hat{s}_{\boldtwo}^{z}
+\hat{\bolds}_{\boldone}\cdot\hat{\bolds}_{\boldtwo}),\label{eq:H2nd12-7}\\
&-\frac{2}{9}
\frac{(t_{1}+t_{2}+t_{3})^{2}+t_{\textrm{odd}}^{2}}{U^{\prime}+J_{\textrm{H}}}
(\frac{1}{4}\hat{n}_{\boldone}\hat{n}_{\boldtwo}
-\hat{\bolds}_{\boldone}\cdot\hat{\bolds}_{\boldtwo}),\label{eq:H2nd12-8}\\
&-\frac{2}{9}
\frac{(t_{1}-t_{2}-t_{3})^{2}+9t_{\textrm{odd}}^{2}}{U^{\prime}-J_{\textrm{H}}}
(\frac{1}{4}\hat{n}_{\boldone}\hat{n}_{\boldtwo}
-2\hat{s}_{\boldone}^{z}\hat{s}_{\boldtwo}^{z}
+\hat{\bolds}_{\boldone}\cdot\hat{\bolds}_{\boldtwo}),\label{eq:H2nd12-9}
\end{align}
and
\begin{align}
&-\frac{2}{9}
\frac{(t_{1}+t_{2}+t_{3})^{2}+t_{\textrm{odd}}^{2}}{U^{\prime}+J_{\textrm{H}}}
(\frac{1}{4}\hat{n}_{\boldone}\hat{n}_{\boldtwo}
-\hat{\bolds}_{\boldone}\cdot\hat{\bolds}_{\boldtwo}),\label{eq:H2nd12-10}
\end{align}
and the contributions from $|\boldi;\Gamma,g_{\Gamma}\rangle
=|\boldi;T_{1},\zeta_{\pm}\rangle$, $|\boldi;T_{1},\zeta_{0}\rangle$ are zero 
for $(\hat{H}_{\textrm{2nd}})_{\boldone \boldtwo}$. 
Combining the contributions from all $|\boldi;\Gamma,g_{\Gamma}\rangle$ 
for $\boldi=\boldone$, $\boldtwo$ with Eq. (\ref{eq:H2nd}), 
we obtain Eq. (\ref{eq:superexchange}) with Eqs. (\ref{eq:JAF}) and (\ref{eq:JFM}).  

\section{Derivation of Eq. (\ref{eq:Hexch-explicit})}
In this appendix, 
we derive Eq. (\ref{eq:Hexch-explicit}). 
This derivation is similar to the derivation of Eq. (\ref{eq:superexchange}), 
described in Appendix A, 
because Eq. (\ref{eq:Hexch}) can be rewritten as 
\begin{align}
&(\hat{H}_{\textrm{exch}}^{a_{1g}\textrm{-}e_{g}^{g_{e}}})_{\boldone\boldtwo}
\approx 
\sum\limits_{\boldi=\boldone,\boldtwo}
\sum\limits_{\Gamma}
\sum\limits_{g_{\Gamma}}
\sum\limits_{n_{2}}
\sum\limits_{s_{3},s_{4}}\notag\\
&\times
\frac{\langle a_{1g}^{s_{3}},a_{1g}^{s_{4}}|\hat{H}_{\textrm{KE}}|\
\boldi;\Gamma,g_{\Gamma}\rangle 
\langle \boldi;\Gamma,g_{\Gamma}|
\hat{H}_{\textrm{KE}}|n_{2}\rangle}
{-E_{\Gamma}}
|a_{1g}^{s_{3}},a_{1g}^{s_{4}}\rangle 
\langle n_{2}|,\label{eq:Hexch-rewrite}
\end{align} 
with $\textstyle\sum_{n_{2}}=
\textstyle\sum_{s_{1},s_{2}}
\textstyle\sum_{|n_{2}\rangle=|a_{1g}^{s_{1}},e_{g}^{g_{e};s_{2}}\rangle,|e_{g}^{g_{e};s_{1}},a_{1g}^{s_{2}}\rangle}$. 
Namely, 
we should calculate the contributions 
from every $|\boldi;\Gamma,g_{\Gamma}\rangle$ to 
$(\hat{H}_{\textrm{exch}}^{a_{1g}\textrm{-}e_{g}^{g_{e}}})_{\boldone\boldtwo}$ 
in the similar way for the second-order perturbation terms 
except taking care of the difference 
between $|a_{1g}^{s_{1}},a_{1g}^{s_{2}}\rangle$ 
and $|a_{1g}^{s_{1}},e_{g}^{g_{e};s_{2}}\rangle$ 
or 
between $|a_{1g}^{s_{1}},a_{1g}^{s_{2}}\rangle$ 
and $|e_{g}^{g_{e};s_{1}},a_{1g}^{s_{2}}\rangle$. 
The calculated result for $|\boldi;\Gamma,g_{\Gamma}\rangle=|\boldi;A_{1}\rangle$ becomes
\begin{widetext} 
\begin{align}
&-\frac{4}{27}
\frac{(t_{1}+2t_{2}+2t_{3})[\omega^{n(g_{e})}(t_{1}-t_{2}-t_{3})]}
{U+2J^{\prime}}
[(\frac{1}{4}\hat{n}_{\boldone}\hat{o}_{\boldtwo}^{g_{e}}
-\hat{\bolds}_{\boldone}\cdot \hat{\boldso}_{\boldtwo}^{g_{e}})
+(\frac{1}{4}\hat{o}_{\boldone}^{g_{e}}\hat{n}_{\boldtwo}
-\hat{\boldso}_{\boldone}^{g_{e}}\cdot \hat{\bolds}_{\boldtwo})]\notag\\
&+\frac{4}{9}
\frac{t_{\textrm{odd}}(t_{1}+2t_{2}+2t_{3})\omega^{n(g_{e})}}
{U+2J^{\prime}}
[(\frac{1}{4}\hat{n}_{\boldone}\hat{o}_{\boldtwo}^{g_{e}}
-\hat{\bolds}_{\boldone}\cdot \hat{\boldso}_{\boldtwo}^{g_{e}})
-(\frac{1}{4}\hat{o}_{\boldone}^{g_{e}}\hat{n}_{\boldtwo}
-\hat{\boldso}_{\boldone}^{g_{e}}\cdot \hat{\bolds}_{\boldtwo})]\label{eq:H3rd-KLK-A1}.
\end{align}
\end{widetext}
As the case with the derivation of Eq. (\ref{eq:superexchange}), 
we first calculated the finite terms of 
$\langle \boldi;A_{1}|\hat{H}_{\textrm{KE}}|a_{1g}^{s_{1}},e_{g}^{g_{e};s_{2}}\rangle$ 
and $\langle \boldi;A_{1}|\hat{H}_{\textrm{KE}}|e_{g}^{g_{e};s_{1}},a_{1g}^{s_{2}}\rangle$
for $\boldi=\boldone, \boldtwo$ and $s_{1},s_{2}=\uparrow,\downarrow$, 
and then expressed the contribution 
in terms of those matrix elements and the operators. 
Then,  
the calculated result for $\Gamma=E$ is 
\begin{widetext}
\begin{align}
&
-[\frac{1}{27}
\frac{(t_{1}-t_{2}-t_{3})[\omega^{n(g_{e})}(2t_{1}-2t_{2}+t_{3})-6t_{2}]
-9\omega^{m(g_{e})}t_{\textrm{odd}}^{2}}{U-J^{\prime}}\notag\\
&+\frac{1}{9}
\frac{(t_{1}-t_{2}-t_{3})[\omega^{n(g_{e})}(2t_{1}+2t_{2}+t_{3})+2t_{2}]
+3(\omega^{n(g_{e})}-1)t_{\textrm{odd}}^{2}}{U-J^{\prime}}]
[(\frac{1}{4}\hat{n}_{\boldone}\hat{o}_{\boldtwo}^{g_{e}}
-\hat{\bolds}_{\boldone}\cdot \hat{\boldso}_{\boldtwo}^{g_{e}})
+(\frac{1}{4}\hat{o}_{\boldone}^{g_{e}}\hat{n}_{\boldtwo}
-\hat{\boldso}_{\boldone}^{g_{e}}\cdot \hat{\bolds}_{\boldtwo})]\notag\\
&
-[\frac{1}{9}
\frac{t_{\textrm{odd}}[\omega^{m(g_{e})}(t_{1}-t_{2})+(3-4\omega^{m(g_{e})})t_{3}]}
{U-J^{\prime}}
+\frac{1}{9}
\frac{t_{\textrm{odd}}[(1-\omega^{n(g_{e})})(t_{1}-t_{2})+(2\omega^{m(g_{e})}-5)t_{3}]}
{U-J^{\prime}}]\notag\\
&\times 
[(\frac{1}{4}\hat{n}_{\boldone}\hat{o}_{\boldtwo}^{g_{e}}
-\hat{\bolds}_{\boldone}\cdot \hat{\boldso}_{\boldtwo}^{g_{e}})
-(\frac{1}{4}\hat{o}_{\boldone}^{g_{e}}\hat{n}_{\boldtwo}
-\hat{\boldso}_{\boldone}^{g_{e}}\cdot \hat{\bolds}_{\boldtwo}),\label{eq:H3rd-a1g-eg-E}
\end{align}
\end{widetext}
the calculated result for $\Gamma=T_{1}$ is 
\begin{widetext}
\begin{align}
&
-
[\frac{1}{9}
\frac{(t_{1}-t_{2}-t_{3})[-\omega^{m(g_{e})}(t_{1}-t_{2})+t_{3}]
+3(1-\omega^{m(g_{e})})t_{\textrm{odd}}^{2}}
{U^{\prime}-J_{\textrm{H}}}
+\frac{1}{9}
\frac{(t_{1}-t_{2}-t_{3})[-(t_{1}-t_{2})+\omega^{m(g_{e})}t_{3}]
+3(\omega^{m(g_{e})}-1)t_{\textrm{odd}}^{2}}
{U^{\prime}-J_{\textrm{H}}}
]\notag\\
&\times 
[(\frac{3}{4}\hat{n}_{\boldone}\hat{o}_{\boldtwo}^{g_{e}}
+\hat{\bolds}_{\boldone}\cdot \hat{\boldso}_{\boldtwo}^{g_{e}})
+(\frac{3}{4}\hat{o}_{\boldone}^{g_{e}}\hat{n}_{\boldtwo}
+\hat{\boldso}_{\boldone}^{g_{e}}\cdot \hat{\bolds}_{\boldtwo})]\notag\\
&
-
[\frac{1}{9}
\frac{t_{\textrm{odd}}[\omega^{n(g_{e})}(2t_{1}+4t_{2}+7t_{3})-(5t_{1}+t_{2}-5t_{3})]}
{U^{\prime}-J_{\textrm{H}}}
+\frac{1}{9}
\frac{t_{\textrm{odd}}[\omega^{n(g_{e})}(7t_{1}+5t_{2}+2t_{3})
+(5t_{1}+t_{2}-5t_{3})]}
{U^{\prime}-J_{\textrm{H}}}]\notag\\
&\times 
[(\frac{3}{4}\hat{n}_{\boldone}\hat{o}_{\boldtwo}^{g_{e}}
+\hat{\bolds}_{\boldone}\cdot \hat{\boldso}_{\boldtwo}^{g_{e}})
-(\frac{3}{4}\hat{o}_{\boldone}^{g_{e}}\hat{n}_{\boldtwo}
+\hat{\boldso}_{\boldone}^{g_{e}}\cdot \hat{\bolds}_{\boldtwo})],\label{eq:H3rd-a1g-eg-T1}
\end{align}
\end{widetext}
and 
the calculated result for $\Gamma=T_{2}$ is 
\begin{widetext}
\begin{align}
&
-
\{\frac{2}{9}
\frac{\omega^{n(g_{e})}[-2(t_{2}+t_{3})^{2}+t_{\textrm{odd}}^{2}]}
{U^{\prime}+J_{\textrm{H}}}
+\frac{1}{9}
\frac{(t_{1}+t_{2}+t_{3})
[-\omega^{m(g_{e})}(t_{1}+t_{2})-t_{3}]+(2-3\omega^{n(g_{e})})t_{\textrm{odd}}^{2}}
{U^{\prime}+J_{\textrm{H}}}\notag\\
&+\frac{1}{9}
\frac{(t_{1}+t_{2}+t_{3})[-(t_{1}+t_{2})-\omega^{m(g_{e})}t_{3}]
-(2+5\omega^{n(g_{e})})t_{\textrm{odd}}^{2}}
{U^{\prime}+J_{\textrm{H}}}\}
[(\frac{1}{4}\hat{n}_{\boldone}\hat{o}_{\boldtwo}^{g_{e}}
-\hat{\bolds}_{\boldone}\cdot \hat{\boldso}_{\boldtwo}^{g_{e}})
+(\frac{1}{4}\hat{o}_{\boldone}^{g_{e}}\hat{n}_{\boldtwo}
-\hat{\boldso}_{\boldone}^{g_{e}}\cdot \hat{\bolds}_{\boldtwo})]\notag\\
&
+\{\frac{2}{3}
\frac{\omega^{n(g_{e})}t_{\textrm{odd}}(t_{2}+t_{3})}
{U^{\prime}+J_{\textrm{H}}}
-\frac{1}{9}
\frac{t_{\textrm{odd}}[-\omega^{n(g_{e})}(2t_{2}+3t_{3})-(3t_{1}+t_{2}+3t_{3})]}
{U^{\prime}+J_{\textrm{H}}}
-\frac{1}{9}
\frac{t_{\textrm{odd}}[\omega^{n(g_{e})}(3t_{1}-t_{2})+(3t_{1}+t_{2}+3t_{3})]}
{U^{\prime}+J_{\textrm{H}}}\}\notag\\
&\times 
[(\frac{1}{4}\hat{n}_{\boldone}\hat{o}_{\boldtwo}^{g_{e}}
-\hat{\bolds}_{\boldone}\cdot \hat{\boldso}_{\boldtwo}^{g_{e}})
-(\frac{1}{4}\hat{o}_{\boldone}^{g_{e}}\hat{n}_{\boldtwo}
-\hat{\boldso}_{\boldone}^{g_{e}}\cdot \hat{\bolds}_{\boldtwo}).\label{eq:H3rd-a1g-eg-T2}
\end{align}
\end{widetext}
Combining all the above contributions with Eq. (\ref{eq:Hexch-rewrite}) 
and carrying out some algebra 
with the equality, 
such as $\omega^{m(g_{e})}+\omega^{n(g_{e})}=-1$, 
we finally obtain Eq. (\ref{eq:Hexch-explicit}) 
with Eqs. (\ref{eq:J-AFS}){--}(\ref{eq:J-FMA}).

\section{Argument for the importance of the multiorbital properties 
and mirror-mixing effect} 
In this appendix, 
we argue the importance of the multiorbital properties 
and mirror-mixing effect 
in the DM interaction. 
For that argument, 
we focus on the first term of Eq. (\ref{eq:HDM-a1g-eg}), 
i.e. $-D(\hat{S}_{\boldone}^{y}\hat{S}_{\boldtwo}^{z}-\hat{S}_{\boldone}^{z}\hat{S}_{\boldtwo}^{y})$. 
This term shows two differences in comparison with the Heisenberg-type interaction: 
one is about whether the interaction is odd about some coordinates 
or even about all; 
the other is about whether the interaction is antisymmetric or symmetric. 
Those two properties distinguish the DM interaction from the Heisenberg-type interaction 
in general. 
To obtain the odd and antisymmetric superexchange interaction, 
the multiorbital properties become vital 
because only the interorbital hopping integral becomes 
odd about some coordinates 
(the intraorbital hopping integrals are even about all the coordinates). 
For the case of the first term of Eq. (\ref{eq:HDM-a1g-eg}), 
we need the hopping integral behaving as an odd function about $y$ and $z$ 
and an even function about $x$. 
Such hopping integral is obtained by the hopping integral 
between the $d_{xz}$ and $d_{xy}$ orbitals because 
the $d_{xz}$ and $d_{xy}$ orbitals behave like $xz$ and $xy$, respectively. 
Then, 
the odd dependence of this hopping integral can hold in the superexchange interactions 
if we use this odd number of times. 
Since we should use another hopping integral to put the electron moved by this 
back into the initial site, 
the combination of the odd-mirror hopping integral and even-mirror hopping integral 
is necessary. 
Furthermore, 
we need the SOC to put the electron back into the initial (ground-state) orbital 
(for the case considered in this paper, 
move from the $e_{g}^{\pm}$ orbital to the $a_{1g}$ orbital). 
Thus, 
the multiorbital properties and mirror-mixing effect is important to obtain 
the DM interaction. 

\section{Details of the rough estimation of $\frac{D}{J^{\textrm{FM}}-J^{\textrm{AF}}}$}
In this appendix, 
we see the details of how to obtain 
the results shown in Figs. \ref{figAd2}(a){--}(d). 
The results are obtained as follows. 
First, 
by setting 
$A=t_{1}+t_{2}+t_{3}$, $B=t_{2}+t_{3}$, $J^{\prime}=J_{\textrm{H}}$ and $U^{\prime}=U-2J_{\textrm{H}}$ 
in Eq. (\ref{eq:D-a1g-eg}) 
and expanding $D$ as the power series of $\frac{J_{\textrm{H}}}{U}$ 
within the $O(\frac{J_{\textrm{H}}}{U})$ terms, 
we express $D$ as 
\begin{align}
D=\frac{12\lambda_{LS}t_{\textrm{odd}}}{9U\Delta_{\textrm{tri}}}
[(A-2B)+6A\frac{J_{\textrm{H}}}{U}].\label{eq:D-a1g-eg_rough-1} 
\end{align}
Second, 
by combining this equation with Eq. (\ref{eq:superexchange-simpler}), 
we obtain
\begin{align}
\frac{D}{J^{\textrm{FM}}-J^{\textrm{AF}}}
=
\frac{3\lambda_{LS}t_{\textrm{odd}}}{\Delta_{\textrm{tri}}}
\frac{(A-2B)+6\frac{J_{\textrm{H}}}{U}}
{(A+B)^{2}-2\frac{J_{\textrm{H}}}{U}[(A-2B)^{2}+9t_{\textrm{odd}}^{2}]}.
\end{align}
Third, 
to understand the rough dependence of $\frac{D}{J^{\textrm{FM}}-J^{\textrm{AF}}}$ 
on $\frac{J_{\textrm{H}}}{U}$ and $\frac{t_{\textrm{odd}}}{A}$ or $\frac{t_{\textrm{odd}}}{B}$, 
we consider four cases about $A$ and $B$, 
i.e., $A\gg B$, $A\ll B$, $A\sim B$, and $A\sim -B$. 
Fourth, 
by calculating the leading terms of 
$\frac{D}{J^{\textrm{FM}}-J^{\textrm{AF}}}$ in each case, 
the leading terms in $A\gg B$, $A\ll B$, $A\sim B$, and $A\sim -B$ are given by
\begin{align}
&\frac{D}{J^{\textrm{FM}}-J^{\textrm{AF}}}
=\frac{3\lambda_{LS}}{\Delta_{\textrm{tri}}}
\frac{t_{\textrm{odd}}}{A}
\frac{1+6\frac{J_{\textrm{H}}}{U}}
{1-2\frac{J_{\textrm{H}}}{U}(1+9\frac{t_{\textrm{odd}}^{2}}{A^{2}})},\label{eq:D-a1g-eg_rough-case1}\\
&\frac{D}{J^{\textrm{FM}}-J^{\textrm{AF}}}
=-\frac{6\lambda_{LS}}{\Delta_{\textrm{tri}}}
\frac{t_{\textrm{odd}}}{B}
\frac{1}
{1-2\frac{J_{\textrm{H}}}{U}(4+9\frac{t_{\textrm{odd}}^{2}}{A^{2}})},\label{eq:D-a1g-eg_rough-case2}\\
&\frac{D}{J^{\textrm{FM}}-J^{\textrm{AF}}}
=-\frac{3\lambda_{LS}}{2\Delta_{\textrm{tri}}}
\frac{t_{\textrm{odd}}}{B}
\frac{1-6\frac{J_{\textrm{H}}}{U}}
{2-\frac{J_{\textrm{H}}}{U}(1+9\frac{t_{\textrm{odd}}^{2}}{A^{2}})},\label{eq:D-a1g-eg_rough-case3}
\end{align}
and 
\begin{align}
&\frac{D}{J^{\textrm{FM}}-J^{\textrm{AF}}}
=\frac{\lambda_{LS}}{6\Delta_{\textrm{tri}}}
\frac{t_{\textrm{odd}}}{B}
\frac{1+2\frac{J_{\textrm{H}}}{U}}
{\frac{J_{\textrm{H}}}{U}(1+\frac{t_{\textrm{odd}}^{2}}{A^{2}})},\label{eq:D-a1g-eg_rough-case4}
\end{align}
respectively. 
By using those equations 
and setting $\lambda_{LS}=0.03$ eV and $\Delta_{\textrm{tri}}=1$ eV, 
we obtain the results shown in Figs. \ref{figAd2}(a){--}(d).

\section{Derivation of Eq. (\ref{eq:HDM-CoulombMultiplet})} 
In this appendix, 
we derive Eq. (\ref{eq:HDM-CoulombMultiplet}). 
For this derivation, 
we first rewrite the third term of Eq. (\ref{eq:3rd-total}) 
for sublattices $1$ and $2$ as 
\begin{align}
&(\hat{H}_{\textrm{3rd}}^{\textrm{KE}-LS-\textrm{KE}})_{\boldone\boldtwo}
\approx 
\sum\limits_{\boldi=\boldone,\boldtwo}
\sum\limits_{\Gamma,\Gamma^{\prime}(\neq \Gamma)}
\sum\limits_{g_{\Gamma},g_{\Gamma^{\prime}}}
\sum\limits_{s_{1},s_{2},s_{3},s_{4}}\notag\\
&\times 
\frac{\langle a_{1g}^{s_{3}},a_{1g}^{s_{4}}|\hat{H}_{\textrm{KE}}
|\boldi;\Gamma,g_{\Gamma}\rangle}
{-E_{\Gamma}}
\langle \boldi;\Gamma,g_{\Gamma}|\hat{H}_{LS}
|\boldi;\Gamma^{\prime},g_{\Gamma^{\prime}}\rangle \notag\\
&\times 
\frac{\langle \boldi;\Gamma^{\prime},g_{\Gamma^{\prime}}|
\hat{H}_{\textrm{KE}}|a_{1g}^{s_{1}},a_{1g}^{s_{2}}\rangle}
{-E_{\Gamma^{\prime}}}
|a_{1g}^{s_{3}},a_{1g}^{s_{4}}\rangle
\langle a_{1g}^{s_{1}},a_{1g}^{s_{2}}|.\label{eq:H3rd-KLK-rewrite}
\end{align}
This shows that 
$(\hat{H}_{\textrm{3rd}}^{\textrm{KE}-LS-\textrm{KE}})_{\boldone\boldtwo}$ 
for our model 
can be derived by combining 
the contributions from every term of $|\boldi;\Gamma^{\prime},g_{\Gamma^{\prime}}\rangle$. 
Those contributions 
can be calculated 
by using the matrix elements, 
$\langle a_{1g}^{s_{3}},a_{1g}^{s_{4}}|\hat{H}_{\textrm{KE}}
|\boldi;\Gamma,g_{\Gamma}\rangle$ and 
$\langle \boldi;\Gamma^{\prime},g_{\Gamma^{\prime}}|
\hat{H}_{\textrm{KE}}|a_{1g}^{s_{1}},a_{1g}^{s_{2}}\rangle$, 
which have been calculated 
in the second-order perturbation terms, 
and the matrix element, 
$\langle \boldi;\Gamma,g_{\Gamma}|\hat{H}_{LS}
|\boldi;\Gamma^{\prime},g_{\Gamma^{\prime}}\rangle$, 
which is obtained by using the relation, 
such as Eqs. (\ref{eq:HLS-Hint-1}){--}(\ref{eq:HLS-Hint-4}). 
Namely, 
the contribution from $|\boldi;\Gamma^{\prime},g_{\Gamma^{\prime}}\rangle
=|\boldi;A_{1}\rangle$ is 
\begin{widetext}
\begin{align}
\frac{2}{9}
\frac{\lambda_{LS}t_{\textrm{odd}}
(t_{1}+2t_{2}+2t_{3})}
{(U^{\prime}-J_{\textrm{H}})(U+2J^{\prime})}
\{
(1-i)
[(\frac{1}{2}\hat{n}_{\boldone}+\hat{s}_{\boldone}^{z})\hat{s}_{\boldtwo}^{+}
-s_{\boldone}^{+}(\frac{1}{2}\hat{n}_{\boldtwo}+\hat{s}_{\boldtwo}^{z})]
+(1+i)
[\hat{s}_{\boldone}^{-}(\frac{1}{2}\hat{n}_{\boldtwo}-\hat{s}_{\boldtwo}^{z})
-(\frac{1}{2}\hat{n}_{\boldone}-\hat{s}_{\boldone}^{z})\hat{s}_{\boldtwo}^{-}]
\},\label{eq:H3rd-intemulti-A1}
\end{align}
the contribution from $|\boldi;\Gamma^{\prime},g_{\Gamma^{\prime}}\rangle
=|\boldi;E,u\rangle$ is 
\begin{align}
&\frac{1}{9}
\frac{i\lambda_{LS}t_{\textrm{odd}}(t_{1}-t_{2}-t_{3})}
{(U^{\prime}-J_{\textrm{H}})(U-J^{\prime})}
[
(\frac{1}{2}\hat{n}_{\boldone}+\hat{s}_{\boldone}^{z})\hat{s}_{\boldtwo}^{+}
-\hat{s}_{\boldone}^{+}(\frac{1}{2}\hat{n}_{\boldtwo}+\hat{s}_{\boldtwo}^{z})
+(\frac{1}{2}\hat{n}_{\boldone}-\hat{s}_{\boldone}^{z})\hat{s}_{\boldtwo}^{-}
-\hat{s}_{\boldone}^{-}(\frac{1}{2}\hat{n}_{\boldtwo}-\hat{s}_{\boldtwo}^{z})
]\notag\\
&+\frac{2}{9}
\frac{\lambda_{LS}t_{\textrm{odd}}(t_{1}-t_{2}-t_{3})}
{(U^{\prime}-J_{\textrm{H}})(U-J^{\prime})}
[
(\frac{1}{2}\hat{n}_{\boldone}+\hat{s}_{\boldone}^{z})\hat{s}_{\boldtwo}^{+}
-\hat{s}_{\boldone}^{+}(\frac{1}{2}\hat{n}_{\boldtwo}+\hat{s}_{\boldtwo}^{z})
-(\frac{1}{2}\hat{n}_{\boldone}-\hat{s}_{\boldone}^{z})\hat{s}_{\boldtwo}^{-}
+\hat{s}_{\boldone}^{-}(\frac{1}{2}\hat{n}_{\boldtwo}-\hat{s}_{\boldtwo}^{z})
],\label{eq:H3rd-intemulti-Eu}
\end{align}
the contribution from $|\boldi;\Gamma^{\prime},g_{\Gamma^{\prime}}\rangle
=|\boldi;E,v\rangle$ is 
\begin{align}
-\frac{1}{3}
\frac{i\lambda_{LS}t_{\textrm{odd}}(t_{1}-t_{2}-t_{3})}
{(U^{\prime}-J_{\textrm{H}})(U-J^{\prime})}
[
(\frac{1}{2}\hat{n}_{\boldone}+\hat{s}_{\boldone}^{z})\hat{s}_{\boldtwo}^{+}
-\hat{s}_{\boldone}^{+}(\frac{1}{2}\hat{n}_{\boldtwo}+\hat{s}_{\boldtwo}^{z})
+(\frac{1}{2}\hat{n}_{\boldone}-\hat{s}_{\boldone}^{z})\hat{s}_{\boldtwo}^{-}
-\hat{s}_{\boldone}^{-}(\frac{1}{2}\hat{n}_{\boldtwo}-\hat{s}_{\boldtwo}^{z})
],\label{eq:H3rd-intemulti-Ev}
\end{align}
the contribution from $|\boldi;\Gamma^{\prime},g_{\Gamma^{\prime}}\rangle
=|\boldi;T_{1},\xi_{\pm}\rangle$ is 
\begin{align}
&[\frac{2}{9}
\frac{i\lambda_{LS}t_{\textrm{odd}}(t_{1}+2t_{2}+2t_{3})}{(U+2J^{\prime})(U^{\prime}-J_{\textrm{H}})}
-\frac{1}{9}
\frac{i\lambda_{LS}t_{\textrm{odd}}(t_{1}-t_{2}-t_{3})}{(U-J^{\prime})(U^{\prime}-J_{\textrm{H}})}
+\frac{1}{3}
\frac{i\lambda_{LS}t_{\textrm{odd}}(t_{1}-t_{2}-t_{3})}{(U-J^{\prime})(U^{\prime}-J_{\textrm{H}})}
\pm \frac{1}{9}
\frac{\lambda_{LS}t_{\textrm{odd}}(-t_{1}+4t_{2}+4t_{3})}
{(U^{\prime}+J_{\textrm{H}})(U^{\prime}-J_{\textrm{H}})}]\notag\\
&\times 
[(\frac{1}{2}\hat{n}_{\boldone}\pm \hat{s}_{\boldone}^{z})\hat{s}_{\boldtwo}^{\mp}
-\hat{s}_{\boldone}^{\mp}(\frac{1}{2}\hat{n}_{\boldtwo}\pm\hat{s}_{\boldtwo}^{z})],
\label{eq:H3rd-intemulti-T1-13}
\end{align}
the contribution from $|\boldi;\Gamma^{\prime},g_{\Gamma^{\prime}}\rangle
=|\boldi;T_{1},\eta_{\pm}\rangle$ is 
\begin{align}
&[
\frac{2}{9}
\frac{\lambda_{LS}t_{\textrm{odd}}(t_{1}+2t_{2}+2t_{3})}
{(U+2J^{\prime})(U^{\prime}-J_{\textrm{H}})}
+\frac{2}{9}
\frac{\lambda_{LS}t_{\textrm{odd}}(t_{1}-t_{2}-t_{3})}
{(U-J^{\prime})(U^{\prime}-J_{\textrm{H}})}
\pm
\frac{1}{9}
\frac{i\lambda_{LS}t_{\textrm{odd}}(-t_{1}+4t_{2}+4t_{3})}
{(U^{\prime}+J_{\textrm{H}})(U^{\prime}-J_{\textrm{H}})}
]
[\pm(\frac{1}{2}\hat{n}_{\boldone}\pm \hat{s}_{\boldone}^{z}) 
\hat{s}_{\boldtwo}^{\mp} 
\mp\hat{s}_{\boldone}^{\mp}(\frac{1}{2}\hat{n}_{\boldtwo}\pm \hat{s}_{\boldtwo}^{z})],
\label{eq:H3rd-intemulti-T1-23}
\end{align}
the contribution from $|\boldi;\Gamma^{\prime},g_{\Gamma^{\prime}}\rangle
=|\boldi;T_{2},\zeta_{0}\rangle$ is 
\begin{align}
\frac{1}{9}
\frac{\lambda_{LS}t_{\textrm{odd}}(-t_{1}+4t_{2}+4t_{3})}
{(U^{\prime}-J_{\textrm{H}})(U^{\prime}+J_{\textrm{H}})}
\{
(1-i)[
(\frac{1}{2}\hat{n}_{\boldone}+\hat{s}_{\boldone}^{z})\hat{s}_{\boldtwo}^{+}
-
\hat{s}_{\boldone}^{+}(\frac{1}{2}\hat{n}_{\boldtwo}+\hat{s}_{\boldtwo}^{z})]
+
(1+i)[
\hat{s}_{\boldone}^{-}(\frac{1}{2}\hat{n}_{\boldtwo}-\hat{s}_{\boldtwo}^{z})
-
(\frac{1}{2}\hat{n}_{\boldone}-\hat{s}_{\boldone}^{z})\hat{s}_{\boldtwo}^{-}]\},
\label{eq:H3rd-intemulti-T2-12}
\end{align}
the contribution from $|\boldi;\Gamma^{\prime},g_{\Gamma^{\prime}}\rangle
=|\boldi;T_{1},\xi_{0}\rangle$ is 
\begin{align}
\frac{1}{9}
\frac{i\lambda_{LS}t_{\textrm{odd}}(2t_{1}+t_{2}+t_{3})}
{(U^{\prime}+J_{\textrm{H}})(U^{\prime}-J_{\textrm{H}})}
[(\frac{1}{2}\hat{n}_{\boldone}+\hat{s}_{\boldone}^{z})
(\frac{1}{2}\hat{n}_{\boldtwo}-\hat{s}_{\boldone}^{z})
+\hat{s}_{\boldone}^{+}\hat{s}_{\boldtwo}^{-}
-\hat{s}_{\boldone}^{-}\hat{s}_{\boldtwo}^{+}
-(\frac{1}{2}\hat{n}_{\boldone}-\hat{s}_{\boldone}^{z})
(\frac{1}{2}\hat{n}_{\boldtwo}+\hat{s}_{\boldtwo}^{z})],
\label{eq:H3rd-intemulti-T1-13s0}
\end{align}
the contribution from $|\boldi;\Gamma^{\prime},g_{\Gamma^{\prime}}\rangle
=|\boldi;T_{2},\xi_{0}\rangle$ is 
\begin{align}
\frac{1}{9}
\frac{i\lambda_{LS}t_{\textrm{odd}}(2t_{1}+t_{2}+t_{3})}
{(U^{\prime}+J_{\textrm{H}})(U^{\prime}-J_{\textrm{H}})}
[(\frac{1}{2}\hat{n}_{\boldone}+\hat{s}_{\boldone}^{z})
(\frac{1}{2}\hat{n}_{\boldtwo}-\hat{s}_{\boldone}^{z})
-\hat{s}_{\boldone}^{+}\hat{s}_{\boldtwo}^{-}
+\hat{s}_{\boldone}^{-}\hat{s}_{\boldtwo}^{+}
-(\frac{1}{2}\hat{n}_{\boldone}-\hat{s}_{\boldone}^{z})
(\frac{1}{2}\hat{n}_{\boldtwo}+\hat{s}_{\boldtwo}^{z})],
\label{eq:H3rd-intemulti-T2-13}
\end{align}
the contribution from $|\boldi;\Gamma^{\prime},g_{\Gamma^{\prime}}\rangle
=|\boldi;T_{1},\eta_{0}\rangle$ is 
\begin{align}
-\frac{1}{9}
\frac{i\lambda_{LS}t_{\textrm{odd}}(2t_{1}+t_{2}+t_{3})}
{(U^{\prime}+J_{\textrm{H}})(U^{\prime}-J_{\textrm{H}})}
[(\frac{1}{2}\hat{n}_{\boldone}+\hat{s}_{\boldone}^{z})
(\frac{1}{2}\hat{n}_{\boldtwo}-\hat{s}_{\boldone}^{z})
+\hat{s}_{\boldone}^{+}\hat{s}_{\boldtwo}^{-}
-\hat{s}_{\boldone}^{-}\hat{s}_{\boldtwo}^{+}
-(\frac{1}{2}\hat{n}_{\boldone}-\hat{s}_{\boldone}^{z})
(\frac{1}{2}\hat{n}_{\boldtwo}+\hat{s}_{\boldtwo}^{z})],
\label{eq:H3rd-intemulti-T1-23s0}
\end{align}
the contribution from $|\boldi;\Gamma^{\prime},g_{\Gamma^{\prime}}\rangle
=|\boldi;T_{2},\eta_{0}\rangle$ is 
\begin{align}
-\frac{1}{9}
\frac{i\lambda_{LS}t_{\textrm{odd}}(2t_{1}+t_{2}+t_{3})}
{(U^{\prime}+J_{\textrm{H}})(U^{\prime}-J_{\textrm{H}})}
[(\frac{1}{2}\hat{n}_{\boldone}+\hat{s}_{\boldone}^{z})
(\frac{1}{2}\hat{n}_{\boldtwo}-\hat{s}_{\boldone}^{z})
-\hat{s}_{\boldone}^{+}\hat{s}_{\boldtwo}^{-}
+\hat{s}_{\boldone}^{-}\hat{s}_{\boldtwo}^{+}
-(\frac{1}{2}\hat{n}_{\boldone}-\hat{s}_{\boldone}^{z})
(\frac{1}{2}\hat{n}_{\boldtwo}+\hat{s}_{\boldtwo}^{z})],
\label{eq:H3rd-intemulti-T2-23}
\end{align}
\end{widetext}
and the contributions from $|\boldi;\Gamma^{\prime},g_{\Gamma^{\prime}}\rangle
=|\boldi;T_{1},\zeta_{\pm}\rangle, |\boldi;T_{1},\zeta_{0}\rangle$ are zero 
for $(\hat{H}_{\textrm{3rd}}^{\textrm{KE}-LS-\textrm{KE}})_{\boldone\boldtwo}$. 
Combining those results with Eq. (\ref{eq:H3rd-KLK-rewrite}),  
we obtain Eq. (\ref{eq:HDM-CoulombMultiplet}) with Eq. (\ref{eq:D-CoulombMultiplet}).

\end{document}